\documentclass[11pt,twoside]{book}
\usepackage[T1]{fontenc}
\usepackage{geometry}
\usepackage{amsmath}
\usepackage{graphics}

\makeatletter

\newcommand{\LyX}{L\kern-.1667em\lower.25em\hbox{Y}\kern-.125emX\spacefactor1000}
\newcommand{\noun}[1]{\textsc{#1}}
\let\SF@@footnote\footnote
\def\footnote{\ifx\protect\@typeset@protect
    \expandafter\SF@@footnote
  \else
    \expandafter\SF@gobble@opt
  \fi
}
\expandafter\def\csname SF@gobble@opt \endcsname{\@ifnextchar[
  \SF@gobble@twobracket
  \@gobble
}
\edef\SF@gobble@opt{\noexpand\protect
  \expandafter\noexpand\csname SF@gobble@opt \endcsname}
\def\SF@gobble@twobracket[#1]#2{}

\usepackage{amssymb}
\makeatother

\begin{document}


\vspace*{5mm}

{\noindent \centering {\large UNIVERSIT\`A DEGLI STUDI DI BOLOGNA } \par}

\vspace*{0.5 cm}

{\noindent \centering {\large UNIVERSITY OF BOLOGNA } \par}

\vspace*{2.5cm}

{\noindent \centering {\large TESI DI DOTTORATO DI RICERCA} \par}

\vspace*{0.5cm}

{\noindent \centering {\large PhD THESIS } \par}

\vspace*{2.5cm}

{\noindent \centering \textbf{\Huge FINITE VOLUME SPECTRUM }\Huge \par}

\vspace{5mm}

{\noindent \centering \textbf{\Huge OF SINE-GORDON MODEL}\Huge \par}

\vspace{5mm}

{\noindent \centering \textbf{\Huge AND ITS RESTRICTIONS}\Huge \par}

\vspace{25mm}

{\noindent \centering {\large Dr. Giovanni FEVERATI \footnotemark } \par}

\vspace{25mm}

{\noindent \centering JANUARY 2000 \par}

\footnotetext{e-mail: feverati@bo.infn.it} 

\thispagestyle{empty}

\newpage

\vspace*{15 cm}

\newpage

\tableofcontents

\chapter{SOME GENERAL FACTS }

After a short introduction, the most important known facts about sine-Gordon
and massive Thirring models are exposed. It is also explained their connection
with the \( c=1 \) free boson.

\section{Introduction}

As a very large number of papers in literature, this thesis principally deals
with the sine-Gordon model, which has been well known at the classical level
for the late fifty years and plays also an important role in quantum theory,
thanks to its particular properties of non-linearity and integrability. It has
been successfully applied in very different sectors of Mathematics and Physics,
from partial differential equation theory to particle physics or solid state
physics. Recent applications of the classical model are related to nonlinear
optics (resonant dielectric media) and optical fibers, magnetic properties of
polymers, propagation of waves in crystals and so on. Interesting applications
of the quantum model are related to Kondo effect and to the thermodynamics of
some chemical compound, as can be found in \cite{copper_benzoate} (see also
section \ref{section:conclusione}). At the same time, the quantum theory shows
a phenomenology that is similar to the Skyrme model used before QCD era to describe
barions and strong interactions. 

The most relevant properties of the model are

\begin{itemize}
\item at a classical level, all the solutions of the equations of motion are known
(exact integrability via inverse scattering method)
\item the classical solutions describe solitons, antisolitons and bound states (breathers)\footnotemark{}
; in a scattering process this solutions are transparent (it is the mathematical
meaning of ``soliton'')
\item it admits, both at a classical and at the quantum level, a countable infinite
set of conserved charges
\item the quantization of the theory describes an interacting particle with its antiparticle
and, in a certain (attractive) regime, bound states
\item the S matrix has been exactly determined; only elastic scattering processes
can take place (i.e. no particle production), that is the quantum analog of
the classical transparency of solitons
\end{itemize}
\footnotetext{
The so called mesonic solutions are excluded in this analysis.
}%

\subsection*{Finite size effects}

Finite size effects are widely recognized to play an important role in modern
statistical mechanics and quantum field theory. From a statistical point of
view, it is known that no phase transitions take place in a finite volume system.
For example, specific heat \( c(T) \), that is divergent at the critical point,
if the system has finite size looses it divergence; one observes only a rounded
peak, in the plot \( c(T) \) versus \( T \). Moreover, there is only an interval
around critical temperature \( T_{c} \) where the finite size effects are relevant.
Out of this interval, they are negligible (because only near \( T_{c} \) the
correlation length can be comparable with the size of the system). The interesting
fact is that specific heat (and other critical quantities) have a scaling behaviour
(i.e. varying the size \( L \)) that is fixed by the (infinite size) critical
exponents (see \cite{christe_henkel}). This is a general fact: as argued in
\cite{cardy86}, the UV behaviour of the scaling functions (see later) is fixed
by the conformal dimensions of the operators that belong to the universality
class of the critical point (i.e. the CFT describing the critical point of the
statistical system).

\begin{figure}[  htbp]
{\centering \includegraphics{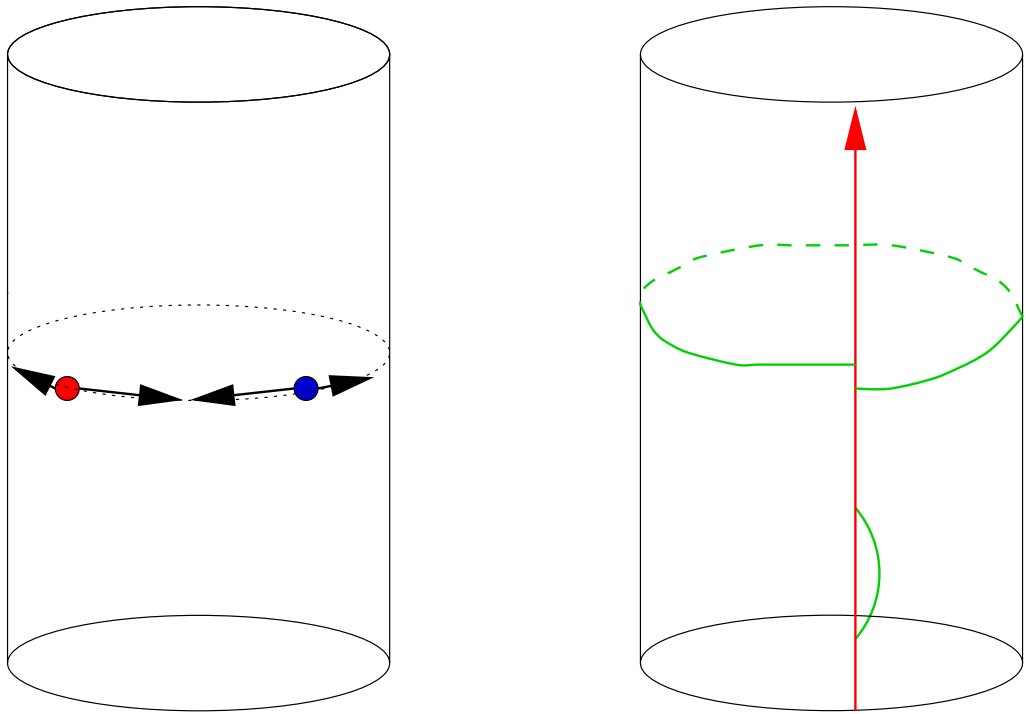} \par}

\caption{\small Finite size effects\label{F.S.E..eps}}
\end{figure}

Also in quantum field theory interesting phenomena appear. If the space-time
geometry is a cylinder of circumference \( L \), Casimir effects change the
energy of a two body interaction, because particles interact in the two possible
directions, as shown in figure \ref{F.S.E..eps} (looking forward one can see
his own back). Also new radiative corrections to a propagating particle may
appear because of the closed geometry.A system on a finite volume has discrete energy and momentum spectra. L\"uscher
\cite{luscher_1, luscher_2} has show that the corrections to the free system
eigenvalues of \( E,\: P \) depends on the scattering amplitudes (i.e. on the
\( S \) matrix). This general fact can be used to extract informations on an
infinite space QFT from a numerical simulation, that obviously is affected by
finite size effects. 

Notice that a cylinder geometry can also be interpreted as a finite temperature
field theory (temperature is \( L^{-1} \)). Such a theory is related to the
phase transition between confined and deconfined phase in QCD. 

In general, a physical quantity (for example the free energy) that depends on
the renormalization scale \( L \) (it parameterizes the renormalization flow)
is a \emph{scaling function}. The energy and momentum computed in (\ref{energia},
\ref{momento}) are scaling functions, because in the NLIE the scale \( L \)
appears. 

There are many methods of investigation of the finite size effects. One of them,
that has been very fruitful even for theories which are not integrable, is the
\emph{truncated conformal space approach} (see \cite{yurov-zam} and section
\ref{section:TCSA}), which is an intrinsically non-perturbative approximation
method. It has problems of principal nature, coming from the fact that one does
not have an analytic control of the spectrum, and of practical nature, because
to reach a certain precision in the resulting energy levels one has sometimes
to resort to very high truncation levels and introduce enormous matrices to
diagonalize. 

For integrable QFT, there also exist \emph{exact analytic} methods to compute
the finite size effects, like e.g. the \emph{Thermodynamic Bethe Ansatz} (TBA),
which was used to calculate the vacuum (Casimir) energy \cite{zam_TBA}. The
method was later extended to include ground states of charged sectors \cite{Fendley}.
More recently, using analytic properties of the TBA equations extended for complex
values of the volume parameter, an approach to get excited states was proposed
in \cite{dorey-tateo}. Their method to get excited states sheds light on the
analytic structure of the dependence of scaling functions on the spatial volume
and up to now was the only method developed to deal with excited states in perturbations
of minimal models. Its main drawback is that to obtain the equation for a given
excited state one has to do analytic continuation for each case separately,
and a major part of this continuation can only be carried out numerically. Because
of the complications of the analytic continuation, this method is limited at
present to simple cases of integrable perturbations of Virasoro minimal models
and some other perturbed conformal field theories. Similar results were obtained
in \cite{BLZ3, susy}. 

This thesis reports on a novel approach to the excited states of IQFTs in finite
volume, based on the \emph{nonlinear integral equation} (NLIE) method, which
has its origin in the so-called \emph{light-cone lattice Bethe Ansatz} approach
to regularize integrable QFTs. It was argued in \cite{ddv 87} that sine-Gordon
theory can be regularized using an inhomogeneous \( 6 \)-vertex model (or equivalently,
an inhomogeneous \( XXZ \) chain). The NLIE was originally developed in this
framework to describe the ground state scaling function (Casimir energy) in
sine-Gordon theory in \cite{ddv 95} and it was shown that in the ultraviolet
limit it reproduces the correct value of the central charge \( c=1 \). Similar
methods were independently introduced in Condensed Matter Physics by other authors
\cite{klumper}. 

The NLIE was first extended to excited states in \cite{fioravanti} where the
spectrum of states containing only solitons (and no antisolitons/breathers)
has been described. Using an idea by Zamolodchikov \cite{polymer} they also
showed that a twisted version of the equation was able to describe ground states
of unitary Virasoro minimal models perturbed by the operator \( \Phi _{(1,3)} \).
A framework for generic excited states of even topological charge in sine-Gordon
theory was outlined by Destri and de Vega in \cite{ddv 97}. However, there
has been a contradiction between the results of the two papers, which was resolved
in \cite{noi PL1, noi NP} where it was shown that it was related to the locality
and the operator content of limiting ultraviolet conformal field theory (CFT).
Besides that, strong evidence for the correctness of the predicted spectrum
was given by comparing it to predictions coming from the truncated conformal
space (TCS) method, pioneered by Yurov and Zamolodchikov in \cite{yurov-zam}
and extended to \( c=1 \) theories in \cite{noi PL1, noi NP}. Later a modification
of the NLIE to describe the states of sine-Gordon/massive Thirring theory with
odd topological charge was conjectured \cite{noi PL2}.

The NLIE for sine-Gordon theory was generalized to models built on general simply-laced
algebras of \( ADE \) type in \cite{mariottini} for the case of the vacuum.
More recently, in \cite{zinn-justin} P. Zinn-Justin extended the method to
the spectrum of excited states for these models and he also made a first attempt
to describe perturbations of minimal models of CFT. A general framework for
describing general excited states of minimal models perturbed by \( \Phi _{(1,3)} \)
(only massive case) can be found in \cite{noi NP2}, where is stated the correct
form of NLIE equation to deal with this case, and also the simplest examples
of the resulting excited states are checked (see chapter \ref{chapter:analisis_NLIE}).

In the following chapters, the general setup of NLIE will be presented, with
the most relevant examples. 

Chapter 1 is devoted to summarize some well known facts that will be used in
the following. 

In Chapter 2 the light-cone lattice is introduced and the Bethe equations for
the 6 vertex model are written. 

Chapter 3 is devoted to obtain an integral equation equivalent to Bethe Ansatz,
but that allows a continuum limit procedure. The analysis of the so obtained
continuum theory is in Chapter 4.

\section{c=1 CFT: free boson\label{section:free_boson}}

To fix some conventions and to define certain objects which are used later,
a brief summary is given of the \( c=1 \) free boson with a target space of
a circle of radius \( R \). The Lagrangian of this CFT is taken to be

\begin{equation}
\label{bosone_libero}
{\cal L}=\displaystyle\frac{1}{8\pi }\displaystyle\int ^{L}_{0}\partial _{\mu }\varphi (x,\, t)\partial ^{\mu }\varphi (x,\, t)dx\: ,\: x\in [0,L]\: ,
\end{equation}
where \( L \) is the spatial volume (i.e. the theory is defined on a cylindrical
spacetime with circumference \( L \)). In the sequel often the complex Euclidean
coordinates will be used \( z=e^{2\pi (t-ix)/L},\: \bar{z}=e^{2\pi (t+ix)/L} \).
The superselection sectors are classified by the \( \widehat{U(1)}_{L}\times \widehat{U(1)}_{R} \)
Kac-Moody symmetry algebra, generated by the currents

\[
J(z)=i\partial _{z}\varphi \, ,\quad \bar{J}(\bar{z})=i\partial _{\bar{z}}\varphi \, .\]
The left/right moving energy-momentum tensor is given by
\[
T(z)=\frac{1}{8\pi }\partial _{z}\varphi \partial _{z}\varphi =\sum ^{\infty }_{k=-\infty }L_{k}z^{-k-2}\quad ,\quad \bar{T}(\bar{z})=\frac{1}{8\pi }\partial _{\bar{z}}\varphi \partial _{\bar{z}}\varphi =\sum ^{\infty }_{k=-\infty }\bar{L}_{k}\bar{z}^{-k-2}\]
The coefficients \( L_{n} \) and \( \bar{L}_{n} \) of the Laurent expansion
of these fields generate two mutually commuting Virasoro algebras. If the (quasi)periodic
boundary conditions are required

\[
\varphi (x+L,\, t)=\varphi (x,\, t)+2\pi mR\, ,\quad m\in \mathbb {Z}\, ,\]
then the sectors are labelled by a pair of numbers \( (n,\, m) \), where \( \frac{n}{R} \)
(\( n \) is half integer because of the locality, see later) is the eigenvalue
of the total field momentum \( \pi _{0} \)

\[
\pi _{0}=\int ^{L}_{0}\pi (x,\, t)dx\, ,\quad \pi (x,\, t)=\frac{1}{4\pi }\partial _{t}\varphi (x,\, t)\: ,\]
and \( m \) is the winding number, i.e. the eigenvalue of the topological charge
\( Q \) defined by

\[
Q=\frac{1}{2\pi R}\int ^{L}_{0}\partial _{x}\varphi (x,\, t)dx\, .\]
In the sector with quantum numbers \( (n,\, m) \), the scalar field is expanded
in modes as follows: 
\[
\begin{array}{rl}
\displaystyle \varphi (x,t)= & \phi (z)+\bar{\phi }(\bar{z})\: ,\\
\phi (z)= & \displaystyle\frac{1}{2}\varphi _{0}-ip_{+}\log z+i\displaystyle\sum _{k\neq 0}\displaystyle\frac{1}{k}a_{k}z^{-k}\: ,\\
\bar{\phi }(\bar{z})= & \displaystyle\frac{1}{2}\varphi _{0}-ip_{-}\log \bar{z}+i\displaystyle\sum _{k\neq 0}\displaystyle\frac{1}{k}\bar{a}_{k}\bar{z}^{-k}\: ,
\end{array}\]
where the left and right moving field momenta \( p_{\pm } \) (which are in
fact the two \( U(1) \) Kac-Moody charges) are given by

\begin{equation}
\label{p+-}
p_{\pm }=\displaystyle\frac{n}{R}\pm \displaystyle\frac{1}{2}mR\, .
\end{equation}
The Virasoro generators take the form

\[
L_{n}=\frac{1}{2}\sum ^{\infty }_{k=-\infty }:a_{n-k}a_{k}:\, ,\quad \bar{L}_{n}=\frac{1}{2}\sum ^{\infty }_{k=-\infty }:\bar{a}_{n-k}\bar{a}_{k}:\, ,\]
where the colons denote the usual normal ordering, according to which the oscillator
with the larger index is put to the right. 

The ground states of the different sectors \( (n,\, m) \) are created from
the vacuum by the (Kac-Moody) primary fields, which are vertex operators of
the form

\begin{equation}
\label{vertex_operators}
V_{(n,\, m)}(z,\overline{z})=:\exp i(p_{+}\phi (z)+p_{-}\bar{\phi }(\overline{z})):\: .
\end{equation}
The left and right conformal weights of the field \( V_{(n,\, m)} \) (i.e.
the eigenvalues of \( L_{0} \) and \( \bar{L}_{0} \)) are given by the formulae

\begin{equation}
\label{delta+-_vertex_op}
\Delta ^{\pm }=\displaystyle\frac{p^{2}_{\pm }}{2}.
\end{equation}
The Hilbert space of the theory is given by the direct sum of the Fock modules
built over the states

\begin{equation}
\label{fock_states}
\left| n,\, m\right\rangle =V_{(n,\, m)}(0,0)\left| vac\right\rangle \: ,
\end{equation}
with the help of the creation operators \( a_{-k}\: ,\: \bar{a}_{-k}\: k>0 \):

\[
{\cal H}=\bigoplus _{(n,\, m)}\{a_{-k_{1}}\ldots a_{-k_{p}}\bar{a}_{-l_{1}}\ldots \bar{a}_{-l_{q}}|n,\, m\rangle ,\, k_{1},\ldots \, k_{p},\, l_{1},\ldots \, l_{q}\in \mathbb {Z}_{+}\}\]
The boson Hamiltonian on the cylinder is expressed in terms of the Virasoro
operators as

\begin{equation}
\label{cylinder_hamilt}
H_{CFT}=\displaystyle\frac{2\pi }{L}\left( L_{0}+\overline{L}_{0}-\displaystyle\frac{c}{12}\right) \: ,
\end{equation}
where the central charge is \( c=1 \). The generator of spatial translations
is given by

\begin{equation}
\label{cylinder_moment}
P=\displaystyle\frac{2\pi }{L}\left( L_{0}-\bar{L}_{0}\right) \; .
\end{equation}
The operator \( L_{0}-\bar{L}_{0} \) is the conformal spin which has eigenvalue
\( nm \) on the primary field \( V_{(n,\: m)} \).

One can also introduce twisted sectors using the operator \( {\cal T} \) that
performs spatial translations by \( L \): \( x\rightarrow x+L \). The primary
fields \( V_{(n,\, m)} \) as defined above satisfy the periodicity condition
\( {\cal T}V_{(n,\, m)}=V_{(n,\, m)}. \) If the more general twisted boundary
condition labelled by a real parameter \( \nu  \) is required 
\[
{\cal T}V_{(n,\, m)}=\exp \left( i\nu Q\right) V_{(n,\, m)}\, ,\]
then it is possible to generate superselection sectors for which \( n\in \mathbb {Z}+\frac{\nu }{2\pi } \). 

It is important to stress that a particular \( c=1 \) CFT is specified by giving
the spectrum of the quantum numbers \( (n,m) \) (and the compactification radius
\( R \)) such that the corresponding set of vertex operators (and their descendants)
forms a \emph{closed and local} operator algebra. The locality requirement is
equivalent to the fact that the operator product expansions of any two such
local operators is single valued in the complex plane of \( z \). This condition,
which is weaker than the modular invariance of the CFT, is the adequate one
since the theory is considered on a space-time cylinder and do not wish to define
it on higher genus surfaces.

By this requirement of locality, it was proved in \cite{kl-me} that there are
only two maximal local subalgebras of vertex operators: \( {\cal A}_{b} \)
generated by the vertex operators 
\[
\{V_{(n,\, m)}:\, n,\, m\in \mathbb {Z}\}\, ,\]
and \( {\cal A}_{f} \) generated by
\[
\{V_{(n,\, m)}:\, n\in \mathbb {Z},\, m\in 2\mathbb {Z}\: or\: n\in \mathbb {Z}+\frac{1}{2},\, m\in 2\mathbb {Z}+1\}\, .\]
Other sets of vertex operators can be built, but the product of two of them
gives a nonlocal expression.\begin{figure}[  htbp]
{\centering \includegraphics{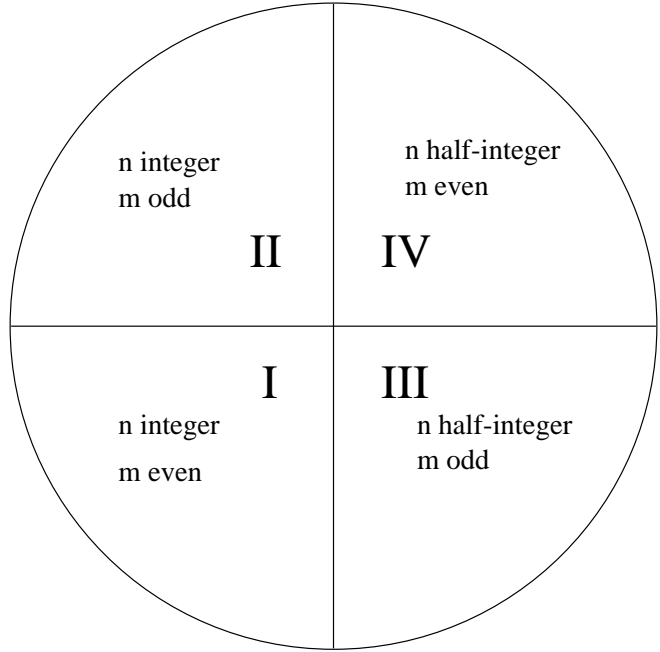} \par}

\caption{\small The family of vertex operators \protect\( V_{(n,\, m)}\protect \) with \protect\( n\in \mathbb Z/2\protect \)
and \protect\( m\in \mathbb Z\protect \). Sector \textbf{I} is the common subspace.
\textbf{I} and \textbf{III} are \protect\( {\cal A}_{f}\protect \), that defines
the UV behaviour of massive Thirring; \textbf{I} and \textbf{II} are \protect\( {\cal A}_{b}\protect \)
that is the UV of sine-Gordon, \textbf{IV} is a sector of non mutually local
vertex operators.\label{4sectors.eps}}
\end{figure}

\section{Sine-Gordon/massive Thirring field theory\label{section:sG-mT}}

The minkowskian lagrangian of sine-Gordon theory is given by\footnote{
Integration sign without any specification means integration on the whole real
axis.
}

\begin{equation}
\label{sG_Lagrangian}
{\cal L}_{sG}=\displaystyle\int \left( \displaystyle\frac{1}{2}\partial _{\nu }\Phi \partial ^{\nu }\Phi +\displaystyle\frac{\mu ^{2}}{\beta ^{2}}:\cos \left( \beta \Phi \right) :\right) dx\, ,
\end{equation}
where \( \Phi  \) denotes a real scalar field, while that of the massive Thirring
theory is of the following form:

\begin{equation}
\label{mTh_Lagrangian}
{\cal L}_{mTh}=\displaystyle\int \left( \bar{\Psi }(i\gamma _{\nu }\partial ^{\nu }+m_{0})\Psi -\displaystyle\frac{g}{2}\bar{\Psi }\gamma ^{\nu }\Psi \bar{\Psi }\gamma _{\nu }\Psi \right) dx\, ,
\end{equation}
describing a current-current selfinteraction of a Dirac fermion \( \Psi  \).
It is known that the two theories are deeply related provided their coupling
constants satisfy 
\[
\frac{\beta ^{2}}{4\pi }=\frac{1}{1+g/\pi }\, .\]
For comparison with the Destri-de Vega nonlinear integral equation, it is important
to deal with cylindrical sine-Gordon and massive Thirring, i.e. the integrals
in (\ref{sG_Lagrangian}, \ref{mTh_Lagrangian}) must be taken in the interval
\( [0,\, L] \). 

The \( \cos  \) term in (\ref{sG_Lagrangian}) can be considered as a perturbation
of the \( c=1 \) free boson compactified on a cylinder, as described in section
\ref{section:free_boson}. Similarly the massless (\( m_{0}=0 \)) Thirring
model is a \( c=1 \) conformal field theory and the mass term plays the role
of a perturbation. From Coleman's paper \cite{s-coleman} it is known that correlation
functions of the perturbing fields \( \bar{\Psi }\Psi  \) and \( :\cos \beta \Phi : \)
are identical, then both models can be considered as the perturbations of a
 \( c=1 \) compactified boson by a potential \( V \) :

\begin{equation}
\label{hamilt_perturb}
H_{sG/mTh}=H_{CFT}+V\quad ,\quad V=\lambda \displaystyle\int _{0}^{L}\left( V_{(1,0)}(z,\overline{z})+V_{(-1,0)}(z,\overline{z})\right) dx\: ,
\end{equation}
which is related to the bosonic lagrangian (\ref{sG_Lagrangian}) by the following
redefinitions of the field and the parameters:

\begin{equation}
\label{rinomina}
\varphi =\sqrt{4\pi }\Phi \: ,\quad R=\displaystyle\frac{\sqrt{4\pi }}{\beta }\: ,\quad \lambda =\displaystyle\frac{\mu ^{2}}{2\beta ^{2}}\: .
\end{equation}
For later convenience, a new parameter \( p \) can be defined by

\begin{equation}
\label{parametro_p}
p=\displaystyle\frac{\beta ^{2}}{8\pi -\beta ^{2}}=\displaystyle\frac{1}{2R^{2}-1}\: .
\end{equation}
The point \( p=1 \) (i.e. \( g=0 \)) is the free fermion point, corresponding
to a massive Dirac free fermion. The particle spectrum of sG for \( p>1 \)
is composed by the soliton (\( s \)) and its antiparticle, the antisoliton
(\( \bar{s} \)). It is known as repulsive regime because no bound states can
take place. \( p<1 \) is the attractive regime, because \( s \) and \( \bar{s} \)
can form bound states that are known as breathers. The values \( p=\displaystyle\frac{1}{k}\; ,\; k=1,2,\ldots  \)
are the thresholds where a new bound state appears. The potential term becomes
marginal when \( \beta ^{2}=8\pi  \) which corresponds to \( p=\infty  \).
The perturbation conserves the topological charge \( Q \), which can be identified
with the usual topological charge of the sG theory and with the fermion number
of the mTh model.

Mandelstam \cite{mandelstam} showed that a fermion operator satisfying the
massive Thirring equation of motion can be constructed as a nonlocal functional
of a pseudoscalar field (boson) satisfying the sine-Gordon equation. But the
fermion and the boson are not relatively local and then do not create the same
particle (the two theories are not equivalent). 

The difference between them is that they correspond to the perturbation by the
same operator of the two \emph{different local} c=1 CFTs \( {\cal A}_{b} \)
and \( {\cal A}_{f} \) as in \ref{section:free_boson}. The short distance
behaviour of the sG theory is described by the local operator algebra \( {\cal A}_{b} \),
while the primary fields of the UV limit of mTh theory are \( {\cal A}_{f} \). 

Note that the two algebras share a common subspace with even values of the topological
charge, generated by \( \{V_{(n,\, m)}:\, n\in \mathbb {Z},\, m\in 2\mathbb {Z}\} \),
where the massive theories described by the lagrangians (\ref{sG_Lagrangian})
and (\ref{mTh_Lagrangian}) are identical. Exactly in this subspace holds the
well known proof by Coleman about the equivalence of the two theories \cite{s-coleman}.
The figure \ref{4sectors.eps} shows the four sectors where all the vertex operators
live.

\section{Truncated Conformal space at \protect\( c=1\protect \)\label{section:TCSA}}

In this section is given a brief description of Truncated Conformal Space (TCS)
method for \( c=1 \) theories. It will be usefull, in chapter \ref{chapter:analisis_NLIE}
to have a comparison of the data obtained from the non-linear integral equation
that will be introduced.

The TCS method was originally created to describe perturbations of Virasoro
minimal models in finite spatial volume \cite{yurov-zam}. Here is presented
an extension of the method to study perturbations of a \( c=1 \) compactified
boson, more closely the perturbation corresponding to sine-Gordon theory.

The Hilbert space of the \( c=1 \) theory (on a cylinder) can be split into
sectors labelled by the values of \( P \) and \( Q \), which are quantised
by integers. The numerical computations shall be restricted to the \( P=0 \)
sector (it is not expected that any relevant new information would come from
considering \( P\neq 0 \)). The TCS method consists of retaining only those
states in such a sector for which the eigenvalue of \( H_{CFT} \) is less than
a certain upper value \( E_{cut} \), so the truncated space is defined as

\begin{equation}
\label{spazio_troncato}
{\cal H}_{TCS}(s,m,E_{cut})=\left\{ \left| \Psi \right\rangle :\: P\left| \Psi \right\rangle =s\left| \Psi \right\rangle ,Q\left| \Psi \right\rangle =m\left| \Psi \right\rangle ,H_{CFT}\left| \Psi \right\rangle \leq E_{cut}\left| \Psi \right\rangle \right\} \: .
\end{equation}

For a given value of \( s \), \( m \) and \( E_{cut} \) this space is always
finite dimensional. In this space, the Hamiltonian, represented on the basis
of the eigenstates of the unperturbed \( c=1 \) Hamiltonian (i.e. eigenstates
of \( \widehat{L_{0}}+\widehat{\overline{L}_{0}} \))\footnote{
In this section, an hat on a letter means matrix
}, is a finite size matrix whose eigenvalues can be computed using a numerical
diagonalization method. The explicit form of this matrix is the following:

\begin{equation}
\label{hamilt._tronca}
\widehat{H}=\displaystyle\frac{2\pi }{L}\left( \widehat{L_{0}}+\widehat{\overline{L}_{0}}-\displaystyle\frac{c}{12}\widehat{I}+\lambda \displaystyle\frac{L^{2-h}}{\left( 2\pi \right) ^{1-h}}\widehat{B}\right) \: ,
\end{equation}
where \( \widehat{L_{0}} \) and \( \widehat{\overline{L_{0}}} \) are diagonal
matrices with their diagonal elements being the left and right conformal weights,
\( \widehat{I} \) is the identity matrix,

\begin{equation}
\label{esponente}
h=\Delta ^{+}_{V}+\Delta ^{-}_{V}=\displaystyle\frac{\beta ^{2}}{4\pi }=\displaystyle\frac{2p}{p+1}
\end{equation}
is the scaling dimension of the perturbing potential \( V \) and the matrix
elements of \( \widehat{B} \) are

\begin{equation}
\label{perturbazione}
\widehat{B}_{\Phi ,\Psi }=\displaystyle\frac{1}{2}\left\langle \Phi \right| V_{(1,0)}(1,1)+V_{(0,1)}(1,1)\left| \Psi \right\rangle \: .
\end{equation}
The units are chosen in terms of the soliton mass \( {\cal M} \) which is related
to the coupling constant \( \lambda  \) by the mass gap formula\footnote{
Notice the analogy with both s-G/mTh that are obtained perturbing the free boson
theory by a ``mass term''.
} obtained from TBA in \cite{mass_scale}:
\begin{equation}
\label{mass_gap}
\lambda =\kappa {\cal M}^{2-h},
\end{equation}
where
\begin{equation}
\label{TBA_costante}
\kappa =\displaystyle\frac{2\Gamma (h/2)}{\pi \Gamma (1-h/2)}\left( \displaystyle\frac{\sqrt{\pi }\Gamma \left( \displaystyle\frac{1}{2-h}\right) }{2\Gamma \left( \displaystyle\frac{h}{4-2h}\right) }\right) ^{2-h}\, .
\end{equation}
In what follows the energy scale is normalized by taking \( {\cal M}=1 \);
the dimensionless volume \( {\cal M}L \) is denoted by \( l \). For numerical
computations, the dimensionless Hamiltonian
\begin{equation}
\label{dimlessham}
\widehat{h}=\displaystyle\frac{\widehat{H}}{{\cal M}}=\displaystyle\frac{2\pi }{l}\left( \widehat{L_{0}}+\widehat{\overline{L}_{0}}-\displaystyle\frac{c}{12}\widehat{I}+\kappa \displaystyle\frac{l^{2-h}}{\left( 2\pi \right) ^{1-h}}\widehat{B}\right) 
\end{equation}
will be used. The usefulness of the TCS method lies in the fact that it provides
a nonperturbative method of numerically obtaining the spectrum (the mass gap,
the mass ratios and the scattering amplitudes) of the theory. Therefore it can
serve as a tool to check the exact results obtained for integrable field theories
and get a picture of the physical behaviour even for the nonintegrable case.
The systematic error introduced by the truncation procedure is called the \emph{truncation
error}; it increases with the volume \( L \) and can be made smaller by increasing
the truncation level (at the price of increasing the size of the matrices, which
is bound from above by machine memory and computation time).

Let us make some general remarks on how the TCS method applies to \( c=1 \)
theories. First note that the Hilbert space (even after specifying the sector
by the eigenvalues of \( P \) and \( Q \)) consists of infinitely many Verma
modules labelled by the quantum number \( n \). At any finite value of \( E_{cut} \)
only finitely many of such Verma modules contribute, but their number increases
with \( E_{cut} \). As a result one has to deal with many more states than
in the case of minimal models. The results of TCS are supposed to approach the
exact results in the limit \( E_{cut}\: \rightarrow \: \infty  \). The convergence
can be very slow, while the number of states rises faster than exponentially
with the truncation level. The perturbing operator has scaling dimension which
ranges between \( 0 \) and \( 2 \), becoming more relevant in the attractive
regime, while the number of states corresponding to a given value of \( E_{cut} \)
becomes larger as moving towards \( p=0 \), which affects the convergence just
the other way around. 

Generally, the energy of any state goes with the volume \( L \) as

\begin{equation}
\label{scalingfun}
\displaystyle\frac{E_{\Psi }(L)}{{\cal M}}=-\displaystyle\frac{\pi \left( c-12\left( \Delta _{\Psi }+\overline{\Delta }_{\Psi }\right) \right) }{6l}+Bl+\displaystyle\sum ^{\infty }_{k=1}C_{k}\left( \Psi \right) l^{k(2-h)}\: ,
\end{equation}
where \( \Delta _{\Psi } \) (\( \overline{\Delta }_{\Psi } \)) are the left
(right) conformal dimensions of the state in the ultraviolet limit, \( B \)
is the universal bulk energy constant (the vacuum energy density) and the infinite
sum represents the perturbative contributions from the potential \( V \). 

The bulk energy constant has also been predicted from TBA and reads \cite{mass_scale}

\begin{equation}
\label{bulk}
B=-\displaystyle\frac{1}{4}\tan \left( \displaystyle\frac{p\pi }{2}\right) 
\end{equation}
(the same result was obtained from the NLIE approach in \cite{ddv 95}). This
is a highly nonanalytic function of \( p \) and it becomes infinite at the
points where \( p \) is an odd integer. In fact, at these points there is a
value of \( k \) for which \( k(2-h)=1 \), and \( C_{k}\left( \Psi \right) \: \rightarrow \: \infty  \).
The infinite parts of \( B \) and \( C_{k}\left( \Psi \right)  \) exactly
cancel, leaving a logarithmic (proportional to \( l\: \log l \)) and a finite
linear contribution to the energy, by a sort of a resonance mechanism. All of
these ``logarithmic points'' are in the repulsive regime. However, due to
UV problems in the repulsive regime we are not able to check numerically the
logarithmic corrections to the bulk energy.

The origin of UV divergences can be understood from \emph{conformal perturbation
theory} (CPT). It is known that when the scaling dimension \( h \) of the perturbing
potential exceeds \( 1 \), CPT suffers from ultraviolet divergences which should
be removed by some renormalization procedure. The TCS method is something very
similar to CPT: it operates in the basis of the UV wave functions as well, but
computes the energy levels using the variational approach and therefore could
be called ``\emph{conformal variation theory}'' \emph{}(CVT). As a result,
it is expected that there could be UV divergences for the range of couplings
where \( h>1 \) which is exactly the repulsive regime \( p>1 \) \cite{kl-me2}.
The numerical analysis has in fact shown that in the repulsive regime the TCS
energy eigenvalues did not converge at all when increasing the truncation level. 

Fortunately, there exists a way out: since it is expected to find a sensible
quantum field theory when the UV cutoff is removed, it should be the case that
the \emph{relative energy levels} \( {\cal E}_{\Psi }(L)=E_{\Psi }(L)-E_{vac}(L) \)
converge to some limit. This is exactly the behaviour that has been observed.
Consequently, in the repulsive regime one can only trust the \textit{relative}
scaling functions produced by the TCS method, while in the attractive regime
even the \textit{absolute} energy values can be obtained (including the predicted
bulk energy constant (\ref{bulk}), which is completely analytic for \( p<1 \)
and thus logarithmic corrections are absent as well). 

Many numerical results show that the smaller the value of \( p \) is the faster
the convergence is (with the understanding that in the repulsive regime by convergence
of TCS we mean the convergence of the energies relative to the vacuum). On the
other hand, even in the attractive regime the convergence is so slow that to
get reliable results (which means errors of order \( 10^{-3}-10^{-2} \) for
the volume \( l \) ranging from \( 0 \) to somewhere between \( 5 \) and
\( 10 \)) requires to work with matrix of dimensions around \( 4000 \). This
means that the TCS for \( c=1 \) theories is far less convergent than the one
for minimal models (in the original Lee-Yang example the authors of \cite{yurov-zam}
took a \( 17 \) dimensional Hilbert space (!) and arrived to very accurate
results).

\chapter{LIGHT-CONE LATTICE QFT}

In this chapter, the most important tools to deal with Destri-de Vega equation
are introduced, with reference to the original papers \cite{ddv 87, devega 89}.
Particular attention will be taken to indicate the path that has been done from
the light-cone lattice until the definition of a particular system (the 6 vertex
model, alias XXZ chain), whose Hilbert space is quite completely known.

\section{Kinematics on light-cone\label{light-cone.section}}

It is a usual way to regularize quantum field theories by defining them on a
space-like ``hamiltonian'' lattice (where time is continuous and space discrete)
or space and time-like ``euclidean'' lattice (when both space and time are
discrete). In statistical mechanics this is not just a regularization method
but can be a right microscopic way to describe physical systems. In two dimensions,
the most known approach is to define a rectangular lattice with axis corresponding
to space and time directions and associate to each site an interaction depending
only on the nearest neighbouring sites. In this case the partition function
can be expressed in terms of a transfer matrix. 

\begin{figure}[  htbp]
{\centering \rotatebox{270}{\includegraphics{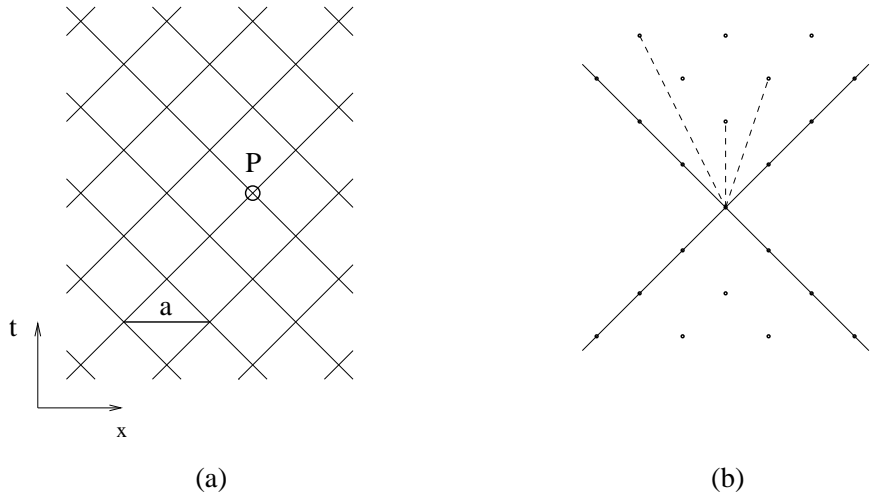}} \par}

\caption{\small (a) light-cone lattice; (b) double light-cone emanating from a point and high-order
interactions (dashed)\label{lcl1.eps}}
\end{figure}

In what follows, a different approach is adopted: Minkowski and Euclidean space-time
can, in fact, be discretized along light-cone directions. Light-cone coordinates
are:
\[
x_{\pm }=x\pm t\]
and the choice 
\[
{\cal M}=\{x_{\pm }=\frac{a}{\sqrt{2}}n_{\pm },\quad n_{\pm }\in \mathbb Z\}\]
defines a light-cone lattice of ``events'' as in figure \ref{lcl1.eps} (a)
. They are spaced by \( a \) in the space and time directions and by \( a/\sqrt{2} \)
in light-cone directions. At every event \( P\in {\cal M} \) there is associated
a double light-cone (in the past and in the future) and only events within this
light-cone can be causally connected (see fig. \ref{lcl1.eps} (b)). Then, any
rational and not greater than \( 1 \) speed is permitted for particles, in
an infinite lattice. The shortest displacement of the particle (one lattice
spacing) is realized at speed \( \pm 1 \) and corresponds, from the statistical
point of view, to nearest neighbours interactions. Smaller speeds can be obtained
with displacements longer than the fundamental plaquette, and correspond to
high order neighbours interactions. In quantum field theory, these are nonlocal
interactions. 

In the following, only the local case (nearest neighbours) is treated. The nearest
neighbours of the event \( P \) are the four nearest points in the light-cone
directions. This implies that particles can have only the speed of light \( \pm 1 \)
and are massless. They are called right-movers (R) and left-movers (L). 

In this case, it is possible to introduce a useful language for connection with
statistical mechanics associating a particle to a link. Consider the two links
in the future that come out from a event \( P \). Particles R and L in \( P \),
by definition, are respectively associated to the right-oriented link and to
the left-oriented link. In this way, the state of a link is defined to be the
state of the point where it begins (also the opposite choice, of connecting
a link with the site where it ends, can be done; it is simply a matter of convention).
For example, if in a point \( O \) there is a particle R, one tells that the
``right-oriented'' link outcoming from it is occupied by R. This correspondence
of points and links is possible because only local interactions are assumed,
and it is useful because the counting of states is simpler. But the physically
correct interpretation is that particles live on events, not on links. Links
are the possible world lines for particles.\begin{figure}[  htbp]
{\centering \rotatebox{270}{\includegraphics{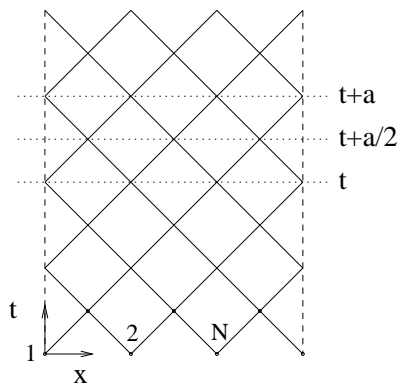}} \par}

\caption{\small Periodicity in space direction; in link language the lines of time must be
intended as shifted up by a little bit, as in figure \ref{upm.eps}.\label{lcl2.eps}}
\end{figure} 

In the following, the lattice is assumed of a finite extent \( L=aN \) in space
direction (\( N \) is the number of sites, counted in the space direction),
with periodic boundary conditions, but infinite in time direction, as shown
in figure \ref{lcl2.eps}. In this way a cylinder topology is defined for space
time. The Hilbert space of states in an event \( P \) is the tensor product
\[
{\cal H}={\cal H}_{L}\otimes {\cal H}_{R}\]
of R and L space of states. The fact that particles can be classified in left
and right does not mean, in general, that the two dynamics are independent,
as happens for example in conformal field theory. In what follows, exactly the
interacting case will be treated. 

Call \( \left| \alpha _{Li},\alpha _{Ri}\right\rangle  \) the generic vector
of a basis of \( {\cal H}_{i} \) where \( i=1,...,N \) labels the sites. The
notation 
\[
\left| \alpha _{2i-1},\alpha _{2i}\right\rangle =\left| \alpha _{Li},\alpha _{Ri}\right\rangle \]
is useful and not ambiguous (even number refers to right, odd number refers
to left). The total Hilbert space is: 
\[
{\cal H}_{N}=\bigotimes ^{N}_{i=1}{\cal H}_{i}\]
and a basic vector can be represented by 
\[
\left| \alpha _{1},\alpha _{2}\right\rangle \otimes ...\otimes \left| \alpha _{2N-1},\alpha _{2N}\right\rangle =\left| \alpha _{1},\alpha _{2},...,\alpha _{2N}\right\rangle \in {\cal H}_{N}.\]
Note that in a \( N \) sites lattice, due to light-cone, \( 2N \) labels are
required. 

\begin{figure}[  htbp]
{\centering \rotatebox{270}{\includegraphics{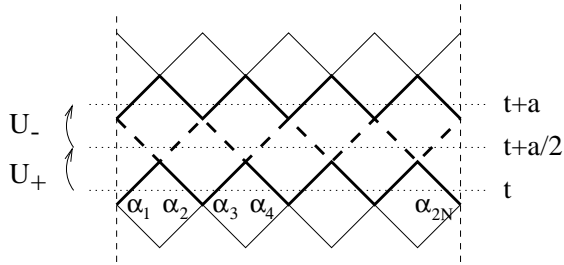}} \par}

\caption{\small Partial evolution operators. The convention used here for the labeling of states
will be used also in the following.\label{upm.eps}}
\end{figure}

If at a given time \( t \) there is a line of sites, the particular characteristic
of the light-cone is that at time \( t+a/2 \) there is another line of sites,
but not equivalent to the previous one, because it is shifted. Only at \( t+a \)
there is an equivalent line (see figure \ref{lcl2.eps}). Then two different
evolution operators can be defined, depending on the initial state:
\begin{equation}
\label{upm}
\begin{array}{c}
U_{+}\left| \alpha _{1},\alpha _{2},...,\alpha _{2N},t\right\rangle =\left| \alpha _{1}',\alpha _{2}',...,\alpha _{2N}',t+a/2\right\rangle \\
\\
U_{-}\left| \alpha _{1}',\alpha _{2}',...,\alpha _{2N}',t+a/2\right\rangle =\left| \alpha _{1}'',\alpha _{2}'',...,\alpha _{2N}'',t+a\right\rangle 
\end{array}
\end{equation}
where the initial states are chosen as in figure \ref{upm.eps}. Schr\"odinger form of equations of motion is used, for a state in Hilbert space.\footnote{
In Schrodinger form, if \( \left| \alpha \right\rangle  \) is a state vector,
its time evolution is given by \( \left| \alpha ,t\right\rangle =U\left| \alpha ,0\right\rangle  \)
where \( U \) satisfies motion's equations: \( U=Te^{-i\displaystyle\int dtH} \) (Dyson's
series)
} The global time operator can be chosen as
\[
U=U_{+}U_{-\textrm{ }}\: \: \textrm{ or }\: \quad U'=U_{-}U_{+}\]
depending on the initial state. For a consistent quantum theory, both this operators
must be unitary. This can be guaranteed if the assumption \( U^{\dagger }_{+}U_{+}=U^{\dagger }_{-}U_{-}=1 \)
is made, that is the elementar operators themselves must be unitary.

\begin{figure}[  htbp]
{\centering \rotatebox{270}{\includegraphics{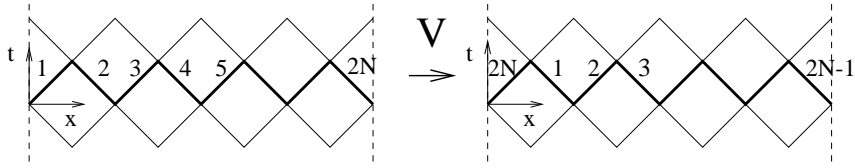}} \par}

\caption{\small Half-shift operator\label{mezzo-shift.eps}}
\end{figure}

\begin{figure}[  htbp]
{\centering \rotatebox{270}{\includegraphics{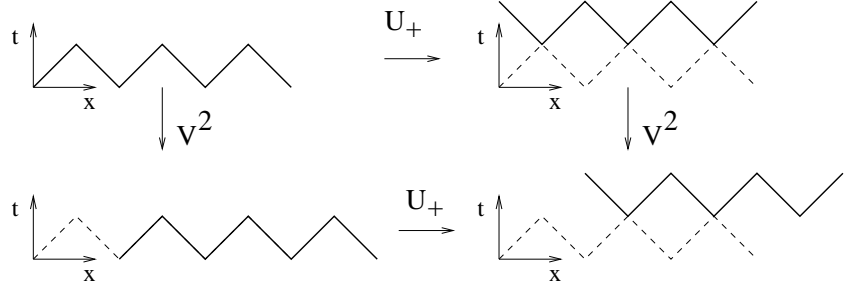}} \par}

\caption{\small Commutation rule: \protect\( V^{2}U_{+}=U_{+}V^{2}\protect \)\label{commutation.eps}}
\end{figure}

\begin{figure}[  htbp]
{\centering \rotatebox{270}{\includegraphics{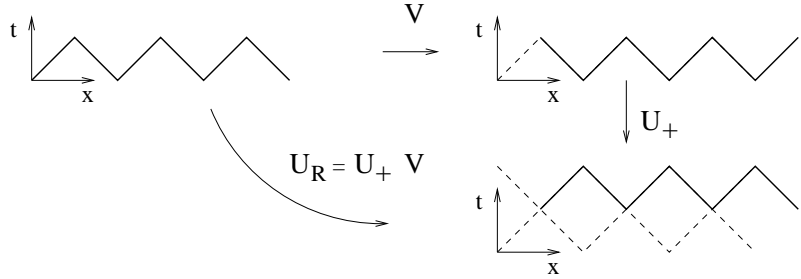}} \par}

\caption{\small Right-oriented shift operator\label{UR.eps}}
\end{figure}

Another operator plays an important role and is defined as follows (the states
are at a certain fixed time): 
\begin{equation}
\label{mezzoshift}
V\left| \alpha _{1},\alpha _{2},...,\alpha _{2N},t\right\rangle =\left| \alpha _{2N},\alpha _{1},...,\alpha _{2N-1},t\right\rangle 
\end{equation}
it corresponds to an half-space shift in the space direction, with exchange
of right and left states(see the figure \ref{mezzo-shift.eps}). Two applications of \( V \) give a shift by an entire lattice spacing, then
\( V^{2} \) is the lattice space evolution operator. It is possible to write
an expression for \( V \) in terms of the permutation operator: 
\[
P_{nm}\left( \left| \alpha _{n}\right\rangle \otimes \left| \alpha _{m}\right\rangle \right) =\left| \alpha _{m}\right\rangle \otimes \left| \alpha _{n}\right\rangle ,\qquad \left| \alpha _{n}\right\rangle \otimes \left| \alpha _{m}\right\rangle \in {\cal H}_{n}\otimes {\cal H}_{m}.\]
that looks like: 
\begin{equation}
\label{mezzo-shift-P}
V=P_{12}P_{23}...P_{2N-1,\, 2N}
\end{equation}
The verification is direct: 
\[
\begin{array}{c}
P_{12}P_{23}...P_{2N-1,\, 2N}\left| \alpha _{1},\alpha _{2},...,\alpha _{2N}\right\rangle =P_{12}P_{23}...P_{2N-2,\, 2N-1}\left| \alpha _{1},...,\alpha _{2N-2},\alpha _{2N},\alpha _{2N-1}\right\rangle =\\
\\
=P_{12}P_{23}...P_{2N-3,\, 2N-2}\left| \alpha _{1},...,\alpha _{2N},\alpha _{2N-2},\alpha _{2N-1}\right\rangle =...=\left| \alpha _{2N},\alpha _{1},...,\alpha _{2N-1}\right\rangle .
\end{array}\]
Also \( V \) is a unitary operator. To show this, one can use the fact that
the permutation is self-adjoint. A more complete list of known properties of
\( P \) is:
\begin{equation}
\label{permutatore}
P=P^{-1}=P^{\dagger },\qquad P^{2}=1
\end{equation}
Then, the adjoint of \( V \) is:
\[
V^{\dagger }=P_{2N-1,2N}...P_{23}P_{12}\]
and \( VV^{\dagger }=1 \) simply follows. The operators defined up to now have
the following commutation rules:
\begin{equation}
\label{comm-uv}
\begin{array}{cc}
\left[ V^{2},U_{\pm }\right] =0;\qquad U_{\pm }=VU_{\mp }V^{\dagger }.
\end{array}
\end{equation}
In figure \ref{commutation.eps}, the first case is proved, by showing the equivalence
of the two paths \( V^{2}U_{+} \) and \( U_{+}V^{2} \). ``Mutatis mutandis'',
all the other cases can be simply obtained. Consequently, the two principal evolution operators, \( V^{2} \) and \( U \),
are commuting:
\[
\left[ V^{2},U\right] =\left[ U,V^{2}\right] =0.\]
As previously shown, they are also unitary. Thanks to all these properties,
it is very natural to identify them as the exponential of the hamiltonian operator,
and the exponential of the linear momentum: 
\begin{equation}
\label{unitari}
\begin{array}{c}
U=e^{-iaH}\\
V^{2}=e^{-iaP}.
\end{array}
\end{equation}
There are other two important operators, defined as:
\begin{equation}
\label{unitari-conoluce}
\begin{array}{c}
U_{R}=U_{+}V\\
U_{L}=U_{+}V^{\dagger }
\end{array}
\end{equation}
As clearly shown in figure \ref{UR.eps} for one of them, they correspond to
one step evolution in light-cone directions. They are commuting and give the expressions:
\[
\begin{array}{cc}
\quad \quad U=U_{R}U_{L}\quad \quad  & V^{2}=U_{R}U^{\dagger }_{L}\\
\left[ U_{R},U_{L}\right] =0 & U_{R}^{\dagger }U_{R}=U_{L}^{\dagger }U_{L}=1
\end{array}\]
then, using also (\ref{unitari}) yields:
\begin{equation}
\label{unit-lightcone}
U_{R}=e^{-i\displaystyle\frac{a}{2}(H+P)},\qquad U_{L}=e^{-i\displaystyle\frac{a}{2}(H-P)}.
\end{equation}

\section{Dynamics on light-cone\label{s-matrix.section}}

A dynamics can be defined by giving all the amplitudes of the different processes
that can take place. The fundamental assumption is that in every site a whole
process can happen, in the sense that if \( \left| \alpha _{L},\alpha _{R}\right\rangle _{in} \)
and \( \left| \beta _{L},\beta _{R}\right\rangle _{out} \) are the incoming
and outgoing states in a certain site, they can be considered asymptotic states
and the transition amplitude is an S-matrix element:
\begin{equation}
\label{ampiezza}
\displaystyle _{out}\left\langle \beta _{R},\beta _{L}\right. \left| \alpha _{R},\alpha _{L}\right\rangle _{in}=S_{(\alpha _{R},\alpha _{L})_{in},(\beta _{R},\beta _{L})_{out}}
\end{equation}
 This, in general, is an \( m\, \rightarrow \, n \) scattering. The system
is then defined via its microscopic amplitudes.

From the definition (\ref{upm}) of evolution operators, the following expression
holds (the ``in'' state is at time \( t \), the ``out'' state at \( t+a/2 \),
with reference at the figure \ref{upm.eps}:
\begin{equation}
\label{up-s-matrix}
\begin{array}{c}
\displaystyle \left\langle \alpha _{1},\alpha _{2},...,\alpha _{2N},t\right| U_{+}\left| \alpha _{1}',\alpha _{2}',...,\alpha _{2N}',t\right\rangle =_{in}\left\langle \alpha _{1},\alpha _{2},...,\alpha _{2N}\right| \left. \alpha _{1}',\alpha _{2}',...,\alpha _{2N}'\right\rangle _{out}=\\
\\
=_{in}\left\langle \alpha _{1},\alpha _{2}\right. \left| \alpha _{1}',\alpha _{2}'\right\rangle _{out}..._{in}\left\langle \alpha _{2N-1},\alpha _{2N}\right. \left| \alpha _{2N-1}',\alpha _{2N}'\right\rangle _{out}=\\
\\
=\prod _{i=1,...,N}S^{\dagger }_{(\alpha _{2i-1},\alpha _{2i})_{in},(\alpha _{2i-1}',\alpha _{2i}')_{out}}
\end{array}
\end{equation}
The operator \( U_{-} \) can be obtained from (\ref{comm-uv}). In the last
line, the product is on all the sites at a given time. 

It is interesting to observe that there is a powerful connection with statistical
mechanics, at this point. Assume, for simplicity, that the lattice is euclidean
and has a very large (thermodynamics) but finite extension in the ``time''
direction. In this way there is an initial time (with an initial state \( \left| \alpha \right\rangle _{in} \))
and a final time (with a final state \( \left| \beta \right\rangle _{out} \)).
The total transition amplitude is given by the product on all sites of the amplitudes,
and the sum on all the internal states (\( \gamma ',\, \gamma '' \)):
\[
_{out}\left\langle \beta \right| \left. \alpha \right\rangle _{in}=\sum _{int\, states}\; \prod _{sites}S_{\gamma ',\gamma ''}\equiv Z\]
The important fact is that if \( S_{\gamma ',\gamma ''} \) is real and positive,
it plays the role of a Boltzmann weight for a statistical system whose partition
function is given by the previous expression, that was conveniently called \( Z \).
It depends only on the external states. The first appearance of this correspondence
is in \cite{zam90}.

\begin{figure}[  htbp]
{\centering \rotatebox{270}{\includegraphics{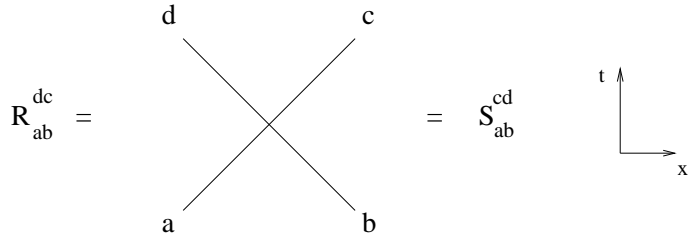}} \par}

\caption{\small Pictorial interpretation of R and S matrix \label{R-S matrix}}
\end{figure}

In both the statistical and the particle interpretation, the integrability (that
will be always assumed in what follows) can be obtained by requiring that the
scattering amplitude \( S_{\gamma ,\gamma '} \) or the Boltzmann weight satisfy
the Yang-Baxter equation (see \cite{zam79,baxter}). In this case a general
\( m\, \rightarrow \, n \) particles scattering can be factorised in the product
of ``elementary'' \( 2\, \rightarrow \, 2 \) particles scattering. Then at
every site there will be associated one of this ``elementary'' amplitudes:
\begin{equation}
\label{ampiezza_2_part}
_{out}\left\langle \beta _{L},\beta _{R}\right. \left| \alpha _{R},\alpha _{L}\right\rangle _{in}=S^{\beta _{R},\beta _{L}}_{\alpha _{R},\alpha _{L}}
\end{equation}
and ``in'' and ``out'' states are two particle states (as in the previous
equation). In statistical language, usually the Boltzmann weight is indicated
with \( R \) and there is a little change of notation with respect to scattering
case:
\begin{equation}
\label{matriceR}
R^{\beta _{R},\beta _{L}}_{\alpha _{L},\alpha _{R}}=S^{\beta _{R},\beta _{L}}_{\alpha _{R},\alpha _{L}}
\end{equation}
(the lower indices are interchanged). A pictorial interpretation of that is
in figure \ref{R-S matrix}.

The Yang-Baxter equation takes the form: 
\begin{equation}
\label{yang-baxter}
S_{12}S_{13}S_{23}=S_{23}S_{13}S_{12}
\end{equation}
(with the indicated change it holds clearly also for R-matrix). 

\begin{figure}[  htbp]
{\centering \rotatebox{270}{\includegraphics{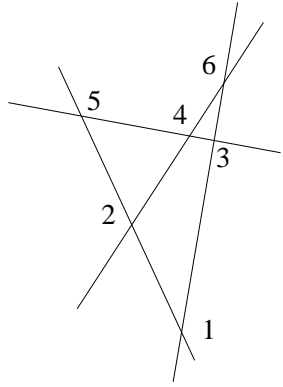}} \par}

\caption{\small Factorizable scattering. Between two interaction points the particles are free.
\label{scattering.eps}}
\end{figure}

Apparently, this ``phenomenological'' approach is quite unusual, because in
traditional lattice quantum field theory at every site is associated one interaction
potential (ex: \( \phi ^{4}(i) \) is the potential on the site \( i \)), not
a whole scattering process. This can appear as a sort of ``macroscopic'' approach,
not based on fundamental interactions. But the properties of factorisable scattering
must be taken into account. Factorization of generic amplitudes in \( 2\, \rightarrow \, 2 \)
particles amplitudes is a sort of quantum superposition principle and the remarkable
fact is that between one scattering and the other, the particles are asymptotic
ones, that means that they are free. For example in figure \ref{scattering.eps}
(1+1 QFT on the continuum) between the point 1 and the point 2 the motion is
free. Every point contain all the interaction. This is what was assumed in the definition
of the lattice. Then it is perfectly justified that every site is connected
with a whole \( 2\, \rightarrow \, 2 \) particles scattering process.\footnote{
In the next paragraphs, it will be explained that, for the particular case of
the 6 vertex R matrix, a more traditional lattice QFT approach can be formulated,
in terms of a fermionic field.
} The previous equation (\ref{up-s-matrix}) can be written now in a more specific
form:
\begin{equation}
\label{up-corretta}
\left\langle \alpha _{1},\alpha _{2},...,\alpha _{2N},t\right| U_{+}\left| \alpha _{1}',\alpha _{2}',...,\alpha _{2N}',t\right\rangle =R^{\alpha _{1}'\alpha _{2}'}_{\alpha _{1}\alpha _{2}}\: R^{\alpha _{3}'\alpha _{4}'}_{\alpha _{3}\alpha _{4}}\: ...\: R^{\alpha _{2N-1}'\alpha _{2N}'}_{\alpha _{2N-1}\alpha _{2N}}
\end{equation}
The physical system dynamics is therefore defined by the assignment of an R
matrix, and integrability is guaranteed. The consistency on quantum mechanical
interpretation is obtained by requiring unitarity, hermitian analyticity and
crossing symmetry for the S matrix obtained by (\ref{matriceR}). 

It is important to obtain an explicit expression for \( U_{L} \) and \( U_{R}: \)
\begin{equation}
\label{ur-corretta}
\begin{array}{c}
\displaystyle \left\langle \alpha _{1},\alpha _{2},...,\alpha _{2N},t\right| U_{R}\left| \alpha _{1}',\alpha _{2}',...,\alpha _{2N}',t\right\rangle =\displaystyle\sum _{\overrightarrow{\gamma }}\left\langle \overrightarrow{\alpha },t\right| U_{+}\left| \overrightarrow{\gamma }\right\rangle \left\langle \overrightarrow{\gamma }\right| V\left| \overrightarrow{\alpha '}t\right\rangle =\\
\\
=\displaystyle\sum _{\overrightarrow{\gamma }}R^{\gamma _{1}\gamma _{2}}_{\alpha _{1}\alpha _{2}}\: R^{\gamma _{3}\gamma _{4}}_{\alpha _{3}\alpha _{4}}\: ...\: \left\langle \gamma _{1},...,\gamma _{2N}\right. \left| \alpha _{2N}',\alpha _{1}',...,\alpha _{2N-1}',t\right\rangle =\\
\\
=R^{\alpha _{2N}'\alpha _{1}'}_{\alpha _{1}\alpha _{2}}\: R^{\alpha _{2}'\alpha _{3}'}_{\alpha _{3}\alpha _{4}}\: ...\: R^{\alpha _{2N-2}'\alpha _{2N-1}'}_{\alpha _{2N-1}\alpha _{2N}}
\end{array}
\end{equation}
and, in a similar way,
\begin{equation}
\label{ul-corretta}
\left\langle \alpha _{1},\alpha _{2},...,\alpha _{2N},t\right| U_{L}\left| \alpha _{1}',\alpha _{2}',...,\alpha _{2N}',t\right\rangle =R^{\alpha _{2}'\alpha _{3}'}_{\alpha _{1}\alpha _{2}}\: R^{\alpha _{4}'\alpha _{5}'}_{\alpha _{3}\alpha _{4}}\: ...\: R^{\alpha _{2N}'\alpha _{1}'}_{\alpha _{2N-1}\alpha _{2N}}
\end{equation}
This expression is consistent with the figure \ref{UR.eps}. The important fact
is that this operators can be expressed in terms of the transfer matrix of an
inhomogeneous lattice model, that will be defined in the next section.

\section{Euclidean transfer matrix}

\begin{figure}[  htbp]
{\centering \rotatebox{270}{\includegraphics{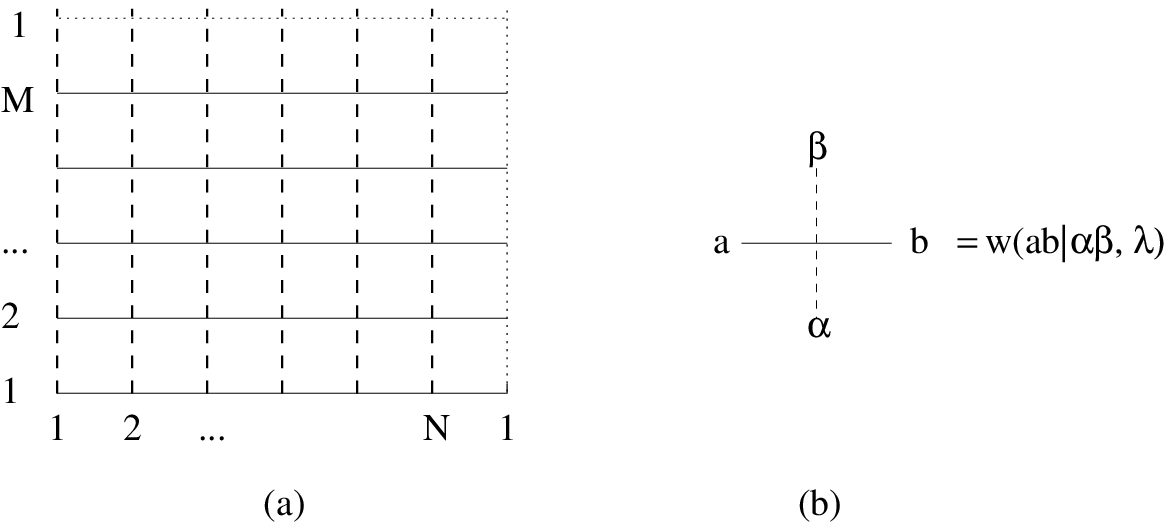}} \par}

\caption{\small (a) periodic (toroidal) square lattice; (b) Boltzmann weight for a site; the
labels are:}

\( a,b\in {\cal A},\quad \alpha ,\beta \in {\cal V} \)\label{square lattice.eps} 
\end{figure}

Consider a two dimensional euclidean square lattice, with periodic boundary
conditions, in both the directions. The links are the physical objects of the
system. They can be in different states belonging to the vector spaces \( {\cal A} \)
and \( {\cal V} \). In principle, this vector states can be different, but
in the following they will be identified. At every site can be associated a
Boltzmann weight depending on the four links crossing at this site and on a
spectral parameter \( \lambda  \) (see the figure \ref{square lattice.eps}).  The simplest case is to take the same Boltzmann weights in all the sites,
but more general configurations are possible. In the following an inhomogeneity
\( \lambda _{i} \) will be assumed, where \( i \) is the column index (all
the sites on a column have the same inhomogeneity), in the sense that the Boltzmann
weight in the column \( i \) is taken to be:
\[
w(ab|\alpha \beta ,\, \lambda -\lambda _{i}).\]
Also for the boundary conditions it is possible to assume more general configurations
than the simplest one (i.e. toroidal b.c.). Assume that between the column N
and the N+1 (that is 1) there is a nontrivial seam line, in such a way that
the Boltzmann weights on the column N (with respect to the normal ones) are
given by:
\begin{equation}
\label{twist-Boltz.}
w^{(N)}(ab|\alpha \beta ,\, \lambda )=e^{i\omega b}w(ab|\alpha \beta ,\, \lambda ).
\end{equation}
This choice is made because only the link \( b \) cross the seam line. This
are called \emph{twisted boundary conditions} (obviously the choice \( \omega =0 \)
reproduces the periodic case).

The partition function for such a model is given by (see the figure \ref{square lattice.eps}):
\begin{equation}
\label{partition-f}
Z=\displaystyle\sum _{link\, states}\prod _{rows}w(ab|\alpha \beta ,\, \lambda -\lambda _{1})...w(ab|\alpha \beta ,\, \lambda -\lambda _{N-1})w^{(N)}(ab|\alpha \beta ,\, \lambda -\lambda _{N})
\end{equation}
(observe the twisted Boltzmann weight in the last column). It is simple to show
that it can be expressed, thanks to periodicity, in terms of a row-to-row transfer
matrix. To obtain that, it is convenient to define an operator \( t \) that
acts on both the horizontal and the vertical spaces:
\begin{equation}
\label{t-operator}
\begin{array}{c}
t(\lambda ):\; {\cal A}\: \rightarrow \: {\cal A}\\
t_{ab}(\lambda )=\left\langle a\right| t(\lambda )\left| b\right\rangle 
\end{array}
\end{equation}
defines the horizontal action; \( t_{ab} \) depends on the horizontal links
and acts on the vertical space as:
\begin{equation}
\label{t-ab-operator}
\begin{array}{c}
t_{ab}(\lambda ):\; {\cal V}\: \rightarrow \: {\cal V}\\
\left\langle \alpha \right| t_{ab}(\lambda )\left| \beta \right\rangle =\left[ t_{ab}(\lambda )\right] _{\alpha \beta }=w(ab|\alpha \beta ,\, \lambda ).
\end{array}
\end{equation}
The twist can be obtained with an appropriate operator that changes the horizontal
states by a phase after the seam line:
\begin{equation}
\label{twist-vettore}
\left| a_{N+n}\right\rangle =e^{i\omega A}\left| a_{n}\right\rangle ,\qquad \left| a_{i}\right\rangle \in {\cal A}_{i}
\end{equation}
where \( \omega  \) is a real parameter (twist), \( A \) is a self-adjoint
operator and \( n \) is an horizontal index (in the vertical direction there
is no twist). \( A \) must be chosen diagonal on the whole horizontal space
\( {\cal A}, \) therefore it acts as a phase. Observe that the lattice euclidean
structure is a normal periodic one, because the twist acts only on the quantum
mechanical structure. For a generic operator \( B_{n} \) acting on the site
\( n \) the consistency with (\ref{twist-vettore}) requires that the twist
acts as:
\begin{equation}
\label{twistoperatore}
B_{N+n}=e^{i\omega A}B_{n}e^{-i\omega A}.
\end{equation}
This expression can be used for the determination of the \( t_{ab} \) operator
acting on the last column: 
\begin{equation}
\label{twistsut}
\begin{array}{c}
_{N}\left\langle a\right| t^{(N)}(\lambda )\left| b\right\rangle _{N+1}=_{N}\left\langle a\right| t^{(N)}(\lambda )e^{i\omega A}\left| b\right\rangle _{1}=\\
=e^{i\omega b}\, _{N}\left\langle a\right| t^{(N)}(\lambda )\left| b\right\rangle _{1}
\end{array}
\end{equation}
where the ``positions'' of the states and operators are emphasized; in particular
observe that only \( b \) crosses the seam line. Using the identification (\ref{t-ab-operator})
the expression (\ref{twist-Boltz.}) can be simply obtained. 

The row-to-row transfer matrix is an operator \( t^{(N)}(\lambda ,\{\lambda _{i}\},\omega ) \)
that acts on the vertical space:
\[
{\cal V}^{(N)}=\bigotimes ^{N}_{i=1}{\cal V}\]
in the following way:
\begin{equation}
\label{transfermatrix}
t^{(N)}(\lambda ,\{\lambda _{i}\},\omega )=\displaystyle\sum _{a_{1}}T^{(N)}_{a_{1}a_{1}}=\displaystyle\sum _{a_{1},...a_{N}}t_{a_{1}a_{2}}(\lambda -\lambda _{1})\otimes t_{a_{2}a_{3}}(\lambda -\lambda _{2})\otimes ...\otimes t_{a_{N}a_{1}}(\lambda -\lambda _{N})e^{i\omega b}
\end{equation}
 The \( T_{ab} \) is called monodromy matrix. The expression for the partition
function is: 
\[
Z=Tr_{V^{(N)}}\left[ t^{(N)}(\lambda ,\{\lambda _{i}\},\omega )\right] ^{M}.\]
Define now a matrix acting on \( {\cal A}\otimes {\cal A} \) given by:
\begin{equation}
\label{def-r-matrix}
R_{a\alpha }^{\beta b}(\lambda )=\left[ t_{ab}(\lambda )\right] _{\alpha \beta }=w(ab|\alpha \beta ,\, \lambda ).
\end{equation}
(observe that in the operator in (\ref{t-ab-operator}) the Latin and Greek
index belong to different spaces, while in R matrix they belong to the same
space and this differentiation of notation is redundant). An obvious generalization
is needed for the last column. This definition (\ref{def-r-matrix}) shows that
the R matrix is defined as a Boltzmann weight, as required in the section \ref{s-matrix.section}.
In the following will be explained the importance of (\ref{def-r-matrix}) for
the integrability.

It is possible to show, at this point, that the transfer matrix (\ref{transfermatrix})
can be used to express the operators defined in (\ref{ur-corretta}, \ref{ul-corretta}).
Consider the following form for the inhomogeneity:
\begin{equation}
\label{inomog}
\lambda _{i}=(-1)^{i+1}\Theta 
\end{equation}
 where \( \Theta  \) is a positive real number, and calculate a matrix element
of the transfer matrix on vectors of \( V^{(2N)} \) (clearly, in this case,
the periodicity requires an even number of horizontal sites):
\[
\begin{array}{c}
\displaystyle \left\langle \alpha _{1},\alpha _{2},...,\alpha _{2N}\right| t^{(2N)}(\Theta ,\{\lambda _{i}\})\left| \alpha _{1}',\alpha _{2}',...,\alpha _{2N}'\right\rangle =\\
\\
=\displaystyle\sum _{a_{1}...a_{2N}}t_{a_{1}a_{2}}(0)_{\alpha _{1}\alpha _{1}'}\, t_{a_{2}a_{3}}(2\Theta )_{\alpha _{2}\alpha _{2}'}\, ...\, t_{a_{2N}a_{1}}(2\Theta )_{\alpha _{2N}\alpha _{2N}'}=\\
\\
=R^{\alpha _{2}'\alpha _{3}'}_{\alpha _{1}\alpha _{2}}\: R^{\alpha _{4}'\alpha _{5}'}_{\alpha _{3}\alpha _{4}}\: ...\: R^{\alpha _{2N}'\alpha _{1}'}_{\alpha _{2N-1}\alpha _{2N}}
\end{array}\]
that is exactly the expression (\ref{ul-corretta}). In an operatorial form
this means that:
\begin{equation}
\label{trans-ul}
t^{(2N)}(\Theta ,\{\lambda _{i}\})=U_{L}.
\end{equation}
A similar picture holds for the operator \( U_{R} \), but to take into account
this case an assumption on R is necessary. Assume that R matrix is hermitian
analytic, that means:
\[
R_{cd}^{ab}(\lambda )^{*}=R_{ab}^{cd}(-\lambda )\]
 that in S matrix language is \( S^{\dagger }(s)=S(s^{*}). \) This is a well
known property of an S matrix and, as explained in the section \ref{s-matrix.section},
the same requirement is necessary for R. Then, as for \( U_{L} \), the following
expression can be obtained:
\begin{equation}
\label{trans-ur}
t^{(2N)}(-\Theta ,\{\lambda _{i}\})^{\dagger }=U_{R}.
\end{equation}
Here the adjoint conjugation is only on the vertical space.\footnote{
The definition of adjoint is \( \left\langle \beta \right| t_{ab}\left| \alpha \right\rangle ^{*}=\left\langle \alpha \right| t^{\dagger }_{ab}\left| \beta \right\rangle  \)
} It is given by:
\[
\left[ t_{ab}(\lambda )^{\dagger }\right] _{\alpha \beta }=w(ab|\beta \alpha ,\, \lambda )^{*}.\]
Transfer matrix can express the evolution operators.

As introduced in the section \ref{s-matrix.section}, if the R matrix satisfies
the Yang-Baxter equation (\ref{yang-baxter}), then it is possible to show that
the transfer matrix commute with itself at different values of the spectral
parameter. In this case the system is integrable. The same conclusion holds
also for the inhomogeneous and twisted models. Assume that the R matrix satisfies
the Yang-Baxter equation, that in the explicit form is:
\begin{equation}
\label{esplicita-yb}
R^{c_{1}c_{2}}_{a_{2}a_{1}}(\vartheta _{1}-\vartheta _{2})R^{b_{1}c_{3}}_{a_{3}c_{1}}(\vartheta _{1}-\vartheta _{3})R^{b_{2}b_{3}}_{c_{3}c_{2}}(\vartheta _{2}-\vartheta _{3})=R^{c_{2}c_{3}}_{a_{3}a_{2}}(\vartheta _{2}-\vartheta _{3})R^{c_{1}b_{3}}_{c_{3}a_{1}}(\vartheta _{1}-\vartheta _{3})R^{b_{1}b_{2}}_{c_{2}c_{1}}(\vartheta _{1}-\vartheta _{2}).
\end{equation}
Using (\ref{def-r-matrix}) it is possible to obtain another form, specific
for the \( t \) operator:
\begin{equation}
\label{ybpert}
R^{ef}_{ab}(\vartheta -\vartheta ')\left[ t_{ec}(\vartheta )\right] _{\alpha \gamma }\left[ t_{fd}(\vartheta ')\right] _{\gamma \beta }=\left[ t_{ae}(\vartheta ')\right] _{\alpha \gamma }\left[ t_{bf}(\vartheta )\right] _{\gamma \beta }R^{cd}_{ef}(\vartheta -\vartheta ')
\end{equation}
For consistency, the R matrix must be of the regular type, that means that at
the origin it is the unit operator (or proportional to) on \( {\cal A}\otimes {\cal A} \):
\[
R_{a\alpha }^{\beta b}(0)=\delta _{a\beta }\delta _{\alpha b}.\]
This equation fixes a normalization for the R matrix. Then, to fit with the
equations (\ref{trans-ul}, \ref{trans-ur}) a normalization factor is required.
It is fixed by the unitarity requirement. The (\ref{ybpert}) holds also if
an identical shift is performed on both the spectral parameters: 
\[
\vartheta \, \rightarrow \, \vartheta -\alpha \]
(the same for \( \vartheta ' \)) because only the difference appears in the
R term. This shift is an inhomogeneity for the \( t \) operator. 

Consider now a group of matrices \( g\in {\cal G} \) of dimension \( \dim {\cal A}\times \dim {\cal A} \)
and determinant one, such that they commute with R:
\begin{equation}
\label{gg-R}
\left[ g\otimes g,R\right] =0.
\end{equation}
Then define a transformed (``gauged'' or twisted) \( t \) operator: 
\[
t_{(g)ab}=g_{ac}t_{cb}.\]
The twist introduced in (\ref{twistsut}) is the special case of \( {\cal G}=U(1)^{\dim {\cal A}}. \)
It is simple to verify that also the gauged \( t_{(g)} \) satisfies the same
(\ref{ybpert}). This is because of the commutation relation (\ref{gg-R}).
The group \( {\cal G} \) is a symmetry group for the Yang-Baxter equation and
for the vertex model defined in this way. With some simple algebra, using many
times eq. (\ref{ybpert}), it is possible to obtain the form for the monodromy
matrix, defined as in (\ref{transfermatrix}) but with a possibly different
gauge \( g_{i} \) in every site of a row (all the column has the same gauge)
and clearly an inhomogeneity. If the following synthetic notation is introduced,
\( G=(g_{1},...,g_{N}) \), then the final result is:
\[
R(\vartheta -\vartheta ')\left[ T_{G}^{(N)}(\vartheta )\otimes _{{\cal A}}T_{G}^{(N)}(\vartheta ')\right] =\left[ T_{G}^{(N)}(\vartheta ')\otimes _{{\cal A}}T_{G}^{(N)}(\vartheta )\right] R(\vartheta -\vartheta ').\]
Taking the trace on the horizontal space \( {\cal A} \) yields:
\begin{equation}
\label{integraliprimi}
\left[ t^{(N)}_{G}(\vartheta ,\{\lambda _{i})),t^{(N)}_{G}(\vartheta ',\{\lambda _{i}))\right] =0
\end{equation}
(observe that the gauge must be exactly the same). At the various values of
\( \vartheta  \), \( t^{(N)}_{G}(\vartheta ,\{\lambda _{i})) \) describes
an infinite family of conserved charges. Then the system is integrable. This
holds in particular in the inhomogeneous and the twisted cases. 

The interesting fact, at this point, is that in some cases the transfer matrix
can be exactly diagonalized by Bethe Ansatz methods. This gives an exact expression
for eigenstates and eigenvalues of the operator \( U \).

\section{6 vertex model: main results\label{section:6vertici_varie}}

The theory developed until now is general and not referred to a specific model.
The simplest non trivial case to take into account in the previous framework
is the \( 4\times 4 \) R matrix, corresponding to the choice \( {\cal A}={\cal V} \).
As shown in \cite{baxter}, the most general solution is the so called 8 vertex
model. This name means that only 8, between the 16 entries of the R matrix,
are nonzero. A special case is the 6 vertex model, for which many results have
been obtained in the light-cone description: this will be the principal object
of this dissertation. The R matrix has the form (lower index are rows and upper
index are columns)
\begin{equation}
\label{6vRmatrix}
R(\vartheta ,\gamma )=\left( \begin{array}{cccc}
a &  &  & \\
 & c & b & \\
 & b & c & \\
 &  &  & a
\end{array}\right) 
\end{equation}
 As Boltzmann weights, all this nonvanishing entries must be real and nonnegative.
But Yang-Baxter equation is solved for all the values of this variables.\footnote{
Observe that the upper and lower entries are the same; this is the assumption
of parity invariance.
} As explained in (\ref{s-matrix.section}), the scattering interpretation is
possible only assuming the properties indicated in that section. 

There is a well known mapping between vertex models and spin chains (see \cite{baxter}),
i.e. the transfer matrix is the exponential of the quantum hamiltonian of the
chain:
\[
t^{(N)}=e^{-H}.\]
In the case of 8 vertex model, the hamiltonian is the XYZ(1/2) chain, while
in the special case of 6 vertex, is the XXZ(1/2) chain. In what follows, this
identification can be useful to interpret some facts connected with Bethe Ansatz.
The XXZ(1/2) chain hamiltonian is given by:
\[
H=\sum ^{N}_{i=1}\left[ \sigma _{x}^{i}\sigma _{x}^{i+1}+\sigma _{y}^{i}\sigma _{y}^{i+1}+(1-\cos \gamma )\sigma _{z}^{i}\sigma _{z}^{i+1}\right] \]
and \( \gamma  \) is the anisotropy. The \( \sigma  \) are Pauli matrices.
The total z-component of the spin will play an important role in Bethe Ansatz. 

This is a statistical approach, but it is possible to give a particle interpretation
to the same R matrix (\ref{6vRmatrix}) on the light-cone lattice. 

The simplest case of particles obeying Pauli exclusion principle (fermions)
and without internal degrees of freedom (color number) is assumed. This means
that in an event only one particle of type R and one L at most can take place.
In other words, one link has two states: empty or occupied. At every point there
are four links, that means 16 possible configurations associated to it. In terms
of events, these are the possible configurations that connect a point with the
nearest neighbours in the future. 

\begin{figure}[  htbp]
{\centering \rotatebox{270}{\includegraphics{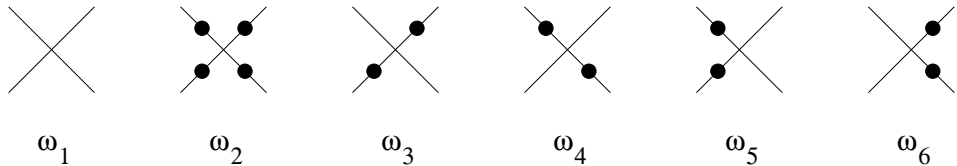}} \par}

\caption{\small 6 permitted amplitudes. Dots are particles.\label{amplitude.eps}}
\end{figure}

Assume now that only amplitudes that conserves the total number of particles
(R+L) are nonvanishing. This reduces to 6 the permitted configurations, as it
is shown in the figure \ref{amplitude.eps}. This is simply the 6 vertex model
whose R matrix is written in (\ref{6vRmatrix}). The assumption of integrability for this amplitudes gives the general six-vertex
model. The requirement of symmetry under parity transformation implies that
\( \omega _{3}=\omega _{4} \) and \( \omega _{5}=\omega _{6} \). The convention
adopted in the figure \ref{square lattice.eps} and in (\ref{t-ab-operator},
\ref{def-r-matrix}, \ref{6vRmatrix}) shows that
\[
b=\omega _{3}=\omega _{4},\quad c=\omega _{5}=\omega _{6}.\]
 Now, this R matrix can be written in an operatorial form, by defining a lattice
chiral fermion \( \psi _{A,n} \), with \( A=R,L \), and \( n \) labels the
sites. The anticommutation rules are the canonical ones: 
\begin{equation}
\label{fermion}
\left\{ \psi _{A,n},\psi _{B,m}\right\} =0,\qquad \left\{ \psi _{A,n},\psi ^{\dagger }_{B,m}\right\} =\delta _{AB}\delta _{nm}.
\end{equation}
This fermion has some interesting properties, that are exposed in the paper
\cite{ddv 87}, and are sketched in the following list:

\begin{enumerate}
\item the R matrix and all the other operators \( U_{any} \) can be written in an
operatorial form in terms of the fermion; 
\item the lattice hamiltonian, in the free case \( \omega _{1}=\omega _{2}=b=1,\; c=0 \),
can be explicitly written; by this, the following dispersion relation can be
obtained: \( E=\pm k \); this is the dispersion relation for a free massless
particle; the unusual fact is that it is monotonous, then there is no doubling
of fermions; this is a consequence of the nonlocality of the hamiltonian
\item the lattice hamiltonian admits a continuum limit \( N\, \rightarrow \, \infty  \)
and \( a\, \rightarrow \, 0 \) but with \( L=Na \) fixed; the locality is
recovered in this limit; the continuum equations of motion are those of the
massive Thirring model. The space is compactified on a cylinder.
\end{enumerate}
In the continuum limit, the massive Thirring model emerges as the field theory
characterising the scaling behavior of the dynamics on the lattice (remember
that \( L \) is finite; the scaling behaviour is understood in terms of this
\( L \)).

\section{6 vertex model: Bethe Ansatz\label{section:6_vertex_BA}}

The assumptions of unitarity and hermitian analyticity will be taken into account,
for the R matrix, and this requires that the variables in (\ref{6vRmatrix})
must have the specific form:
\begin{equation}
\label{R_matrix_entries}
a=a(\vartheta ,\gamma )=\sinh (i\gamma -\vartheta ),\quad b=b(\vartheta ,\gamma )=\sinh \vartheta ,\quad c=c(\vartheta ,\gamma )=i\sin \gamma 
\end{equation}
The transfer matrix defined by this R matrix (\ref{transfermatrix}, \ref{def-r-matrix})
can be diagonalized with Bethe Ansatz method. In terms of the spin chain, this
means that there are two operators, usually indicated by \( B(\vartheta ) \)
and \( C(\vartheta ) \), whose expression is known, and there is a ``reference
state''\footnote{
The ``reference state'' at this point is only a mathematical object. Physically
speaking, it corresponds to the ferromagnetic state with all the spins up.
} \( \left| \Omega \right\rangle  \) such that:
\[
C(\vartheta )\left| \Omega \right\rangle =0\]
and 
\begin{equation}
\label{base}
B(\vartheta _{1})...B(\vartheta _{M})\left| \Omega \right\rangle ,
\end{equation}
 for appropriate values of \( \vartheta _{j} \) is an eigenstate of the transfer
matrix. The ``appropriate values'' of \( \vartheta _{j} \) can be obtained
as  the solution of a set of \( M \) coupled nonlinear equations, that are
called Bethe Ansatz equations. In general, because of (\ref{integraliprimi}),
the transfer matrix contains all the conserved charges, in particular the hamiltonian.
Then the Bethe Ansatz eigenstates are also eigenstates of the hamiltonian. The
Hilbert space of the theory and the action of conserved charges on it are then
perfectly know. 

All this computations for the 6 vertex model were obtained in \cite{baxter, fadd-takt, devega 89};
the final results are written here for the eigenvalues of the inhomogeneous
and twisted transfer matrix: 
\[
\begin{array}{c}
\displaystyle \tau (\vartheta ,\Theta ,\omega )=e^{i\omega }\left[ a(\vartheta -\Theta )\, a(\vartheta +\Theta )\right] ^{N}\prod ^{M}_{j=1}\displaystyle\frac{\sinh \displaystyle\frac{\gamma }{\pi }\left[ i\displaystyle\frac{\pi }{2}+\vartheta _{j}+\vartheta \right] }{\sinh \displaystyle\frac{\gamma }{\pi }\left[ i\displaystyle\frac{\pi }{2}-\vartheta _{j}-\vartheta \right] }+\\
+e^{-i\omega }\left[ b(\vartheta -\Theta )\, b(\vartheta +\Theta )\right] ^{N}\prod ^{M}_{j=1}\displaystyle\frac{\sinh \displaystyle\frac{\gamma }{\pi }\left[ i\displaystyle\frac{3\pi }{2}-\vartheta _{j}-\vartheta \right] }{\sinh \displaystyle\frac{\gamma }{\pi }\left[ -i\displaystyle\frac{\pi }{2}+\vartheta _{j}+\vartheta \right] }
\end{array}\]
and the values of \( \vartheta _{j} \) are defined by the set of coupled nonlinear
equations called Bethe Ansatz equations: 
\begin{equation}
\label{bethe}
\left( \displaystyle\frac{\sinh \displaystyle\frac{\gamma }{\pi }\left[ \vartheta _{j}+\Theta +\displaystyle\frac{i\pi }{2}\right] \sinh \displaystyle\frac{\gamma }{\pi }\left[ \vartheta _{j}-\Theta +\displaystyle\frac{i\pi }{2}\right] }{\sinh \displaystyle\frac{\gamma }{\pi }\left[ \vartheta _{j}+\Theta -\displaystyle\frac{i\pi }{2}\right] \sinh \displaystyle\frac{\gamma }{\pi }\left[ \vartheta _{j}-\Theta -\displaystyle\frac{i\pi }{2}\right] }\right) ^{N}=-e^{2i\omega }\prod _{k=1}^{M}\displaystyle\frac{\sinh \displaystyle\frac{\gamma }{\pi }\left[ \vartheta _{j}-\vartheta _{k}+i\pi \right] }{\sinh \displaystyle\frac{\gamma }{\pi }\left[ \vartheta _{j}-\vartheta _{k}-i\pi \right] }
\end{equation}
where \( 2N \) is the length of the chain and \( N \) the number of sites
in a row of the light-cone lattice. The \( \vartheta _{j} \) are \emph{called
Bethe roots}, and in principle, can take any complex value. But there is a periodicity
in their imaginary part:
\begin{equation}
\label{periodicita'}
\vartheta _{j}\, \rightarrow \, \vartheta _{j}+\displaystyle\frac{\pi ^{2}}{\gamma }i
\end{equation}
then only a strip around the real axis must be taken into account for the Bethe
roots:
\begin{equation}
\label{strisciafisica}
\vartheta _{j}\in \mathbb {R}\times i\, \left] -\displaystyle\frac{\pi ^{2}}{2\gamma },\displaystyle\frac{\pi ^{2}}{2\gamma }\right] .
\end{equation}
Moreover, only the range 
\[
0<\gamma <\pi \]
will be examined. 

From Bethe Ansatz it is known that the whole spectrum of the theory can be obtained
using all the Bethe configurations having \( M\leq N \) and \( \vartheta _{j}\neq \vartheta _{k} \)
for every \( j\neq k. \) In general for a state with \( M \) roots the third
component of the spin of the chain is 
\begin{equation}
\label{spin-chain}
S=N-M,
\end{equation}
because every operator \( B(\vartheta _{i}) \) counts as \( -1 \) spin. 

This XXZ(1/2) chain has 2 states in every site, then \( 2^{2N} \) states. Then
the energy spectrum is upper and lower bounded. Changing the sign of the hamiltonian
gives another permitted physical system. This means that there are two possible
vacua. The first one is the so called ferromagnetic ground state, corresponding
to \( M=0 \) that is the reference state \( \left| \Omega \right\rangle . \)
It has spin \( S=N \). The second one is the antiferromagnetic ground state,
that can be obtained with \( M=N \) and all the roots \( \vartheta _{i} \)
real. It has spin \( S=0 \). 

In what follows, only the antiferromagnetic ground state will be considered,
because it has one important property: in the thermodynamic limit (\( N\, \rightarrow \, \infty  \))
it can be interpreted as a Dirac vacuum (a sea of particles created by \( B \))
and the excitations on this vacuum behave as particles. 

The energy \( E \) and momentum \( P \) of a state can be read out by the
transfer matrix eigenvalues by use of the equations (\ref{trans-ul}, \ref{trans-ur}).
The final form is:

\begin{equation}
\label{autovalori}
\displaystyle e^{i\displaystyle\frac{a}{2}(E\pm P)}=e^{\pm i\omega }\prod ^{M}_{j=1}\displaystyle\frac{\sinh \displaystyle\frac{\gamma }{\pi }\left[ i\displaystyle\frac{\pi }{2}-\Theta \pm \vartheta _{j}\right] }{\sinh \displaystyle\frac{\gamma }{\pi }\left[ i\displaystyle\frac{\pi }{2}+\Theta \mp \vartheta _{j}\right] }.
\end{equation}
Other integrals of motion can be obtained in a similar way by transfer matrix
(see (\ref{integraliprimi})). Observe that the second term in the transfer
matrix expression vanishes because \( b(0)=0 \).

For future analysis, it is more convenient to express the coupling constant
\( \gamma  \) in terms of a different variable \( p \):
\[
p=\frac{\pi }{\gamma }-1,\qquad 0<p<\infty .\]
Then all the expressions will be in terms of \( p. \)

\chapter{A NONLINEAR EQUATION FOR BETHE ANSATZ}

In this chapter the fundamental nonlinear integral equation (on the lattice)
for the Bethe Ansatz is derived. In literature it is known as Destri-de Vega
equation. It was obtained first in \cite{ddv 95, ddv 97} and, in the final
form, in \cite{noi NP}. On this equation the continuum limit is performed.

\section{Counting function\label{section:count-funct}}

It is possible to write the Bethe equations (\ref{bethe}) in terms of a counting
function, that will be called \( Z_{N}(\vartheta ). \) To do that, one important
preliminar definition is needed:
\[
\phi (\vartheta ,\nu )=i\log \frac{\sinh \frac{1}{p+1}(i\pi \nu +\vartheta )}{\sinh \frac{1}{p+1}(i\pi \nu -\vartheta )},\qquad \phi (-\vartheta ,\nu )=-\phi (\vartheta ,\nu )\]
where the primary interest is in values \( \nu =1/2,\, 1 \) and the oddness
on the analyticity strip around the real axis defines a precise logarithmic
branch: the so called fundamental branch. In the appendix \ref{funzione fi}
the complete structure of the cuts and all the relevant properties of this function
are exposed. The counting function\footnote{
The name counting function will became clear in the next paragraphs, where the
counting equation will be obtained. 
} can be defined by 
\begin{equation}
\label{def.Zn}
Z_{N}(\vartheta )=N[\phi (\vartheta +\Theta ,\displaystyle\frac{1}{2})+\phi (\vartheta -\Theta ,\displaystyle\frac{1}{2})]-\displaystyle\sum _{k=1}^{M}\phi (\vartheta -\vartheta _{k},1)+2\omega 
\end{equation}
used to express the logarithm of the Bethe equations, one obtains simply:
\begin{equation}
\label{quantum}
Z_{N}(\vartheta _{j})=2\pi I_{j}\, ,\qquad I_{j}\in \mathbb {Z}+\displaystyle\frac{1+\delta }{2},\qquad \delta =(M)_{mod\, 2}=(N-S)_{mod\, 2}\in \left\{ 0,1\right\} .
\end{equation}
The number \( I_{j} \) plays the role of a quantum number for the Bethe basic
vectors (\ref{base}) and it must be chosen depending on the value of \( \delta . \)
Notice that \( \delta  \) and \( \omega  \) play a similar role, because both
produce a shift in the quantum numbers \( I_{j} \) (if \( \omega  \) is absorbed
in the definition of \( I_{j} \)): in the first case the shift is exactly \( \pi , \)
in the second case it is a real (possibly complex) number. This means that the
variable \( \delta  \) can be absorbed in \( \omega  \) but the most convenient
choice is to use them both.

Observe that Bethe roots can be obtained as zeros of the equation:
\begin{equation}
\label{zero-condition}
1+(-1)^{\delta }e^{iZ_{N}(\vartheta _{j})}=0
\end{equation}

\section{Classification of Bethe roots\label{classificazione}}

From Bethe Ansatz it is known that a Bethe state (\ref{base}) is uniquely characterized
by the set of quantum numbers \( \left\{ I_{j}\right\} _{j=1,...,M}\: ,\quad M\leq N \)
that appear in (\ref{quantum}). Notice that \( M\leq N \) means \( S\geq 0. \)
The values of \( \vartheta _{j} \) to put in (\ref{base}) can be obtained
solving Bethe equations. It is also known that only states with 
\[
\vartheta _{j}\neq \vartheta _{i},\quad \forall \: j\neq i\]
are necessary to form a basis for the space of states. It is a sort of fermionic
character for Bethe states \cite{faddeev 95}. 

As usual in many cases, Bethe roots can be either real or in complex conjugate
pairs (for large \( N \)). In the specific case (\ref{bethe}), there is another
possibility, due to periodicity (\ref{periodicita'}): if a complex solution
has imaginary part \( \Im m\, \vartheta =\frac{\pi }{2}(p+1) \) it appears
as a single (in (\ref{bethe}) it produces the left hand side real, then its
complex conjugate is not required). It is called \emph{self-conjugate root}.
Remember now that the maximal number of real roots (\( M=N \)) describes the
antiferromagnetic ground state. From the point of view of the counting function,
a more precise classification of roots is required:

\begin{itemize}
\item \emph{real roots}; they are real solutions of (\ref{quantum}) that appear in
the vector (\ref{base}); their number is \( M_{R} \);
\item \emph{holes}; real solutions of (\ref{quantum}) that do NOT appear in the vector
(\ref{base}); their number is \( N_{H} \);
\item \emph{special roots/holes} (special objects); they are real roots or holes where
the derivative \( Z_{N}'(\vartheta _{j}) \) is negative\footnotemark{}
; their number is \( N_{S} \); they must be counted both as ``normal'' and
as ``special'' objects;
\item \emph{close pairs}; complex conjugate solutions with imaginary part in the range
\( 0<|\Im m\, \vartheta |<\pi \min (1,p) \); their number is \( M_{C} \);
\item \emph{wide roots in pairs}; complex conjugate solutions with imaginary part
in the range \( \pi \min (1,p)<|\Im m\, \vartheta |<\pi \frac{p+1}{2} \);
\item \emph{self-conjugate roots} (wide roots appearing as single); \( \Im m\, \vartheta =\pi \frac{p+1}{2} \);
their number is \( M_{SC} \).
\end{itemize}
\footnotetext{
The characteristics of this type of solutions will be completely clarified in
the next sections.
}%
The total number of wide roots appearing in pairs or single is \( M_{W} \).
The following notation will be used (sometimes) for later convenience, to indicate
the position of the solutions: \( h_{j} \) for holes, \( y_{j} \) for special
objects, \( c_{j} \) for close roots, \( w_{j} \) for wide roots.

\begin{figure}[  htbp]
{\centering \includegraphics{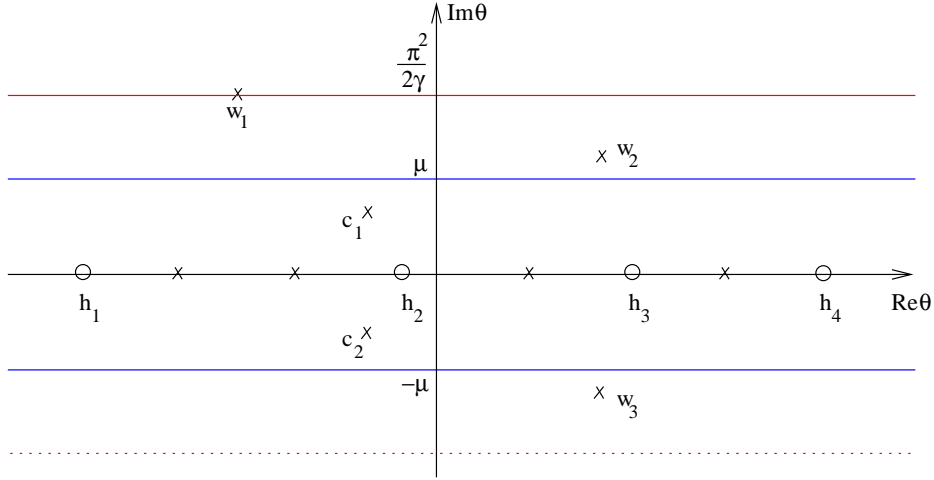} \par}

\caption{\small The different types of roots and holes and their position in the complex plane.
\protect\( \mu \protect \) denotes \protect\( \pi \min \left\{ 1,\, p\right\} \protect \).
The upper line at \protect\( \frac{\pi ^{2}}{2\gamma }=\frac{\pi }{2}(p+1)\protect \)
is the self-conjugate one.\label{radici.eps}}
\end{figure}

Complex roots with imaginary part larger than the self-conjugates are not required
because of the periodicity of Bethe equations. This classification is not at
all academic; it will play an important role in the physical interpretation
of the final equation. A graphical representation of the various types of solutions
is given in figure \ref{radici.eps}. An important remark must be made: from the definition itself of \( Z_{N} \)
(\ref{def.Zn}) it is obvious that only for states without complex roots the
fundamental strip for \( \phi  \) (see (\ref{strisciafond})), that is the
largest strip around the real axis without singularities, is the fundamental
strip for \( Z_{N}. \) In all the other cases the analyticity strip for \( Z_{N} \)
is narrower, and depends on the imaginary parts of the complex roots.

An important property follows from this classification: the \( Z_{N} \) function
is \emph{real analytic} if \( \omega  \) is a real number
\begin{equation}
\label{analiticita_reale}
Z_{N}\left( \vartheta ^{*}\right) =\left( Z_{N}(\vartheta )\right) ^{*}
\end{equation}

\section{Counting equation}

It is possible to obtain an equation relating the numbers of all the various
types of solutions. The path is simple, and makes use of the asymptotic values
obtained in (\ref{phi-limiti}) to calculate the limits \( \vartheta \, \rightarrow \, \pm \infty  \)
in \( Z_{N} \) (\ref{def.Zn}). Observe that the term \( \phi (\vartheta -\vartheta _{j},1) \),
in the case of the wide roots, takes contributions from the horizontal strips
next to the fundamental, as in figure \ref{cuts.eps}. This contributions depend
from \( M_{W\downarrow } \) (number of wide roots below the real axis) and
\( M_{W\uparrow } \) (number of wide roots over the real axis). The asymptotic
limits are then:
\begin{equation}
\label{ZNlimiti}
\begin{array}{c}
\displaystyle Z_{N}(+\infty )=N\pi +\pi \displaystyle\frac{p-1}{p+1}S+2\pi \, \mathrm{sign}(p-1)\, M_{W\downarrow }+2\omega \\
\\
Z_{N}(-\infty )=-N\pi -\pi \displaystyle\frac{p-1}{p+1}S-2\pi \, \mathrm{sign}(p-1)\, M_{W\uparrow }+2\omega 
\end{array}
\end{equation}
Observe that the number of self-conjugate roots is: \( M_{W\uparrow }-M_{W\downarrow }=M_{SC}. \)
On the lattice, all the values for \( N,\, S \) are permitted. But in view
of the continuum limit, \( N \) is expected to be larger than \( S \): \( N\gg S \)
(also the number of holes and complex roots is expected to be much smaller than
\( N \)). Then 
\[
Z_{N}(+\infty )>Z_{N}(-\infty ).\]
The function is globally increasing (but not necessarily monotonous: in some
points it can be a decreasing function). Define now two variables \( \zeta _{\pm } \),
choosing \( k_{\pm } \) in such a way that they take value in the indicated
interval:
\[
\zeta _{\pm }=\pm Z_{N}(\pm \infty )+\pi \delta -2\pi k_{\pm },\qquad -\pi <\zeta _{\pm }\leq \pi \]
Using the previous expressions for \( Z_{N} \) the result is:
\[
\zeta _{\pm }=2\pi \left( -\frac{S}{p+1}\pm \frac{\omega }{\pi }+\left\lfloor \frac{1}{2}+\frac{S}{p+1}\mp \frac{\omega }{\pi }\right\rfloor \right) \]
where the new symbol \( \left\lfloor x\right\rfloor  \) is the integer part
of \( x \), that is the largest integer smaller or equal to \( x \). In this
way, the following expression holds: 
\begin{equation}
\label{Imaxmin}
\begin{array}{c}
Z_{N}(+\infty )=2\pi I_{max}+\pi +\zeta _{+}\\
Z_{N}(-\infty )=2\pi I_{min}-\pi -\zeta _{-}
\end{array}
\end{equation}
where \( I_{max} \) is the largest quantum number satisfying \( I_{max}<\displaystyle\frac{Z_{N}(+\infty )}{2\pi } \)
and \( I_{min} \) is the smallest quantum number satisfying \( I_{min}>\displaystyle\frac{Z_{N}(-\infty )}{2\pi } \).
This very precise definition is required to take into account for the special
roots/holes that can appear in the tails (this can happen when \( Z_{N}'<0 \)
for very large values of \( |\vartheta | \)). 

The total number of real solutions is the sum of real roots and holes, and in
terms of \( I_{max},\: I_{min} \) it can be written as:
\[
M_{R}+N_{H}=I_{max}-I_{min}+1+2N_{S}.\]
Using the equations (\ref{spin-chain}, \ref{Imaxmin}) and the expression (\ref{ZNlimiti})
the result is:
\begin{equation}
\label{latticecounting}
\begin{array}{c}
N_{H}-2N_{S}=2S+M_{C}+2\, \theta (p-1)\, M_{W}+\\
\\
-\left\lfloor \displaystyle\frac{1}{2}+\displaystyle\frac{S}{p+1}+\displaystyle\frac{\omega }{\pi }\right\rfloor -\left\lfloor \displaystyle\frac{1}{2}+\displaystyle\frac{S}{p+1}-\displaystyle\frac{\omega }{\pi }\right\rfloor .
\end{array}
\end{equation}
It is the so called \emph{lattice counting equation}. Remember that \( S \)
is a nonnegative integer. In the case of \( \omega =0 \), it turns out that
\( N_{H} \) is even (remember that \( M_{C} \) is the number of close roots,
and is even).

The most important fact is that in this equation doesn't appear the number of
real roots. This fact, together to what will be explained in the next paragraph,
allows to consider the real roots as a sea of particles (Dirac vacuum) and all
other types of solutions (holes, complex) as excitations on this sea.

\section{Non linear integral equation (I)\label{section:NLIE_1}}

In this section an equation generating \( Z_{N} \) will be obtained. The counting
function (\ref{def.Zn}) can be written in the following way:

\noindent 
\begin{equation}
\label{count}
\begin{array}{c}
\displaystyle Z_{N}(\vartheta )=N\left[ \phi (\vartheta +\Theta ,\displaystyle\frac{1}{2})+\phi (\vartheta -\Theta ,\displaystyle\frac{1}{2})\right] +\displaystyle\sum ^{N_{H}}_{k=1}\phi (\vartheta -h_{k},1)+\\
-\displaystyle\sum ^{M_{C}+M_{W}}_{k=1}\phi (\vartheta -\xi _{k},1)-\displaystyle\sum ^{M_{R}+N_{H}}_{k=1}\phi (\vartheta -x_{k},1)
\end{array}
\end{equation}
In this case \( \xi _{k} \) collects close and wide roots, and \( x_{k} \)
the real roots and holes. Now, it is convenient to deal with the derivative
of this expression:
\begin{equation}
\label{count-derivata}
\begin{array}{c}
\displaystyle Z_{N}'(\vartheta )=N\left[ \phi '(\vartheta +\Theta ,\displaystyle\frac{1}{2})+\phi '(\vartheta -\Theta ,\displaystyle\frac{1}{2})\right] +\displaystyle\sum ^{N_{H}}_{k=1}\phi '(\vartheta -h_{k},1)+\\
-\displaystyle\sum ^{M_{C}+M_{W}}_{k=1}\phi '(\vartheta -\xi _{k},1)-\displaystyle\sum ^{M_{R}+N_{H}}_{k=1}\phi '(\vartheta -x_{k},1)
\end{array}
\end{equation}

\begin{figure}[  htbp]
{\centering \includegraphics{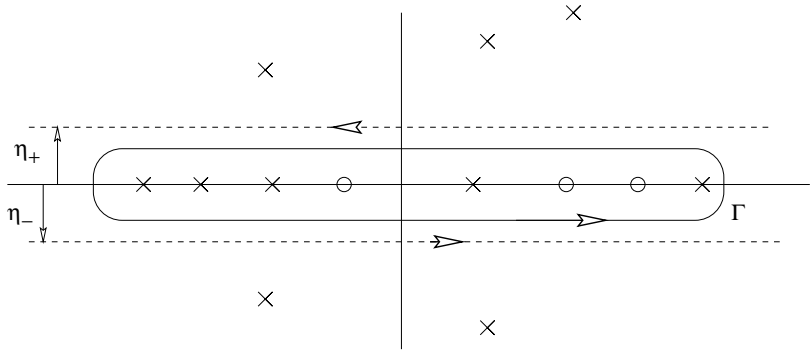} \par}

\caption{\small Contours for the integration. The crosses are roots while the circles are holes.
\label{curvagamma.eps}}
\end{figure}

\noindent Both the previous equations hold for \( \vartheta \in \mathbb C \).
Let \( \hat{x} \) be a real solution of the Bethe equation. Assume \( Z'_{_{N}}(\hat{x})\neq 0 \)
and define a complex neighbour of \( \hat{x} \) by a small \( |\nu |\ll 1 \):
\( \mu =\hat{x}+\nu  \). Clearly \( (-1)^{\delta }e^{iZ_{N}(\hat{x})}=-1 \).
Consider the expression:

\begin{equation}
\label{serie}
1+(-1)^{\delta }e^{iZ_{N}(\hat{x}+\nu )}\approx 1+(-1)^{\delta }e^{iZ_{N}(\hat{x})+i\nu Z_{N}'(\hat{x})}\approx -i\nu Z_{N}'(\hat{x})
\end{equation}
 then the following identity holds:

\begin{equation}
\label{identita}
\displaystyle\frac{1}{\mu -\hat{x}}=\displaystyle\frac{1}{\nu }=\displaystyle\frac{(-1)^{\delta }e^{iZ_{N}(\hat{x}+\nu )}iZ_{N}'(\hat{x}+\nu )}{1+(-1)^{\delta }e^{iZ_{N}(\hat{x}+\nu )}}+...
\end{equation}
(the dots are regular terms in \( \mu -\hat{x} \)). From the Cauchy theorem
and from (\ref{identita}), given an analytic function \( f(x) \) on an appropriate
strip containing the real axis, yields: 
\begin{equation}
\label{cauchy}
f(\hat{x})=\displaystyle\oint _{\Gamma _{\hat{x}}}\displaystyle\frac{d\mu }{2\pi i}\displaystyle\frac{f(\mu )}{\mu -\hat{x}}=\displaystyle\oint _{\Gamma _{\hat{x}}}\displaystyle\frac{d\mu }{2\pi i}f(\mu )\displaystyle\frac{(-1)^{\delta }e^{iZ_{N}(\mu )}iZ_{N}'(\mu )}{1+(-1)^{\delta }e^{iZ_{N}(\mu )}}
\end{equation}
where \( \Gamma _{\hat{x}} \) is a anti-clockwise curve encircling \( \hat{x} \)
and avoiding other singularities of the denominator, i.e. other Bethe solutions
(real or complex). It is always possible to find such a closed curve, because
Bethe solutions are finite in number. An equation like (\ref{cauchy}) can be
written for all the real roots of (\ref{quantum}), \( {x_{k},\: \forall \: k} \).
The derivative \( \phi '(\vartheta ,1) \) is an analytic function, if poles
are avoided, and applying to that the expression (\ref{cauchy}), the last sum
in (\ref{count-derivata}) becomes:

\begin{equation}
\label{integr_gamma}
\begin{array}{c}
\displaystyle \displaystyle\sum ^{M_{R}+N_{H}}_{k=1}\phi '(\vartheta -x_{k},1)=\displaystyle\sum ^{M_{R}+N_{H}}_{k=1}\displaystyle\oint _{\Gamma _{x_{k}}}\displaystyle\frac{d\mu }{2\pi i}\phi '(\vartheta -\mu ,1)\displaystyle\frac{(-1)^{\delta }e^{iZ_{N}(\mu )}iZ_{N}'(\mu )}{1+(-1)^{\delta }e^{iZ_{N}(\mu )}}=\\
=\displaystyle\oint _{\Gamma }\displaystyle\frac{d\mu }{2\pi i}\phi '(\vartheta -\mu ,1)\displaystyle\frac{(-1)^{\delta }e^{iZ_{N}(\mu )}iZ_{N}'(\mu )}{1+(-1)^{\delta }e^{iZ_{N}(\mu )}}
\end{array}
\end{equation}
The sum on the contours was modified to a unique curve \( \Gamma  \) encircling
all the real solutions \( {x_{k}} \), and avoiding the complex Bethe solutions
(this is possible because they are finite in number), as in the figure \ref{curvagamma.eps}. 

Clearly the \( \Gamma  \) curve must be contained in the strip 
\[
0<\eta _{+},\eta _{-}<\min \{\pi ,\pi p,|\Im m\: c_{k}|\: \forall \: k\}\]
Without loss of generality, assume that \( \eta _{+}=\eta _{-}=\eta  \), and
deform \( \Gamma  \) to the contour of the strip characterized by \( \eta  \).
The regions at \( \pm \infty  \) do no contribute because of the vanishing
of \( \phi ' \), then the integral can be performed only on the lines \( \mu =x\pm i\eta  \),
where \( x \) is real\footnote{
\( \int =\int _{-\infty }^{+\infty } \) In all the cases where the limits of
integration are not explicitly specified, the integral is taken on the whole
real axis.
}:

\begin{equation}
\label{integr_striscia}
\begin{array}{c}
\displaystyle \displaystyle\oint _{\Gamma }\displaystyle\frac{d\mu }{2\pi i}\phi '(\vartheta -\mu ,1)\displaystyle\frac{(-1)^{\delta }e^{iZ_{N}(\mu )}iZ_{N}'(\mu )}{1+(-1)^{\delta }e^{iZ_{N}(\mu )}}=\\
\\
=\displaystyle\int \displaystyle\frac{dx}{2\pi i}\phi '(\vartheta -x+i\eta ,1)\displaystyle\frac{(-1)^{\delta }e^{iZ_{N}(x-i\eta )}iZ_{N}'(x-i\eta )}{1+(-1)^{\delta }e^{iZ_{N}(x-i\eta )}}+\\
\\
-\displaystyle\int \displaystyle\frac{dx}{2\pi i}\phi '(\vartheta -x-i\eta ,1)\displaystyle\frac{(-1)^{\delta }e^{iZ_{N}(x+i\eta )}iZ_{N}'(x+i\eta )}{1+(-1)^{\delta }e^{iZ_{N}(x+i\eta )}}
\end{array}
\end{equation}
The first term at the right can be written as:

\begin{equation}
\label{integr}
\begin{array}{c}
\displaystyle \displaystyle\frac{(-1)^{\delta }e^{iZ_{N}(x-i\eta )}iZ_{N}'(x-i\eta )}{1+(-1)^{\delta }e^{iZ_{N}(x-i\eta )}}=\\
\displaystyle\frac{-(-1)^{\delta }e^{-iZ_{N}(x-i\eta )}iZ_{N}'(x-i\eta )}{1+(-1)^{\delta }e^{-iZ_{N}(x-i\eta )}}+iZ_{N}'(x-i\eta )
\end{array}
\end{equation}
Putting this in (\ref{integr}) the integral contribution from \( Z_{N}'(x-i\eta ) \)
is independent from \( \eta  \) because no singularities are crossed if \( \eta \, \rightarrow \, 0 \)
and at the infinity the integrand is zero. The following convenient redefinition
can be introduced (see (\ref{phi_primo})):
\begin{equation}
\label{kappa}
K(x)\equiv \displaystyle\frac{\phi '(x,1)}{2\pi }=\displaystyle\frac{1}{\pi (p+1)}\displaystyle\frac{\sin \displaystyle\frac{2\pi }{p+1}}{\cosh \displaystyle\frac{2\vartheta }{p+1}-\cos \displaystyle\frac{2\pi }{p+1}}
\end{equation}
In general, for every complex value of \( \vartheta  \):

\begin{equation}
\label{integ}
\begin{array}{c}
\displaystyle \displaystyle\int dx\left[ \delta (x)Z_{N}'(\vartheta -x)+K(\vartheta -x)Z_{N}'(x)\right] =N\left[ \phi '(\vartheta +\Theta ,\displaystyle\frac{1}{2})+\phi '(\vartheta -\Theta ,\displaystyle\frac{1}{2})\right] +\\
+\displaystyle\sum ^{N_{H}}_{k=1}\phi '(\vartheta -h_{k},1)-\displaystyle\sum ^{M_{C}+M_{W}}_{k=1}\phi '(\vartheta -\xi _{k},1)+\\
-\displaystyle\int \displaystyle\frac{dx}{2\pi i}\phi '(\vartheta -x+i\eta ,1)\displaystyle\frac{-(-1)^{\delta }e^{-iZ_{N}(x-i\eta )}iZ_{N}'(x-i\eta )}{1+(-1)^{\delta }e^{-iZ_{N}(x-i\eta )}}+\\
+\displaystyle\int \displaystyle\frac{dx}{2\pi i}\phi '(\vartheta -x-i\eta ,1)\displaystyle\frac{(-1)^{\delta }e^{iZ_{N}(x+i\eta )}iZ_{N}'(x+i\eta )}{1+(-1)^{\delta }e^{iZ_{N}(x+i\eta )}}
\end{array}
\end{equation}
The second term on the left takes the form
\[
\int dx\, K(\vartheta -x,1)\, Z_{N}'(x)=\int dx\, K(x,1)\, Z_{N}'(\vartheta -x);\]
that can be obtained by shifting the integration line: \( x\, \rightarrow \, x+\Im m\, \vartheta  \).
But if poles are crossed (this can happen if the imaginary part is sufficiently
large), their contribution must appear in the right hand side of the equation. 

Using Fourier transformations, the convolution in the left hand side of (\ref{integ})
can be simply handled. Observe first that the Fourier transform of \( \left[ \delta (\vartheta -x)+K(\vartheta -x)\right]  \)
is simply \( 1+\tilde{K} \). In the appendix \ref{funzione fi} the form of
\( \tilde{K} \) is given (as \( \widetilde{\phi '} \) in (\ref{Fourierphi'})).
It is obvious that \( 1+\tilde{K} \) is nonvanishing, so it can be reversed.
Call \( \tilde{\Delta } \) the inverse:
\[
\tilde{\Delta }(k)\equiv \frac{1}{1+\tilde{K}(k)}.\]
The inverse Fourier transform of \( \tilde{\Delta } \), indicated as \( \Delta (x) \),
is a distribution, because \( \tilde{\Delta }(k)\, \rightarrow \, 1 \) for
\( k\, \rightarrow \, \pm \infty . \) The following obvious equation holds:
\[
\int dx\, \Delta (x)\, \left( \delta (\theta -x)+K(\vartheta -x)\right) =\delta (\theta )\]
and this allows to invert the convolution in (\ref{integ}). Calling globally
\( {\cal F}(\vartheta ) \) the right hand side of (\ref{integ}) the inversion
is:
\begin{equation}
\label{inversione}
Z_{N}'(\vartheta )=\displaystyle\int dx\, \Delta (\vartheta -x)\, {\cal F}(x).
\end{equation}
The various terms in \( {\cal F} \) give different contributions, that shall
be computed in the following.

For the two terms \( \phi '(\vartheta \pm \Theta ,\displaystyle\frac{1}{2}) \) the convolution
can be completely performed using (\ref{convolution}):
\begin{equation}
\label{termine_cinetico}
\displaystyle\int dx\, \Delta (\vartheta -x)\, \phi '(x\pm \Theta ,\displaystyle\frac{1}{2})=\displaystyle\frac{1}{\cosh (\vartheta \pm \Theta )}.
\end{equation}
 The effect of the inversion on \( \phi '(\vartheta ,1) \) gives an important
object, that will be called \( G: \)
\begin{equation}
\label{kernel_G}
\displaystyle\int dx\, \Delta (\vartheta -x)\, \displaystyle\frac{\phi '(x,1)}{2\pi }=\displaystyle\int dx\, \Delta (\vartheta -x)\, K(x)\equiv G(\vartheta )
\end{equation}
Using Fourier transforms (see \ref{convolution}) for \( \Delta  \) and \( \phi ' \)
(\ref{Fourierphi'}) the following expression can be obtained:
\begin{equation}
\label{funzioneG}
\begin{array}{c}
\displaystyle G(\theta )=\displaystyle\frac{1}{p+1}\displaystyle\int \displaystyle\frac{dk}{2\pi }e^{ik\theta \displaystyle\frac{1}{p+1}}\tilde{K}(k)\displaystyle\frac{1}{1+\tilde{K}(k)}=\\
=\displaystyle\frac{1}{2\pi }\displaystyle\int dk\, e^{ik\theta }\displaystyle\frac{\sinh \displaystyle\frac{\pi (p-1)k}{2}}{2\sinh \displaystyle\frac{\pi pk}{2}\, \cosh \displaystyle\frac{\pi k}{2}}.
\end{array}
\end{equation}
The expression (\ref{Fourierphi'}) used in the previous computation for the
Fourier transform of \( \phi ' \) holds only for \( \theta  \) in the fundamental
strip (\ref{strisciafond}), i.e. for real solutions or close pairs, as in section
\ref{classificazione}. This means that contributions to \( Z_{N}' \) coming
from wide roots require a different approach (see later). Instead the terms
containing holes and close roots are completely arranged in this way. 

The Fourier transform of (\ref{funzioneG}) is obviously
\[
\tilde{G}(k)=\frac{\tilde{K}(k)}{1+\tilde{K}(k)}.\]
This is an even function, real on the real axis (\emph{real analyticity}), positive
if \( p>1 \) and negative in the opposite case. The same properties, clearly,
hold for \( G(\theta ) \). The function \( \tilde{G} \) vanishes exponentially
for large values of \( |k| \):
\[
\displaystyle \tilde{G}\, \sim \, e^{-\frac{\pi \, \min (1,p)}{p+1}|k|}.\]
(consequently also the function \( G(\theta ) \) has a similar asymptotic behaviour).
Because of this, the Fourier transform in (\ref{funzioneG}) at the points
\[
|\Im m\, \theta |=\pm \pi \, \min (1,p)\]
has a singularity. As before, this points correspond exactly to the limit between
close roots and wide roots. This confirms exactly that wide roots require a
different analysis. Call \( source \) the contribution to (\ref{inversione})
by one wide root put in \( w \):
\[
\begin{array}{c}
\displaystyle source=\displaystyle\int dx\, \Delta (x)\, \phi '(\vartheta -x-w,1)=\\
=\displaystyle\frac{1}{p+1}\displaystyle\int dx\, \displaystyle\int \displaystyle\frac{dk}{2\pi }e^{ikx\displaystyle\frac{1}{p+1}}\tilde{\Delta }(k)\, \phi '(\vartheta -x-w,1).
\end{array}\]
Using the known expression (\ref{phi_primo}) for \( \phi ' \) the computation
can be performed into a closed form. The integral in \( x \) must be performed
first. It is a Fourier transform, but because of the \( w \) contribution,
it is different from the Fourier transform of \( \phi ' \) calculated in (\ref{Fourierphi'}),
that holds for \( \vartheta  \) near the real axis. This is because \( \Im m\, w \)
is larger than the position of the first singularity of \( \phi ' \), and the
poles to take into account in the two cases are different. The result can be
written in the following form:
\begin{equation}
\label{Gseconda}
source=2\pi G_{II}(\vartheta -w)
\end{equation}
 where the function \( G_{II} \) is defined by (for \( |\Im m\theta |>\pi \min (1,p) \)):

\begin{equation}
\label{Gseconda_esplicita}
2\pi G_{II}(\theta )=\left\{ \begin{array}{c}
\displaystyle \displaystyle\frac{i}{p}\left[ \coth \displaystyle\frac{\theta }{p}\: \mathrm{sign}\Im m(\theta )+\coth \displaystyle\frac{(i\pi -\theta \, \mathrm{sign}\Im m(\theta ))}{p}\right] \qquad \mathrm{if}\quad p>1\\
\\
\mathrm{sign}\Im m(\theta )\: i\left[ \displaystyle\frac{1}{\sinh \theta }+\displaystyle\frac{1}{\sinh (\theta -i\pi p\, \mathrm{sign}\Im m(\theta ))}\right] \qquad \mathrm{if}\quad p<1
\end{array}\right. 
\end{equation}
A quite singular property, pointed out in \cite{ddv 97}, is that this function
can be interpreted as the so called \emph{second determination} of the previous
function \( G \). The general definition of the second determination is:
\begin{equation}
\label{seconda_det}
f_{II}(\theta )=\left\{ \begin{array}{c}
f(\theta )+f(\theta -i\pi \, \mathrm{sign}\Im m(\theta ))\qquad \mathrm{if}\quad p>1\\
\\
f(\theta )-f(\theta -i\pi p\, \mathrm{sign}\Im m(\theta ))\qquad \mathrm{if}\quad p<1
\end{array}\right. \: \textrm{ for }\: |\Im m\theta |>\pi \min (1,p)
\end{equation}
Applied to (\ref{funzioneG}), it yields exactly (\ref{Gseconda}). 

\begin{figure}[  htbp]
{\centering \includegraphics{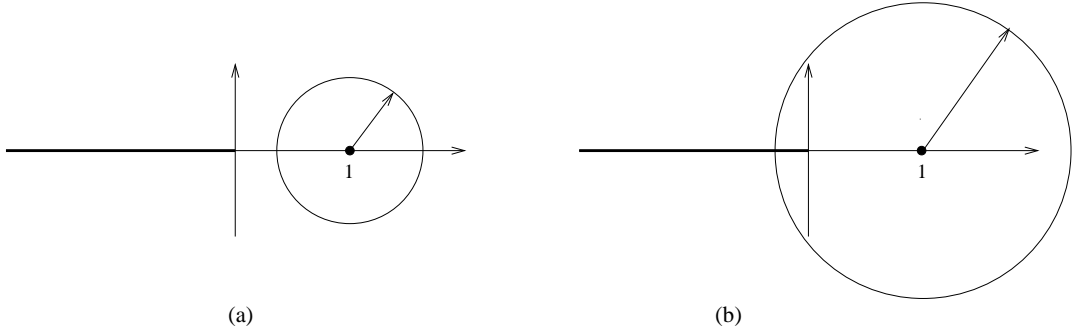} \par}

\caption{\small Plane of the \protect\( \log \protect \). The circles have radius \protect\( \left| e^{iZ_{N}(x+i\eta )}\right| \protect \).
In (a) it is \protect\( <1\protect \), in (b) it is \protect\( >1\protect \).\label{crossing.eps}}
\end{figure}

\begin{figure}[  htbp]
{\centering \includegraphics{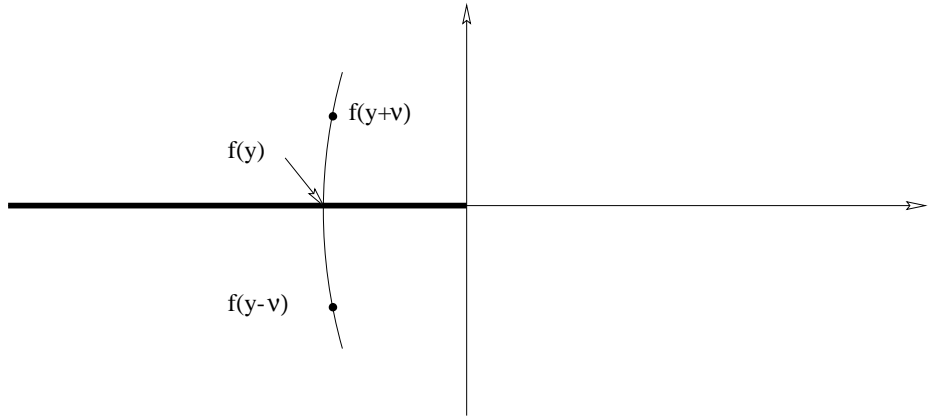} \par}

\caption{\small Jump in the \protect\( \log \protect \): \protect\( \log f(\hat{y}+\nu )-\log f(\hat{y}-\nu )=2\pi i\protect \)\label{special.eps}}
\end{figure}

Taking into account (\ref{termine_cinetico}, \ref{kernel_G}, \ref{Gseconda})
in (\ref{inversione}), the equation (\ref{integ}) takes the form of an integro-differential
equation for \( Z_{N}' \):

\begin{equation}
\label{zetap}
\begin{array}{c}
\displaystyle Z_{N}'(\vartheta )=N\left[ \displaystyle\frac{1}{\cosh (\vartheta +\Theta )}+\displaystyle\frac{1}{\cosh (\vartheta -\Theta )}\right] +\displaystyle\sum ^{N_{H}}_{k=1}2\pi G(\vartheta -h_{k})+\\
-\displaystyle\sum ^{M_{C}}_{k=1}2\pi G(\vartheta -c_{k})-\displaystyle\sum ^{M_{W}}_{k=1}2\pi G_{II}(\vartheta -w_{k})+\\
+\displaystyle\frac{1}{i}\displaystyle\int dxG(\vartheta -x-i\eta )\displaystyle\frac{(-1)^{\delta }e^{iZ_{N}(x+i\eta )}iZ_{N}'(x+i\eta )}{1+(-1)^{\delta }e^{iZ_{N}(x+i\eta )}}+\\
-\displaystyle\frac{1}{i}\displaystyle\int dxG(\vartheta -x+i\eta )\displaystyle\frac{-(-1)^{\delta }e^{-iZ_{N}(x-i\eta )}iZ_{N}'(x-i\eta )}{1+(-1)^{\delta }e^{-iZ_{N}(x-i\eta )}}
\end{array}
\end{equation}
 This equation holds for both the cases of \( \omega  \) real or complex. The
most common case is \( \omega \in \mathbb R \). This allows to write the last
two lines in a more compact form, valid only if \( \vartheta  \) is on the
real axis. The real analyticity of \( Z_{N} \) and of \( G \) is used:
\[
2\Im m\int dxG(\vartheta -x-i\eta )\frac{(-1)^{\delta }e^{iZ_{N}(x+i\eta )}iZ_{N}'(x+i\eta )}{1+(-1)^{\delta }e^{iZ_{N}(x+i\eta )}}\qquad \textrm{if }\vartheta \in \mathbb R.\]
Observe that the second factor in the integral terms seems to be the derivative
of \( \log \left[ 1+(-1)^{\delta }e^{\pm iZ_{N}(x\pm i\eta )}\right]  \), but
this substitution requires some care, because of the polydromy of the logarithmic
function. Call, for the sake of simplicity, the argument of the \( \log  \):

\begin{equation}
\label{funz_f}
f(x)=1+(-1)^{\delta }e^{iZ_{N}(x+i\eta )}
\end{equation}
 and use the fundamental branch as in the appendix \ref{funzione fi}. If \( f(x) \)
cross the cut of the \( \log  \) on the negative real axis, the function \( \log _{FD}f(x) \)
has a jump by \( \pm 2\pi i \). The exact real point \( y_{\uparrow } \) where
this happens (suppose for simplicity only one such point; the extension is trivial)
is given (see the figure \ref{crossing.eps})by the condition \( f(y_{\uparrow })\in \left] -\infty ,0\right[  \), that
is:
\begin{equation}
\label{condizione1_special}
1+(-1)^{\delta }e^{i\, \Re e\, Z_{N}(y_{\uparrow }+i\eta )}=0\qquad \mathrm{and}\qquad \Im mZ_{N}(y_{\uparrow }+i\eta )<0.
\end{equation}
Only the expression 
\[
\log _{FD}f(x)-2\pi i\theta (x-y_{\uparrow })\epsilon \]
 gives rise to a continuous function. \( \epsilon =\pm 1 \) is a sign that
is positive if the cross is in the clockwise direction (from below to up) as
in figure \ref{special.eps} and negative in the opposite case. An important
observation can be made by the following Taylor expansion:
\begin{equation}
\label{sviluppo_realZ}
\Re e\, Z_{N}(y_{\uparrow }+i\eta )=Z_{N}(y_{\uparrow })-\displaystyle\frac{\eta ^{2}}{2}Z_{N}''(y_{\uparrow })+...
\end{equation}

\begin{equation}
\label{sviluppo_imagZ}
\Im m\, Z_{N}(y_{\uparrow }+i\eta )=\eta \left( Z_{N}'(y_{\uparrow })-\displaystyle\frac{\eta ^{2}}{6}Z_{N}'''(y_{\uparrow })+...\right) .
\end{equation}
It is clear that if and only if \( \eta \, \rightarrow \, 0 \) the condition
(\ref{condizione1_special}) becomes the definition of a special solution, i.e.
\( y_{\uparrow } \) is a special object only in the limit of \( \eta  \) going
to zero because in this case the second terms in (\ref{sviluppo_realZ}, \ref{sviluppo_imagZ})
are negligible. Then the condition (\ref{condizione1_special}) becomes \( Z_{N}'(y_{\uparrow })<0 \)
and \( Z_{N}=2\pi I \). The sign must be chosen positive: \( \epsilon =1. \)\footnote{
In what follows, always this case will be assumed.
} For the general case \( \eta \neq 0 \), \( y_{\uparrow } \) is a special
object \( y \) shifted by a little bit.

A completely analogous analysis can be performed for the lower integral, obtaining
a \( y_{\downarrow } \). From (\ref{sviluppo_realZ}), because of the real
analyticity of \( Z_{N} \), the following equation holds: \( y_{\uparrow }=y_{\downarrow }\equiv \hat{y}. \)
The result is:
\begin{equation}
\label{derivata_continua}
\displaystyle\frac{(-1)^{\delta }e^{\pm iZ_{N}(x\pm i\eta )}(\pm i)Z_{N}'(x\pm i\eta )}{1+(-1)^{\delta }e^{\pm iZ_{N}(x\pm i\eta )}}=\displaystyle\frac{d}{dx}\log _{FD}\left( 1+(-1)^{\delta }e^{\pm iZ_{N}(x\pm i\eta )}\right) \mp 2\pi i\delta (x-\hat{y}).
\end{equation}
Using this expression in (\ref{zetap}), a new source term appears every times
there is a such point \( \hat{y} \):
\begin{equation}
\label{zeta_primo_special}
\begin{array}{c}
\displaystyle Z_{N}'(\vartheta )=N\left[ \displaystyle\frac{1}{\cosh (\vartheta +\Theta )}+\displaystyle\frac{1}{\cosh (\vartheta -\Theta )}\right] +\displaystyle\sum ^{N_{H}}_{k=1}2\pi G(\vartheta -h_{k})+\\
-\displaystyle\sum ^{M_{C}}_{k=1}2\pi G(\vartheta -c_{k})-\displaystyle\sum ^{M_{W}}_{k=1}2\pi G_{II}(\vartheta -w_{k})-2\pi \displaystyle\sum ^{N_{S}}_{k=1}\left( G(\vartheta -\hat{y}_{k}+i\eta )+G(\vartheta -\hat{y}_{k}-i\eta )\right) \\
+\displaystyle\frac{1}{i}\displaystyle\int dxG(\vartheta -x-i\eta )\displaystyle\frac{d}{dx}\log _{FD}\left[ 1+(-1)^{\delta }e^{iZ_{N}(x+i\eta )}\right] +\\
-\displaystyle\frac{1}{i}\displaystyle\int dxG(\vartheta -x+i\eta )\displaystyle\frac{d}{dx}\log _{FD}\left[ 1+(-1)^{\delta }e^{-iZ_{N}(x-i\eta )}\right] .
\end{array}
\end{equation}
The sum on \( \hat{y}_{k} \) has been indicated as sum on the specials, even
if they are not exactly in the position of specials, because in the usual computations
made on the NLIE \( \eta  \) takes a small value. The condition to be used
for the correct position is (\ref{condizione1_special}). In the following, we omit the label FD, intending always the \( \log  \) in
the fundamental branch. It is very important to note that special objects are
not completely independent degrees of freedom, unlike the holes and complex
solutions, that are fixed ``a priori''. Special objects appear when the derivative
of \( Z_{N} \) on a root or on a hole becomes negative. A special object then
apparently appears two times in (\ref{zeta_primo_special}), both as a real
root/hole and as a special root/hole.

\section{Non linear integral equation (II)\label{section:NLIE2}}

From a Bethe Ansatz point of view, only the function \( Z_{N} \) (and not the
derivative) has a physical meaning, because it is required to obtain the position
of various types of solution, given the quantum numbers (\ref{quantum}). To
obtain \( Z_{N} \), an integration in \( \vartheta  \) must be performed in
(\ref{zeta_primo_special}). All the source terms must be integrated. For the
case of holes, specials and closes this is given by the function:

\begin{equation}
\label{chi}
\chi (\theta )=\displaystyle\int _{0}^{\theta }dx2\pi G(x)
\end{equation}
It is and odd and real analytic function. It asymptotic value is known by direct
integration:
\begin{equation}
\label{chi_infinito}
\chi _{\infty }=\chi (+\infty )=\displaystyle\frac{\pi }{2}\displaystyle\frac{p-1}{p}
\end{equation}
Because \( G \) has always the same sign for every \( x \), the \( \chi (\theta ) \)
is a monotonous function. 

The integration of the source for wide roots is simply the integral of (\ref{Gseconda_esplicita}).
But this integration produces logarithms, then there is a problem of definition
of the appropriate branch. The \( Z_{N} \) function is continuous in the analyticity
strip, then the most correct choice for the wide sources satisfies this continuity.
The forms that are proposed in the following do so (they have no jumps) for
\( \theta  \) with imaginary part \( \pi \max (1,p)>|\Im m\theta |>\pi \min (1,p) \)
(the \( \log  \) is in the fundamental determination):
\begin{equation}
\label{chi_sec_esplic}
\chi _{II}(\theta )=\left\{ \begin{array}{c}
i\: \textrm{sign}(\Im m\theta )\, \left( \log \sinh \left( \displaystyle\frac{\theta }{p}\right) -\log \sinh \left( \displaystyle\frac{\theta -i\pi \, \textrm{sign}\Im m\theta }{p}\right) \right) \qquad \mathrm{if}\quad p>1\\
\\
i\: \textrm{sign}(\Im m\theta )\, \left( \log \left( -\tanh \displaystyle\frac{\theta }{2}\right) +\log \left( \tanh \displaystyle\frac{\theta -i\pi p\, \textrm{sign}\Im m\theta }{2}\right) \right) \qquad \mathrm{if}\quad p<1
\end{array}\right. 
\end{equation}
The notation \( \chi _{II} \) is used to indicate the integral of the \( G_{II} \).
As in the previous case of \( G \), this is the same as taking the second determination
of \( \chi  \) in (\ref{chi}). The two forms corresponding to the opposite
signs of the variable are related by the oddity:
\[
\chi _{II}(-\theta )=\chi _{II}(\theta )\qquad \textrm{if}\quad |\Im m\theta |>\pi \min (1,p).\]
This function can be continued to smallest values of the imaginary part, but
obviously it has a line of discontinuity. The asymptotic limits are (now \( \vartheta  \)
is real): 
\begin{equation}
\label{limiti_chi}
\lim _{\vartheta \, \rightarrow \, \pm \infty }\chi _{II}(\vartheta -w)=\left\{ \begin{array}{c}
\pm \left( 2\chi _{\infty }-\pi \right) \qquad \mathrm{if}\quad p>1\\
\\
\pm \pi \qquad \mathrm{if}\quad p<1
\end{array}\right. 
\end{equation}
The final form obtained in this way is the so called \textit{fundamental non
linear integral equation,} also known as \emph{Destri-de Vega equation} (see
\cite{ddv 97}) for the counting function:
\begin{equation}
\label{nlie}
\begin{array}{c}
Z_{N}(\vartheta )=2N\arctan \displaystyle\frac{\sinh \vartheta }{\cosh \Theta }+g(\vartheta |\vartheta _{j})+\\
\\
+\displaystyle\int \displaystyle\frac{dx}{i}G(\vartheta -x-i\eta )\log \left( 1+(-1)^{\delta }e^{iZ_{N}(x+i\eta )}\right) +\\
\\
-\displaystyle\int \displaystyle\frac{dx}{i}G(\vartheta -x+i\eta )\log \left( 1+(-1)^{\delta }e^{-iZ_{N}(x-i\eta )}\right) +\alpha 
\end{array}
\end{equation}
where 
\begin{equation}
\label{g_sorgenti}
\begin{array}{c}
g(\vartheta |\vartheta _{k})=\displaystyle\sum ^{N_{H}}_{k=1}\chi (\vartheta -h_{k})-\displaystyle\sum ^{N_{S}}_{k=1}\left( \chi (\vartheta -\hat{y}_{k}+i\eta )+\chi (\vartheta -\hat{y}_{k}-i\eta )\right) +\\
\\
-\displaystyle\sum ^{M_{C}}_{k=1}\chi (\vartheta -c_{k})-\displaystyle\sum ^{M_{W}}_{k=1}\chi _{II}(\vartheta -w_{k})
\end{array}
\end{equation}
Both the previous equations hold for \( \vartheta  \) in the fundamental strip.
Out of that, the analytic continuation of \( Z_{N} \) is required, because
the first singularity of \( G \) is crossed. To obtain an equation expressing
the analytic continuation of \( Z_{N} \) beyond this singularity, the second
determination of the various terms appearing in the right hand side of (\ref{nlie})
must be taken, as defined in (\ref{seconda_det}). Notice that it is not the
second determination of \( Z_{N} \) but its analytic continuation:
\begin{equation}
\label{ZN_sec_determ}
\begin{array}{c}
\textrm{for }|\Im m\, \vartheta |>\min (1,\, p):\qquad Z_{N}(\vartheta )=2N\left[ \arctan \displaystyle\frac{\sinh \vartheta }{\cosh \Theta }\right] _{II}+\\
\\
+g(\vartheta |\vartheta _{k})_{II}+\displaystyle\int \displaystyle\frac{dx}{i}G(\vartheta -x-i\eta )_{II}\log \left( 1+(-1)^{\delta }e^{iZ_{N}(x+i\eta )}\right) +\\
\\
-\displaystyle\int \displaystyle\frac{dx}{i}G(\vartheta -x+i\eta )_{II}\log \left( 1+(-1)^{\delta }e^{-iZ_{N}(x-i\eta )}\right) +\alpha _{II}
\end{array}
\end{equation}

This explicit form of the analytic continuation requires to compute \( g(\vartheta |\vartheta _{k})_{II} \).
Remembering (\ref{g_sorgenti}), it is clear that for holes, specials and close
roots the second determination of \( \chi (\vartheta -\vartheta _{j}) \) (see
(\ref{chi_sec_esplic})) appears. For the wide roots the second determination
of \( \chi _{II}(\vartheta -w_{j}) \) appears. It must be determined considering
that
\[
\chi _{W}(\vartheta ,w_{j})\equiv \chi _{II}(\vartheta -w_{j})\]
is a function of \( \vartheta  \):
\begin{equation}
\label{seconda_doppia_det}
\chi _{II}(\vartheta -w_{j})_{II}\equiv \left( \chi _{W}(\vartheta ,w_{j})\right) _{II}.
\end{equation}
The explicit forms are not given there but they can be simply computed from
the definition of second determination.

The integral of the convolution term is possible because \( G \) vanishes exponentially
at infinity. \( \alpha  \) is the integration constant and must be determined. 

There is a useful way to write the log term in (\ref{nlie}), valid for real
\( x \) and if \( \eta \, \rightarrow \, 0 \):
\begin{equation}
\label{funzione_Q}
\begin{array}{c}
{\cal Q}_{N}(x)\equiv \lim _{\eta \, \rightarrow \, 0}2\Im m\, \log \left( 1+(-1)^{\delta }e^{iZ_{N}(x+i\eta )}\right) =\\
\\
=\lim _{\eta \, \rightarrow \, 0}\displaystyle\frac{1}{i}\log \displaystyle\frac{1+(-1)^{\delta }e^{iZ_{N}(x+i\eta )}}{1+(-1)^{\delta }e^{-iZ_{N}(x-i\eta )}}=\left( Z_{N}(x)+\pi \delta \right) \textrm{ mod }2\pi 
\end{array}
\end{equation}
The expression \( \left( A\right) \textrm{ mod }2\pi  \) means exactly \( -\pi <A\leq \pi . \)
The second line in (\ref{funzione_Q}) holds only if \( Z_{N}'(x)>0 \), because
the branch cut of the \( \log  \) is not crossed and \( \left| \Im m\, \log f(x)\right| <\pi  \)
(the notation (\ref{funz_f}) is used). In the opposite case the branch cut
is crossed and the previous expression can take values larger than \( \pi  \):
\( \left| \Im m\, \log f(x)\right| <2\pi . \) In the limit case \( Z_{N}'(x)=0 \)
(for example this happens when \( x\, \rightarrow \, \infty  \)) the values
permitted are \( -\pi \leq \Im m\, \log f(x)\leq \pi . \) 

At this point it is possible to explicitly calculate, from (\ref{nlie}), the
limit \( Z_{N}(+\infty ) \), using (\ref{nlie}, \ref{latticecounting}, \ref{chi_infinito},
\ref{limiti_chi}). The limit gives:
\begin{equation}
\label{limit_ddv}
\begin{array}{c}
Z_{N}(+\infty )=N\pi +\chi _{\infty }\left( N_{H}-2N_{S}-M_{C}-2\theta (p-1)M_{W}\right) +\\
\\
+\pi \, \textrm{sign}(p-1)\, M_{W}+\displaystyle\int dx\, G(x)\, {\cal Q}_{N}(+\infty )+\alpha .
\end{array}
\end{equation}
The limit \( {\cal Q}_{N}(+\infty ) \) can be calculated by using in (\ref{funzione_Q})
the limit of \( Z_{N} \) (\ref{ZNlimiti}). By a comparison of the (\ref{limit_ddv})
with the asymptotic values of \( Z_{N} \) calculated in (\ref{ZNlimiti}) the
value of \( \alpha  \) can be obtained:
\begin{equation}
\label{alfa}
\alpha =\omega \, \displaystyle\frac{p+1}{p}+\chi _{\infty }\left( \left\lfloor \displaystyle\frac{1}{2}+\displaystyle\frac{S}{p+1}+\displaystyle\frac{\omega }{\pi }\right\rfloor -\left\lfloor \displaystyle\frac{1}{2}+\displaystyle\frac{S}{p+1}-\displaystyle\frac{\omega }{\pi }\right\rfloor \right) 
\end{equation}
In this expression the following term has been omitted: \( -\pi \, \textrm{sign}(p-1)\, M_{SC}. \)
It is a multiple of \( \pi  \) and can be taken into account simply changing
the value of \( \delta  \) from \( 0 \) to \( 1 \) (or viceversa) if \( M_{SC} \)
is odd, and shifting of an integer quantity the quantum numbers
\begin{equation}
\label{delta_corretto}
\delta =(M+M_{SC})_{mod\, 2}=(N-S+M_{SC})_{mod\, 2}\in \left\{ 0,1\right\} .
\end{equation}
This sort of manipulation on the quantum numbers has been explained at the end
of the section \ref{section:count-funct}. 

Observe that the result is exactly \( \alpha =0 \) if there is no twist. There
is another important observation. In (\ref{bethe}) the twist term \( \omega  \)
is invariant for the shift 
\[
\omega \, \rightarrow \, \omega +\pi .\]
The same sort of invariance is required in \( \alpha  \) for the NLIE (which
is equivalent to Bethe equations). It is simple to verify that the expression
for \( \alpha  \) (\ref{alfa}) displays the following symmetry 
\[
\alpha \, \rightarrow \, \alpha +2\pi \, \, \, \textrm{when}\, \, \, \omega \, \rightarrow \, \omega +\pi \, .\]
Note that shifting \( \alpha  \) by \( 2\pi  \) is an invariance of the NLIE
(\ref{nlie}), if an appropriate redefinition of the Bethe quantum numbers is
made: \( I_{j}\, \rightarrow \, I_{j}+1 \). This shift does not affects physical
quantities, that depend only on the variables \( \vartheta _{j} \).

With the given value of the integration constant the equation (\ref{nlie})
is complete. The quantization condition (\ref{quantum}) can now be written
as:
\begin{equation}
\label{quant_1}
Z_{N}(\vartheta _{j})=2\pi I_{j}\, ,\qquad I_{j}\in \mathbb {Z}+\displaystyle\frac{1+\delta }{2}
\end{equation}
for the various solutions. For the wide roots remember the previous warning
about the analytic continuation of the \( Z_{N} \) function.

The previous equation is completely equivalent to Bethe equations. No new physics
has been introduced, until this point. Simply an equation that generates the
counting function \( Z_{N} \) has been obtained (the NLIE).

\section{Energy and momentum}

The expression (\ref{autovalori}) can be written using the function \( \phi  \)
as:
\begin{equation}
\label{E_P}
\displaystyle \begin{array}{c}
\displaystyle E=\displaystyle\frac{1}{a}\displaystyle\sum ^{M}_{j=1}\left( \phi (\Theta -\vartheta _{j},1/2)+\phi (\Theta +\vartheta _{j},1/2)-2\pi \right) \\
\\
P=\displaystyle\frac{1}{a}\displaystyle\sum ^{M}_{j=1}\left( \phi (\Theta -\vartheta _{j},1/2)-\phi (\Theta +\vartheta _{j},1/2)\right) +2\omega 
\end{array}
\end{equation}
The choice of the logarithmic branch in the energy ensures that the contribution
of each real root is negative definite. This is consistent with the known ground
state structure (as in section \ref{section:6_vertex_BA}), that is given by
the maximal number of real roots. It will be clear that excitations give only
positive contributions.

It is possible to relate the previous expressions to the counting function.
To do that, consider first the following quantity:
\begin{equation}
\label{funzione_W}
W(\theta )=\displaystyle\sum _{j=1}^{M}\phi '(\theta -\vartheta _{j},1/2)\qquad \theta \in \mathbb R.
\end{equation}
 It can be integrated to obtain the pieces appearing in (\ref{E_P}). Clearly
the following expressions hold:
\begin{equation}
\label{pezzi_energia}
\begin{array}{c}
\displaystyle \displaystyle\sum ^{M}_{j=1}\phi (\Theta -\vartheta _{j},1/2)=\displaystyle\int _{asymp}^{\Theta }dx\, W(x)\\
\displaystyle\sum ^{M}_{j=1}\phi (\Theta +\vartheta _{j},1/2)=-\displaystyle\sum ^{M}_{j=1}\phi (-\Theta -\vartheta _{j},1/2)=-\displaystyle\int _{asymp}^{-\Theta }dx\, W(x).
\end{array}
\end{equation}
The integration requires to fix a constant. This can be simply done by computing
for \( \Theta \, \rightarrow +\infty  \) the left hand side in (\ref{pezzi_energia})
and imposing the equality with the primitive of \( W \) at the same limit.
This has been indicated by the symbol
\[
\int _{asymp}^{\pm \Theta }\]
The function \( W \) admits an expression in terms of the counting function.
The sum over roots in (\ref{funzione_W}) can be expressed as in (\ref{count})
and using the same notations:
\[
W(\theta )=\sum _{j=1}^{M_{R}+N_{H}}\phi '(\theta -x_{j},1/2)-\sum _{j=1}^{N_{H}}\phi '(\theta -h_{j},1/2)+\sum _{j=1}^{M_{C}+M_{W}}\phi '(\theta -\xi _{j},1/2).\]
 Using now the same trick used for \( Z_{N}' \), as in (\ref{integr_gamma},
\ref{integr_striscia}, \ref{integr}), and the expression (\ref{derivata_continua})
generating the ``special'' contribution, the following expression yields:
\begin{equation}
\label{W_1}
\begin{array}{c}
\displaystyle W(\vartheta )=\displaystyle\int \displaystyle\frac{dx}{2\pi i}\phi '(\vartheta -x+i\eta ,1/2)\displaystyle\frac{d}{dx}\log _{FD}\left[ 1+(-1)^{\delta }e^{-iZ_{N}(x-i\eta )}\right] +\\
-\displaystyle\int \displaystyle\frac{dx}{2\pi i}\phi '(\vartheta -x-i\eta ,1/2)\displaystyle\frac{d}{dx}\log _{FD}\left[ 1+(-1)^{\delta }e^{iZ_{N}(x+i\eta )}\right] +\\
-\displaystyle\sum ^{N_{H}}_{k=1}\phi '(\vartheta -h_{k},1/2)+\displaystyle\sum ^{M_{C}}_{k=1}\phi '(\vartheta -c_{k},1/2)+\displaystyle\sum ^{M_{W}}_{k=1}\phi '(\vartheta -w_{k},1/2)+\\
+\displaystyle\sum ^{N_{S}}_{k=1}\left( \phi '(\vartheta -\hat{y}_{k}+i\eta ,1/2)+\phi '(\vartheta -\hat{y}_{k}-i\eta ,1/2)\right) +\displaystyle\int \displaystyle\frac{dx}{2\pi }\phi '(\vartheta -x,1/2)\, Z_{N}'(x).
\end{array}
\end{equation}
Now the equation (\ref{zeta_primo_special}) can be put in the last integral
in the previous expression. The terms ``similar'' can be collected (i.e. holes
with holes, close roots with close roots, and so on) to obtain the following
form:
\[
\begin{array}{c}
W(\vartheta )=-\displaystyle\sum ^{N_{H}}_{k=1}\left[ \phi '(\vartheta -h_{k},1/2)-\displaystyle\int dx\, \phi '(\vartheta -x,1/2)\, G(x-h_{k})\right] +\\
+\displaystyle\sum ^{M_{C}}_{k=1}\left[ \phi '(\vartheta -c_{k},1/2)-\displaystyle\int dx\, \phi '(\vartheta -x,1/2)\, G(x-c_{k})\right] +\\
+\displaystyle\sum ^{M_{W}}_{k=1}\left[ \phi '(\vartheta -w_{k},1/2)-\displaystyle\int dx\, \phi '(\vartheta -x,1/2)\, G_{II}(x-w_{k})\right] +\\
+\displaystyle\sum ^{N_{S}}_{k=1}\left[ \phi '(\vartheta -\hat{y}_{k}+i\eta ,1/2)+\phi '(\vartheta -\hat{y}_{k}-i\eta ,1/2)\right. +\\
\left. -\displaystyle\int dx\, \phi '(\vartheta -x,1/2)\, \left( G(x-\hat{y}_{k}+i\eta )+G(x-\hat{y}_{k}-i\eta )\right) \right] +\\
+\displaystyle\int \displaystyle\frac{dx}{2}\phi '(\vartheta -x,1/2)N\left[ \displaystyle\frac{1}{\cosh (\vartheta +\Theta )}+\displaystyle\frac{1}{\cosh (\vartheta -\Theta )}\right] +\\
-2\Im m\displaystyle\int \displaystyle\frac{dx}{2\pi }\phi '(\vartheta -x-i\eta ,1/2)\displaystyle\int dy\left[ \delta (x-y)-G(x-y)\right] \displaystyle\frac{d}{dy}\log _{FD}\left[ 1+(-1)^{\delta }e^{iZ_{N}(y+i\eta )}\right] =\\
=W_{H}+W_{C}+W_{W}+W_{S}+W_{bulk}+W_{I}
\end{array}\]
where the notation introduced on the last line is obvious (the meaning of bulk
will be clear in the following; label I is for interaction term or, that is
the same, for integral). 

As in section \ref{section:NLIE_1}, there are different terms to analyze.

First, rearranging a little bit the terms and introducing a delta-function,
it is possible to see that in the case of holes, closes, specials and in the
integral term there is a contribution by:
\[
\delta (x-y)-G(x-y)=\Delta (x-y)\]
(this equality is a simple consequence of the definition of \( \Delta  \)).
The corresponding terms contain the integral:
\[
\int dx\, \phi '(x-\vartheta _{j},1/2)\, \Delta (\vartheta -x)=\frac{1}{\cosh (\vartheta -\vartheta _{j})}\]
 that is well known because it has been used in (\ref{termine_cinetico}). This
holds in the case of holes and closes and \( \vartheta _{j} \) is the corresponding
rapidity. For the integral term and the specials the contribution is exactly:
\[
\frac{1}{\cosh (\vartheta -y-i\eta )}.\]
where \( y \) is the integration variable or the special rapidity. 

The wide computation is more involved. The expression to be computed is
\[
W_{W}(\vartheta )=\sum ^{M_{W}}_{k=1}\left[ \phi '(\vartheta -w_{k},1/2)-\int dx\, \phi '(\vartheta -x,1/2)\, G_{II}(x-w_{k})\right] \]
and the most convenient way to do that is to use the Fourier transformation
of all the terms. For \( \phi '(\vartheta -w_{k},1/2) \) and \( G_{II} \)
it has been used in the computation for (\ref{Gseconda_esplicita}). For \( \phi ' \)
on the real axis it is in (\ref{Fourierphi'}). The result is
\[
W_{W}(\vartheta )=\left\{ \begin{array}{c}
0\\
\left[ \displaystyle\frac{1}{\cosh (\vartheta )}\right] _{II}+O(\vartheta )
\end{array}\right. \begin{array}{c}
\textrm{ for }p>1\\
\textrm{ for }p<1
\end{array}\]
In the case of \( p>1 \) it simply cancels out. In the other case, there is
no such cancellation, for generic wide positions. The explicit form of the contribution
\( O(\vartheta ) \) is quite long and shall not be written out at this point.
The important fact is that it is a fast vanishing contribution for large \( \vartheta  \).
Observe that the second determination (\ref{seconda_det}) of a whatever trigonometric
hyperbolic function, for \( p>1 \), is exactly zero, then the notation \( [1/\cosh (\vartheta )]_{II} \)
will be used also in this case.

The final form is:
\[
\begin{array}{c}
\displaystyle W(\vartheta )=-\displaystyle\sum ^{N_{H}}_{k=1}\displaystyle\frac{1}{\cosh (\vartheta -h_{k})}+\displaystyle\sum ^{N_{S}}_{k=1}\left( \displaystyle\frac{1}{\cosh (\vartheta -\hat{y}_{k}-i\eta )}+\displaystyle\frac{1}{\cosh (\vartheta -\hat{y}_{k}+i\eta )}\right) +\\
+\displaystyle\sum ^{M_{C}}_{k=1}\displaystyle\frac{1}{\cosh (\vartheta -c_{k})}+\displaystyle\sum ^{M_{W}}_{k=1}\left[ \displaystyle\frac{1}{\cosh (\vartheta -w_{k})}\right] _{II}+O(\vartheta )+\\
+\displaystyle\int \displaystyle\frac{dx}{2\pi }\phi '(\vartheta -x,1/2)N\left[ \displaystyle\frac{1}{\cosh (x+\Theta )}+\displaystyle\frac{1}{\cosh (x-\Theta )}\right] +\\
-2\Im m\displaystyle\int \displaystyle\frac{dx}{2\pi }\displaystyle\frac{1}{\cosh (\vartheta -x-i\eta )}\displaystyle\frac{d}{dx}\log _{FD}\left[ 1+(-1)^{\delta }e^{iZ_{N}(x+i\eta )}\right] 
\end{array}\]
Integrating this function, as in (\ref{pezzi_energia}), the energy and momentum
of this lattice system can be obtained. In all the terms appears \( 1/\cosh x \);
its primitive is 
\begin{equation}
\label{identita_trigon}
\displaystyle\int \displaystyle\frac{dx}{\cosh x}=\arctan \sinh x=2\arctan \tanh (x/2).
\end{equation}
 The last form is the most convenient, to calculate the continuum limit. Then:

\begin{equation}
\label{E_P_reticolo}
\begin{array}{c}
\displaystyle a\displaystyle\frac{E\pm P}{2}=-\displaystyle\sum ^{N_{H}}_{k=1}2\arctan \tanh ((\Theta \mp h_{k})/2)+\displaystyle\sum ^{M_{C}}_{k=1}2\arctan \tanh ((\Theta \mp c_{k})/2)+\\
+\displaystyle\sum ^{N_{S}}_{k=1}2\left( \arctan \tanh ((\Theta \mp \hat{y}_{k}-i\eta )/2)+\arctan \tanh ((\Theta \mp \hat{y}_{k}+i\eta )/2)\right) +\\
+\displaystyle\sum ^{M_{W}}_{k=1}2\arctan \tanh ((\Theta \mp w_{k})/2)_{II}+O(\Theta )\mp \displaystyle\int \displaystyle\frac{dx}{2\pi }\displaystyle\frac{1}{\cosh (\pm \Theta -x)}{\cal Q}_{N}(x)+S\, \pi +\\
+\displaystyle\int \displaystyle\frac{dx}{2\pi }\phi (\Theta \mp x,1/2)N\left[ \displaystyle\frac{1}{\cosh (x+\Theta )}+\displaystyle\frac{1}{\cosh (x-\Theta )}\right] \pm \omega -N\, \pi 
\end{array}
\end{equation}
The term \( O(\Theta ) \) vanishes for large \( \Theta  \).

The function (\ref{funzione_Q}) has been used, by putting \( \eta \, \rightarrow \, 0 \),
and observing that \( x \) is real.

\section{Continuum limit\label{section:limite_cont.}}

At the conclusion of the previous long computations there is an important observation:
in the final expressions for energy and momentum there is no explicit contribution
from the real roots. This also happens in the NLIE (\ref{nlie}). On the lattice
it is not a relevant observation, but in this section it shall be shown that
it can justify a continuum limit procedure, \( N\, \rightarrow \, \infty  \)
and \( a\, \rightarrow \, 0 \), and a particle interpretation. In the lattice
energy expression there is a bulk term (see later) i.e. a term increasing with
\( N \) and completely independent from the solutions \( \vartheta _{j} \)
of Bethe equations. This bulk term can be subtracted and the remaining terms
(due only to holes and complex roots) can be chosen in such a way that they
describe particle excitations. 

To do that, in the limit procedure, they must be chosen in finite number (order
\( O(1) \)) and also its ``rapidity'' \( \vartheta _{j} \) must be finite.
From (\ref{quantum}) this means that only real roots can appear in the asymptotic
tails of \( Z_{N} \). All the other lattice states must be discarded (as usual,
lattice theory contains more states than the continuum theory).

A state where holes and complex roots are not considered, is the hamiltonian
vacuum. If they are considered, they give positive contribution to the energy,
behaving as particle excitations on a vacuum state. Real roots are completely
disappeared, in this limit. They can be interpreted (in the limit procedure)
as a sort of Dirac sea that is the hamiltonian vacuum on with holes and complex
roots built particle excitations. 

This is the structure required to have a consistent quantum field theory. 

All this interpretation is possible only because of the antiferromagnetic vacuum
choice, made in section (\ref{section:6_vertex_BA}). 

What will be proved now is that there is a consistent way to do continuum limit
in the NLIE and in the expression of the energy. 

In \cite{ddv 89} it was shown that the correct way to do the continuum limit
for this model on the light-cone is to send \( N,\, \Theta \rightarrow \, \infty  \)
connected in the following way:
\begin{equation}
\label{teta_n}
\Theta \approx \log \displaystyle\frac{4N}{{\cal M}L}.
\end{equation}
\( L=N\, a \) is the spatial dimension of the lattice (as in section \ref{light-cone.section})
and stay fixed in the limit. \( {\cal M} \) is the renormalized physical mass.
As in section \ref{section:6_vertex_BA}, the light-cone lattice is periodic
in space direction. Then its continuum limit yields a compactified space of
length \( L \) (space-time is a cylinder of circumference \( L \)). The lattice
6 vertex model becomes a field theory defined on this cylinder. The interpretation
of what field theory is defined by this continuum limit, at the various values
of the twist, is the job of the next chapter.

To do the limit, consider first the counting equation. It takes the following
form, on the continuum
\begin{equation}
\label{continuum_counting}
N_{H}-2N_{S}=2S+M_{C}+2\, \theta (p-1)\, M_{W}
\end{equation}
because the structures in the tails are extremely simplified. The other terms
appearing in the lattice counting equation (\ref{latticecounting}) in fact
came from the configurations where in the tails there are special root/holes
or self-conjugate of the so called first class (see the original paper \cite{ddv 97}).
Such configurations completely disappear in the continuum limit.

The principal interest is to obtain the limit of the energy and momentum eigenvalues
(\ref{E_P_reticolo}). Clearly, to fix the positions of the roots \( \vartheta _{j} \)
the limit of (\ref{quant_1}) is required. This can be made using a \emph{continuum
counting function} that is defined by:
\begin{equation}
\label{def_Z}
\displaystyle Z(\vartheta )=\lim _{N\, \rightarrow \, \infty }Z_{N}(\vartheta ).
\end{equation}
This limit computation is made on the sequence of counting functions that are
implicitly defined in (\ref{nlie}) at the various values of \( N \); the (\ref{teta_n})
must be also taken onto account. The interesting fact is that this limit can
be done explicitly in all the terms appearing in (\ref{nlie}), i.e. a \emph{continuum
NLIE} can be obtained. To show that, the various terms are analyzed. 

The first term and the integral in (\ref{nlie}) are trivial matter to compute.
The source terms for holes, complex solutions and specials don't have any explicit
dependence from \( N \). Only the positions that appear in this terms must
be determined, obviously, by the continuum NLIE. Then the term \( g(\vartheta |\vartheta _{j}) \)
is unchanged. The result is:
\begin{equation}
\label{nlie-cont}
\begin{array}{c}
\displaystyle Z(\vartheta )={\cal M}L\sinh \vartheta +g(\vartheta |\vartheta _{j})+\\
+\displaystyle\int \displaystyle\frac{dx}{i}G(\vartheta -x-i\eta )\log \left( 1+(-1)^{\delta }e^{iZ(x+i\eta )}\right) +\\
-\displaystyle\int \displaystyle\frac{dx}{i}G(\vartheta -x+i\eta )\log \left( 1+(-1)^{\delta }e^{-iZ(x-i\eta )}\right) +\alpha 
\end{array}
\end{equation}
The quantization conditions (\ref{quantum}, \ref{quant_1}) can now be written
as:
\begin{equation}
\label{cont_quant}
Z(\vartheta _{j})=2\pi I_{j}\, ,\qquad I_{j}\in \mathbb {Z}+\displaystyle\frac{1+\delta }{2}
\end{equation}
for the various solutions. This terminates the limit procedure on NLIE. Observe
that on the continuum only the implicit ``definition'' of \( Z \) by (\ref{nlie-cont})
is available, instead of and explicit expressions as in the case (\ref{def.Zn}).

Consider now in the energy expression (\ref{E_P_reticolo}) the terms with the
form \( 2/a\arctan \, \tanh ((\Theta -\theta )/2) \). Using the following hyperbolic
trigonometric identity
\[
\displaystyle \arctan \, \tanh (x)=\frac{\pi }{4}-\arctan \, e^{-x}\]
the following asymptotic behaviour can be obtained:
\[
\frac{N}{L}\, 2\, \arctan \, \tanh (\Theta -\theta )\approx \frac{N\pi }{L2}-\frac{1}{2}{\cal M}\, e^{\theta }.\]
Then, collecting in the expression for \( (E\pm P)/2 \) the contribution of
this type, yields (remember that there is a \( S\pi  \) in (\ref{E_P_reticolo})):

\begin{equation}
\label{conta_pi}
\displaystyle\frac{\pi }{2}(2S-N_{H}+2N_{S}+M_{C}+2\theta (p-1)M_{W})=0
\end{equation}
(the (\ref{continuum_counting}) has been used). 

One of the two integral contributions is
\[
\begin{array}{c}
\displaystyle \lim _{N\, \rightarrow \, \infty }\mp \displaystyle\frac{N}{L}\displaystyle\int \displaystyle\frac{dx}{2\pi }\displaystyle\frac{1}{\cosh (\pm \Theta -x)}{\cal Q}_{N}(x)=\\
=\mp \displaystyle\frac{{\cal M}}{2}\displaystyle\int \displaystyle\frac{dx}{2\pi }e^{\pm x}{\cal Q}(x).
\end{array}\]
The final result has been obtained by exchanging the limit and the integral. 

The other one is the last line in (\ref{E_P_reticolo}). It can be handled by
shifting the integration variable and observing that one term in the square
brackets is odd and gives a vanishing integral:
\[
\begin{array}{c}
\displaystyle E^{\pm }_{N,\, bulk}=\displaystyle\frac{N^{2}}{L}\left[ \displaystyle\int \displaystyle\frac{dx}{2\pi }\phi (\Theta \mp x,1/2)\left[ \displaystyle\frac{1}{\cosh (x+\Theta )}+\displaystyle\frac{1}{\cosh (x-\Theta )}\right] -\pi \right] \pm \omega =\\
=\displaystyle\frac{N^{2}}{L}\left[ \displaystyle\int \displaystyle\frac{dx}{2\pi }\left[ \displaystyle\frac{\phi (x,1/2)}{\cosh (2\Theta -x)}\right] -\pi \right] \pm \omega \: .
\end{array}\]
This energy contribution is ``source independent''. It has been labeled by
``bulk'' because it will be shown that it diverges as \( N \). It is interesting
to compute the contribution to \( E \) and \( P \), instead of the diverging
contribution to \( (E\pm P)/2 \):
\[
\begin{array}{c}
E_{N\, bulk}=E^{+}_{N,\, bulk}+E^{-}_{N,\, bulk}=\displaystyle\frac{N^{2}}{L}\left[ \displaystyle\int \displaystyle\frac{dx}{\pi }\left[ \displaystyle\frac{\phi (x,1/2)}{\cosh (2\Theta -x)}\right] -2\pi \right] =-\displaystyle\frac{N^{2}}{L}(\pi +\gamma )\\
\\
P_{N\, bulk}=E^{+}_{N\, bulk}-E^{-}_{N\, bulk}=2\omega 
\end{array}\]
 The first integral can be computed observing that \( \phi  \) is a constant
for very large values of \( x \), and can be put out of the integral sign.
The second one contains the difference of two equal contributions and only \( \omega  \)
is not deleted. 

Collecting all the terms follows:

\begin{equation}
\label{E_P_cono_luce}
\begin{array}{c}
\displaystyle \displaystyle\frac{E\pm P}{2}=\displaystyle\frac{{\cal M}}{2}\left( \displaystyle\sum ^{N_{H}}_{k=1}e^{\pm h_{k}}-\displaystyle\sum ^{N_{S}}_{k=1}\left( e^{\pm \hat{y}_{k}+i\eta }+e^{\pm \hat{y}_{k}-i\eta }\right) +\right. \\
\left. -\displaystyle\sum ^{M_{C}}_{k=1}e^{\pm c_{k}}-\displaystyle\sum ^{M_{W}}_{k=1}e_{II}^{\pm w_{k}}\mp \displaystyle\int \displaystyle\frac{dx}{2\pi }e^{x}{\cal Q}_{N}(x)\right) +E^{\pm }_{N\, bulk}
\end{array}
\end{equation}
or, subtracting all the diverging contributions,

\begin{equation}
\label{energia}
\begin{array}{c}
\displaystyle E={\cal M}\left( \displaystyle\sum ^{N_{H}}_{k=1}\cosh h_{k}-\displaystyle\sum ^{N_{S}}_{k=1}\left( \cosh (\hat{y}_{k}+i\eta )+\cosh (\hat{y}_{k}-i\eta )\right) +\right. \\
\left. -\displaystyle\sum ^{M_{C}}_{k=1}\cosh c_{k}-\displaystyle\sum ^{M_{W}}_{k=1}\cosh _{II}w_{k}-\displaystyle\int \displaystyle\frac{dx}{2\pi }\sinh x{\cal Q}(x)\right) 
\end{array}
\end{equation}
and 
\begin{equation}
\label{momento}
\begin{array}{c}
\displaystyle P={\cal M}\left( \displaystyle\sum ^{N_{H}}_{k=1}\sinh h_{k}-\displaystyle\sum ^{N_{S}}_{k=1}\left( \sinh (\hat{y}_{k}+i\eta )+\sinh (\hat{y}_{k}-i\eta )\right) +\right. \\
\left. -\displaystyle\sum ^{M_{C}}_{k=1}\sinh c_{k}-\displaystyle\sum ^{M_{W}}_{k=1}\sinh _{II}w_{k}+\displaystyle\int \displaystyle\frac{dx}{2\pi }\cosh x{\cal Q}(x)\right) 
\end{array}
\end{equation}
Observe that for \( p>1 \) the second determination of hyperbolic functions
vanishes. This means that wide roots contribute to the energy only in an implicit
way, because they contribute to the position of other objects, by (\ref{nlie-cont}).

\section{Physical interpretation\label{section:interpr_fisica}}

The limit procedure described in the previous section is mathematically consistent,
but the question is if from the physical point of view it describes a consistent
quantum theory and allows for a meaningful physical interpretation. 

Before to propose it, an important remark must be made about the allowed values
for \( S \). It is clear from (\ref{spin-chain}) that on the lattice only
integer and nonnegative values can be taken into account for \( S \). But on
the continuum the definition of \( S \) is no more related to the Bethe state
(that is undefined), instead it is given implicitly by (\ref{continuum_counting}).
Then, \emph{``a priori''}, there are no arguments that fixes its values to
be integers. As our group showed in \cite{noi PL2}, the half-integer choice
for \( S \) is necessary (and gives completely consistent results) to describe
odd numbers of particles. 

At this point the following physical interpretation can be proposed. It will
be refined to describe the correspondence with particles. Also it will be supported
by many arguments that will be clarified in the next chapters. 

\textbf{Physical interpretation:}

\begin{itemize}
\item the physical vacuum (hamiltonian ground state) corresponds to absence of sources
(i.e. holes, complex); all the sources are excitations on this vacuum
\item for \( \omega =0 \) and at the various values of \( S \) this theory describes
the sine-Gordon/massive Thirring model on a finite space of size \( L \); \( 2S \)
is the topological charge and can take nonnegative integer values.
\item for the values 
\begin{equation}
\label{omega}
\omega =\displaystyle\frac{k\pi }{s}\, ,\qquad k=1,...,q'-1
\end{equation}
 it describes the quantum reduction of sine-Gordon model, i.e. the massive integrable
theory obtained perturbing the minimal model \( Vir(r,s) \) with the operator
\( \Phi _{(1,3)} \).
\end{itemize}
Observe that it has been assumed that only nonnegative values of \( S \) are
required to describe the whole Hilbert space of the theory. Indeed the theory
is assumed charge-conjugation invariant then negative topological charge states
have the same energy and momentum as their charge conjugate states. The assumption
that all the states can be described by the NLIE is absolutely not trivial,
or better still until this moment it is not available a general proof of this
fact, but only a number of specific cases supports this conjecture. 

All this things are the argument of the next chapters.

\chapter{ANALYSIS OF THE CONTINUUM THEORY\label{chapter:analisis_NLIE}}

\section{Principal questions}

In the next sections, a carefull analysis of the vacuum and of some excited
states from NLIE will be performed at the various scales (various \( L \)),
to completely understand the physical interpretation suggested at the end of
the section \ref{section:limite_cont.}. The principal question is what theory
is described by equations (\ref{nlie-cont}, \ref{energia}, \ref{momento}). 

As scale parameter, as appears in (\ref{nlie-cont}), can be chosen equivalently
the size \( L \) or the \emph{adimensional size} \( l={\cal M}L \), where
\( {\cal M} \) plays the role of a mass scale. The limit of very large \( l \)
can be interpreted both as large size (that reproduces infinite Minkowski space-time)
or large mass scale, that is an \emph{infrared point} in a renormalization flow
(IR). Finite size effects do not appear, in that limit.

At the opposite limit of small \( l \), important finite size effects are expected;
a small value of \( l \) can be obtained with a small mass \( {\cal M} \),
then this case is an \emph{ultraviolet point} in a renormalization flow (UV).
The complete control on the scaling functions can be obtained analyzing the
whole range of values \( l>0 \).

\section{Connection with sine-Gordon/massive Thirring}

The section \ref{section:limite_cont.} ends with a conjecture about the physical
meaning of the model described by the continuum NLIE and the corresponding energy
and momentum expressions. There are many arguments to support this interpretation.
Some of them came from the well known properties of the 6-vertex Bethe Ansatz,
other from the analysis of the NLIE itself. 

Consider the \( R \) matrix in (\ref{6vRmatrix}, \ref{R_matrix_entries}).
It has the same entries as the \( R_{sG} \) matrix that appears by putting
sine-Gordon on a lattice, and obtaining the corresponding Bethe Ansatz equations
(lattice Thirring B.A. gives exactly the same). The equality of \( R \) matrix
and of Bethe Ansatz equations means that there is a common integrable structure
in the two systems.

Moreover, the lattice sine-Gordon B.A. can be obtained as the continuum limit
\( N\, \rightarrow \, \infty  \) (and with \( \Theta =\log \displaystyle\frac{4N}{{\cal M}L} \))
of the inhomogeneous 6-vertex Bethe equations (\ref{bethe}). The Bethe Ansatz
techniques, then, strongly suggest the previous interpretation. 

The 6-vertex model in its thermodynamics limit is critical, as shown in \cite{baxter},
then there is a conformal field theory describing this critical point. From
the eigenvalues of the transfer matrix (obtained by B.A.) it is possible to
extract the conformal properties, as shown in \cite{cardy86}. As shown in \cite{karowsky},
for 6-vertex model they reproduce exactly the equation (\ref{delta+-_vertex_op})
for \( \omega =0 \) (that is the UV structure of sine-Gordon and massive Thirring)
and the minimal models conformal weights for \( \omega  \) chosen as in (\ref{omega}). 

It has been shown, in \cite{ddv 87}, that a fermion satisfying massive Thirring
equations of motion can be constructed, as described in section (\ref{section:6vertici_varie}).
It describes the scaling behaviour of the inhomogeneous 6-vertex light-cone
lattice dynamics. A similar argument can be made on the XXZ chain quantum hamiltonian,
whose exponential is the transfer matrix of 6-vertex model. In \cite{faddeev 95}
a (classical) field satisfying sine-Gordon equations is built from this XXZ
chain. 

There is one more suggestion, that comes from the NLIE itself: the \( \chi  \)
function obtained in the derivation of the NLIE (\ref{chi}) is well known in
literature, because it is exactly the logarithm of the soliton-soliton sine-Gordon
\( S \) matrix:
\begin{equation}
\label{matrice_S_chi}
\chi (\vartheta )=-i\log S^{++}_{++}(\vartheta )
\end{equation}
as in \cite{zam79}. 

This arguments are the starting points for the analysis of NLIE itself that
will be performed in the next sections. They completely confirm the physical
interpretation given in section \ref{section:limite_cont.}.

\section{Some general facts about the IR limit\label{section:IR}}

This IR analysis is important, because it can connect NLIE with the (minkowskian)
scattering theory. The limit \( L\, \rightarrow \, \infty  \) is not at all
a thermodynamic limit, because the size of the system is scaled, but the number
of particles is fixed. Then, the interaction among the physical particles is
expected to cease affecting the energy and momentum, because their density is
vanishing. Hence, \( E \) and \( P \) should approach finite limits equal
to a free massive spectrum. In fact, for large \( l \), the dominant term in
(\ref{nlie-cont}) is the \( l\sinh \vartheta  \). Using it as first iteration
in the convolution terms of (\ref{nlie-cont}, \ref{energia}, \ref{momento}),
they become small (of order \( O\left( e^{-l}\right)  \)) and can be dropped.
The surviving part in NLIE can be seen as a dressed Bethe Ansatz giving constraints
on the asymptotic states:
\begin{equation}
\label{IR_NLIE}
Z(\vartheta )=l\sinh \vartheta +g(\vartheta |\vartheta _{k})+\alpha \qquad \textrm{for}\quad l\, \rightarrow \, \infty .
\end{equation}
 Out of the fundamental strip where this equation holds, a similar expression
can be written, using the second determination. Assume for the moment \( \alpha =0 \).
Taking the exponential of the previous equation and imposing the quantization
condition yields:
\[
e^{il\sinh \vartheta _{j}}e^{ig(\vartheta _{j}|\vartheta _{k})}=\pm 1\]
that has a structure similar to the usual quantization condition of particles
in a box:
\begin{equation}
\label{quantizz-box}
e^{iLP(\lambda _{j})}\prod _{k\neq j}S(\lambda _{j}-\lambda _{k})=\pm 1
\end{equation}
(\( + \) is referred to periodic and \( - \) to antiperiodic boundary conditions).
The equation (\ref{quantizz-box}) is not a Bethe Ansatz equation\footnote{
It is called Dressed Bethe Ansatz because it contains real particles instead
of pseudoparticles appearing in usual B.A.
} , in general, because the \( \lambda _{k} \) that appear in it are real rapidities
of particles, instead of the rapidities of the pseudoparticles (\( \vartheta _{j} \))
required to built Bethe states. But in (\ref{IR_NLIE}) can appear complex ``rapidities''.
The correct correspondence between the box quantization (\ref{quantizz-box})
and the infrared NLIE (\ref{IR_NLIE}) will be carefully analyzed later. The
important fact, for the moment, is that the comparison between this two equations
allows to read out scattering data from NLIE and this can be compared with the
known S-matrix of the models to which the NLIE is referred. 

Another important IR observation is that the derivative of the counting function
is a very large positive number and no one special root/hole can takes place
\[
Z'(\vartheta )=l\cosh \vartheta +g'(\vartheta |\vartheta _{k})\qquad \textrm{for}\quad l\, \rightarrow \, \infty .\]
This suggest that special roots/holes are not physical particles. Instead, as
it is clearly shown in section \ref{section:NLIE_1}, they are a mathematical
artifact of the logarithmic contribution that appears in the convolution term. 

The final important observation in IR analysis comes from quantization conditions
(\ref{quant_1})
\[
2\pi I_{j}=l\sinh \vartheta _{j}+g(\vartheta _{j}|\vartheta _{k})+\alpha \]
The number \( I_{j} \) is real. Distinguishing the real and imaginary part
of this equation gives: 
\[
\begin{array}{c}
2\pi I_{j}=l\sinh \, \Re e\, \vartheta _{j}\, \cos \, \Im m\, \vartheta _{j}+\Re e\, (g(\vartheta _{j}|\vartheta _{k})+\alpha )\\
0=l\cosh \, \Re e\, \vartheta _{j}\, \sin \, \Im m\, \vartheta _{j}+\Im m\, (g(\vartheta _{j}|\vartheta _{k})+\alpha )
\end{array}\]
To deal with the first equation, the real part of the \( \vartheta _{j} \)
for increasing values of the large scale \( l \) moves toward the origin as

\[
\Re e\, \vartheta _{j}\propto \frac{1}{l\cos \, \Im m\, \vartheta _{j}}.\]
From the second equation, the imaginary part must develops a singularity from
\( \Im m\, (g(\vartheta _{j}|\vartheta _{k})) \) to annihilate the rapidly
diverging term \( l\, \sin \, \Im m\, \vartheta _{j} \). Because the positions
of singularities are known, this simple argument can be used to fix (at the
IR) the imaginary part of the sources. Being the real part zero, the source
position at IR is completely fixed.

\section{The intermediate regions\label{section:intermedio}}

The NLIE (\ref{nlie-cont}) contains a source term \( g(\vartheta |\vartheta _{j}) \)
which is specified by giving the root/hole structure of the given state and
depends on the positions of the holes, special roots/holes and complex roots.
These positions in turn are fixed by the Bethe quantization conditions (\ref{cont_quant}).
The NLIE supplemented by the quantization conditions gives a set of coupled
nonlinear equations in the function \( Z(\vartheta ) \) and the variables \( \vartheta _{j} \)
that can be solved numerically by an iterative procedure. First one chooses
a starting position for the sources \( \vartheta _{j} \). Then one iterates
the integral equation (using fast Fourier transform to evaluate the convolution)
to update the counting function \( Z \). Using this new \( Z \), an improved
determination of position of the \( \vartheta _{j} \) can be obtained, and
is fed back into the integral equation for a new iteration cycle. The process
is repeated until the solution is found to a prescribed precision (usually \( 10^{-6} \)).
As explained in section \ref{section:IR}, for large \( l \) the source term
dominates, while the correction coming from the integral term is exponentially
small. Therefore it is reasonable to expect that the further one goes to the
IR regime the faster the iteration converges which is in fact what happens in
the computations. Hence it is preferable to start iterating at the largest desired
value of \( l \) and decrease the volume gradually, always taking as a starting
point at the next value of the volume the solution found at the previous value.

As a preliminar analysis, the general behaviour of the \( Z(x) \) function
is that of \( \sinh x \) for large \( l \). Decreasing the scale, the tangent
in the central portion of \( Z(x) \) for \( x\in \mathbb R \) becomes more and
more horizontal; instead for large \( |x| \) the leading term is always the
hyperbolic sine. This central position broadens to a single or double plateau
system that extends to all \( |x| \)'s smaller than \( \log (2/l), \) and
rapidly disappears for greater values, in favor of the dominant exponential
growth. 

\begin{figure}[  htbp]
{\centering \includegraphics{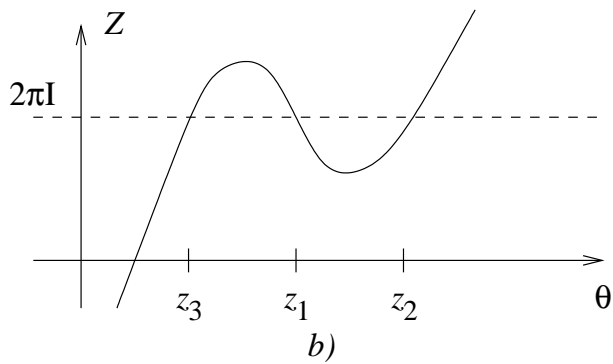} \par}

\caption{\small The typical behaviour of the \protect\( Z\protect \) function when \protect\( z_{1}\protect \)
is a special root/hole \label{oscill._special.eps}}
\end{figure}

It can happens, when decreasing \( l \), that special holes appear. But there
is the so called \emph{number of effective holes}, that is given by 
\begin{equation}
\label{holes_effettive}
N_{H,eff}=N_{H}-2N_{S}
\end{equation}
 that is a constant independent of \( l \). This is a consequence of one simple
fact: when a globally increasing function (as \( Z \) is) makes a small oscillation
as in figure \ref{oscill._special.eps} there are an odd number of points that
intersect the horizontal line corresponding to a certain quantum number. Then to every real solution (when \( Z \)is increasing) correspond three (of
five or seven or ...) real solutions in the case of specials. Of this odd number
of solutions, only one is a real root (no Bethe roots can have the same quantum
numbers). All the other must be holes (specials or normal). Then (as in figure
\ref{oscill._special.eps})
\[
1\textrm{ root for large }l=1\textrm{ root}+2\textrm{ holes}-2\times 1\textrm{ special r}.\textrm{ or h}.\textrm{ for small }l.\]
 The general case is (\ref{holes_effettive}). The physical meaning of the holes,
then, must be related to the effective number \( N_{H,eff} \), not to \( N_{H}. \)various

For the finite values of the scale, it is interesting to made the comparison
of the numerical data from the NLIE predictions with those obtained with the
Truncated Conformal Space Approach (TCSA, see section (\ref{section:TCSA}))
at several values of the parameter \( p \). For illustration, some interesting
cases will be presented. For the various values of the coupling constant, our
group obtained a spectacular agreement between the results obtained by the two
methods, up to deviations of order \( 10^{-4}-10^{-3} \)(see \cite{noi NP, noi PL1, noi PL2}).
The deviation grows with the volume \( L \), exactly as expected for truncation
errors. As also told in (\ref{section:TCSA}), in the attractive regime (\( p<1 \))
the TCSA convergence is fast.  The opposite happens for the NLIE, i.e. it converges
faster in the repulsive regime (\( p>1 \)) and for large volume \( L \), but
in general it is possible to have precisions of order \( 10^{-7} \) (and higher)
in both the regimes.

By studying other parameters such as the mass gap, the breather-soliton mass
ratios and the rate of convergence of the energy levels with increasing the
value of \( E_{cut} \), the small differences between the two methods can be
clearly attributed to the inaccuracy in the TCS data.

\section{The UV limit computation }

The ultraviolet limit of the states described by the NLIE will be examined,
in order to compare it with the known facts about the UV limit of sG/mTh theory
and conformal minimal models, outlined in sections (\ref{section:free_boson},
\ref{section:sG-mT}). 

More explicitly, it is known, from conformal perturbation theory, that the behaviour
of energy and momentum for very small \( L \) are given by:

\begin{equation}
\label{andamento-uv}
\begin{array}{c}
\displaystyle E(L)=-\displaystyle\frac{\pi \tilde{c}(l)}{6L}=-\displaystyle\frac{\pi }{6L}\left( c-12(\Delta +\bar{\Delta })\right) +\ldots \\
P(L)=\displaystyle\frac{2\pi }{L}(\Delta -\bar{\Delta })+\ldots 
\end{array}
\end{equation}
where the ellipsis are most regular terms in the space size \( L \). Therefore,
to determine the UV behaviour, in \( E \) and \( P \) only terms containing
\( \displaystyle\frac{1}{L} \) must be retained. Consider, e.g., the hole contribution to
the energy: \( {\cal M}\cosh h \). In order to have this hole, this term must
behaves as \( 1/L \), that is possible only if 
\begin{equation}
\label{log-uv}
h=\textrm{finite}\pm \log \displaystyle\frac{2}{l}.
\end{equation}
This means that only holes with this rapidity can contribute to UV. The same
argument applies to specials and complex solutions.

The behaviour of the sources for \( l\, \rightarrow \, 0 \) can be classified
by three possibilities: they can remain finite (they are called \emph{central}),
or they can move towards the two infinities as \( \pm \log \displaystyle\frac{2}{l} \) (\emph{right/left-movers}).
The finite parts of their positions can be obtained by extracting the divergent
part: 
\begin{equation}
\label{soluzioni-uv}
\left\{ \vartheta _{j}(l)\right\} \, =\, \left\{ \vartheta _{j}^{\pm }\pm \log \displaystyle\frac{2}{l},\, \vartheta ^{0}\right\} .
\end{equation}
(the \( l \) dependence of the roots has been made manifest). The number of
right/left moving resp. central holes is indicated by \( N^{\pm ,0}_{H} \)
and similarly the numbers \( N^{\pm ,0}_{S} \), \( M^{\pm ,0}_{C} \) and \( M^{\pm ,0}_{W} \)
are introduced. The finite parts in (\ref{soluzioni-uv}) satisfy a modified
version of the NLIE, known as \emph{kink equation}. To obtain that, observe
that \( Z \) has an implicit dependence from \( l \) that can be made manifest
writing \( Z(\lambda ,l). \) Define the \emph{kink functions} with the following
expression:
\begin{equation}
\label{kink_functionm}
Z_{\pm }(\vartheta )=\lim _{l\rightarrow 0}\, Z\left( \vartheta \pm \log \displaystyle\frac{2}{l},l\right) 
\end{equation}
The term ``kink'' has its origin in the fact that the asymptotic form of the
function \( Z(\vartheta ) \) has two plateaus stretching between the central
region and the regions of the left/right movers. The functions \( Z_{\pm }(\vartheta ) \)
describe the interpolation between the plateaus and the asymptotic behaviour
of \( Z(\vartheta ) \) at the corresponding infinity. Note that if there are
no central objects then the two plateaus merge into a single one stretching
from the left movers' region to the right movers' one. Using (\ref{soluzioni-uv},
\ref{kink_functionm}) in the NLIE (\ref{nlie-cont}), the source terms may
behave in different ways. Consider e.g. one hole \( h(l) \) of the type \( h^{+} \).
Its source term for the kink ``+'' is: 
\[
\lim _{l\rightarrow 0}\, \chi \left( \vartheta +\log \frac{2}{l}-h(l)\right) =\chi (\vartheta -h^{+}).\]
There are \( N^{+}_{H} \) terms of this of type. For a \( h^{-} \) hole the
source contribution to the same ``+'' kink is 
\[
\lim _{l\rightarrow 0}\, \chi \left( \vartheta +\log \frac{2}{l}-h(l)\right) =\lim _{l\rightarrow 0}\, \chi \left( \lambda -h^{-}+2\log \frac{2}{l}\right) =\chi _{\infty }\]
(\( \chi _{\infty } \) is defined in (\ref{chi_infinito})). One \( h^{0} \)
hole behaves in the same way. Then there are \( N_{H}-N^{+}_{H}=N_{H}^{-}+N_{H}^{0} \)
such terms. Analogous arguments apply to ``--'' kink and to the other roots.
For wide roots the limits of second determination are given in (\ref{limiti_chi}). 

The following definitions are introduced, according to (\ref{continuum_counting}):

\[
S^{\pm ,0}=\frac{1}{2}[N_{H}^{\pm ,0}-2N_{S}^{\pm ,0}-M_{C}^{\pm ,0}-2M^{\pm ,0}_{W}\theta (p-1)]\]
Observe that \( S \) is always integer, whereas \( S^{\pm ,0} \) can be half
integer; moreover, in some cases, \( S^{0} \) can be negative. All this numbers
are interpreted as the spin of the right, left movers and fixed solutions. Clearly
\[
S=S^{+}+S^{-}+S^{0}.\]
Using all this computations in the definition (\ref{kink_functionm}), the NLIE
for the ``+'' and ``--'' kinks reads:
\begin{equation}
\label{kink}
Z_{\pm }(\vartheta )=\pm e^{\pm \vartheta }+\alpha +g_{\pm }(\vartheta )+\displaystyle\int dxG(\vartheta -x){\cal Q}_{\pm }(x)
\end{equation}
where the following definitions are used:
\begin{equation}
\label{lW+-}
\begin{array}{c}
\displaystyle g_{\pm }(\vartheta )=\lim _{l\rightarrow 0}\, g\left( \left. \vartheta \pm \log \displaystyle\frac{2}{l}\right| \vartheta _{k}\right) =\pm 2\chi _{\infty }(S-S^{\pm })+2\pi l_{W}^{\pm }+\displaystyle\sum ^{N_{H}^{\pm }}_{k=1}\chi (\vartheta -h_{k}^{\pm })+\\
\\
-\displaystyle\sum ^{N_{S}^{\pm }}_{k=1}\left( \chi (\vartheta -\hat{y}_{k}^{\pm }+i\eta )+\chi (\vartheta -\hat{y}_{k}^{\pm }-i\eta )\right) -\displaystyle\sum ^{M_{C}^{\pm }}_{k=1}\chi (\vartheta -c_{k}^{\pm })-\displaystyle\sum _{k=1}^{M_{W}^{\pm }}\chi _{II}(\vartheta -w_{k}^{\pm })\, ;\\
\\
l_{W}^{\pm }=\pm \textrm{sign}(p-1)\, \displaystyle\frac{1}{2}\, (M_{W}-M_{W}^{\pm })
\end{array}
\end{equation}
and the function \( {\cal Q}_{\pm }(x) \) is related to \( Z_{\pm } \) as
\( Z \) to \( {\cal Q} \) in (\ref{funzione_Q}).

This equations allow to write the quantization conditions as:
\begin{equation}
\label{UV_quantum}
Z_{\pm }\left( \vartheta _{j}^{\pm }\right) =2\pi I_{j}^{\pm }.
\end{equation}
It is a matter of convenience to put the apex \( \pm  \) on the quantum numbers.
Indeed, they do not change in the limit procedure: they are exactly the same
as for finite \( l \). 

By simply taking the limit \( l\, \rightarrow \, 0 \) in NLIE reads an equation
for the fixed objects \( \vartheta ^{0} \):
\begin{equation}
\label{fixed_objects}
Z_{0}(\vartheta )=\lim _{l\rightarrow 0}\, Z(\vartheta ,l)=\alpha +g_{0}(\vartheta )+\displaystyle\int dxG(\vartheta -x){\cal Q}_{0}(x)
\end{equation}
where 
\[
\begin{array}{c}
\displaystyle g_{0}(\vartheta )=\lim _{l\rightarrow 0}\, g\left( \vartheta |\vartheta _{k}\right) =2\chi _{\infty }(S^{-}-S^{+})+2\pi l_{W}^{0}+\displaystyle\sum ^{N_{H}^{0}}_{k=1}\chi (\vartheta -h_{k}^{0})+\\
\\
-\displaystyle\sum ^{N_{S}^{0}}_{k=1}\left( \chi (\vartheta -\hat{y}_{k}^{0}+i\eta )+\chi (\vartheta -\hat{y}_{k}^{0}-i\eta )\right) -\displaystyle\sum ^{M_{C}^{0}}_{k=1}\chi (\vartheta -c_{k}^{0})-\displaystyle\sum _{k=1}^{M_{W}^{0}}\chi _{II}(\vartheta -w_{k}^{0})\, ;\\
\\
l_{W}^{0}=\textrm{sign}(p-1)\, \displaystyle\frac{1}{2}\, (M_{W}^{-}-M_{W}^{+})\, .
\end{array}\]
 As in the previous case the function \( {\cal Q}_{0}(x) \) is related to the
corresponding \( Z_{0}(x) \) by the usual expression (\ref{funzione_Q}). 

In the following the asymptotic values of \( Z_{\pm ,0} \) and the corresponding
values for \( {\cal Q} \) will play an important role. Equations (\ref{kink})
simply gives 
\begin{equation}
\label{zeta-infty}
Z_{+}(+\infty )=+\infty =-Z_{-}(-\infty ),
\end{equation}
consequently the value for \( {\cal Q}(x) \) can be obtained using (\ref{funzione_Q}):

\begin{equation}
\label{q-infty}
{\cal Q}_{\pm }(\pm \infty )=0
\end{equation}
(to compute it, before do the limit in \( x \) and later the limit in \( \eta  \),
because the NLIE holds with a finite \( \eta  \)). Moreover it can be shown
that
\[
\begin{array}{c}
g_{+}(-\infty )=g_{0}(+\infty )=2\chi _{\infty }(S-2S^{+})+2\pi k_{W}^{+}\\
g_{-}(+\infty )=g_{0}(-\infty )=-2\chi _{\infty }(S-2S^{-})+2\pi k_{W}^{-}
\end{array}\]
and
\[
k_{W}^{\pm }=\pm \frac{1}{2}\textrm{ sign }(p-1)\: (M_{W}-2M_{W}^{\pm })=\frac{M_{SC}}{2}+\textrm{integer}.\]
A direct consequence of this is that 
\begin{equation}
\label{doppio_plateau}
Z_{+}(-\infty )=Z_{0}(+\infty )\qquad \textrm{and}\qquad Z_{-}(+\infty )=Z_{0}(-\infty )
\end{equation}
because they satisfy the same equation. Remember that for small \( l \) the
counting function gives, in the central region, one or two plateaus that stretch
to the infinities. Then the previous equations simply mean that \( Z_{0}(+\infty ) \)
is the right plateau height and and \( Z_{0}(-\infty ) \) the left plateau
height. If there are no central objects, they take the same value, yielding
so a one plateau system. 

The explicit values can be computed, using the kink equations. The integral
can be handled because the \( {\cal Q} \) in the asymptotic limit takes a constant
value and the kernel \( G \) rapidly converges to zero:
\begin{equation}
\label{plateau}
\begin{array}{c}
Z_{+}(-\infty )=\alpha +g_{+}(-\infty )+\displaystyle\frac{\chi _{\infty }}{\pi }{\cal Q}_{+}(-\infty )\\
Z_{-}(+\infty )=\alpha +g_{-}(+\infty )+\displaystyle\frac{\chi _{\infty }}{\pi }{\cal Q}_{-}(+\infty )\, .
\end{array}
\end{equation}
They are called \emph{plateau equations}. From the definition of \( {\cal Q} \)
in (\ref{funzione_Q}), and making attention to the warning concerning the range
of values allowed for it, it follows that
\[
Z_{\pm }(\mp \infty )={\cal Q}_{\pm }(\mp \infty )+\pi \delta +2\pi k_{\pm }\]
with an appropriate choice of the integers \( k_{\pm } \) such that the condition
\( -\pi \leq {\cal Q}_{+}(-\infty )\leq \pi  \) holds (remember that \( k_{W}^{\pm } \)
can be half-integer).The solution of the plateau equation is then
\begin{equation}
\label{Q+-infinito}
\omega _{\pm }\equiv {\cal Q}_{\pm }(\mp \infty )=\pm 2\pi \displaystyle\frac{p-1}{p+1}(S-2S^{\pm })+2\pi \displaystyle\frac{p}{p+1}\left( \displaystyle\frac{\alpha }{\pi }-\delta +2k_{W}^{\pm }-2k_{\pm }\right) .
\end{equation}
Notice that for \( p>1 \) there can be cases without solution. If instead it
is there, there is no ambiguity in the choice of \( k_{\pm } \), because its
contribution is a multiple of \( 4\pi \displaystyle\frac{p}{p+1}>2\pi  \), that is larger
than the range of values allowed for \( {\cal Q}_{\pm } \). Instead, for \( p<1 \)
the solution always exists, but it can happens that two or more choices of \( k_{\pm } \)
satisfy the condition on the range of \( {\cal Q}_{\pm } \). It will be clear
later, that this phenomenon is related to the fact that wide roots become excitations
independent of the other one's, which is manifestly evident from the fact that
they no longer contribute to the spin of the state \( S \) and that the asymptotic
value of their source contribution \( \chi _{II} \) in the attractive regime
is simply a \( \pi  \) (see \ref{limiti_chi}).

Before to continue the general computation, the kink equations for the wide
roots will be explicitly written. They can be obtained by using the same procedure
used for (\ref{kink}), applied to the correct expression for \( Z(\vartheta ) \)
given in (\ref{ZN_sec_determ}):
\begin{equation}
\label{Z_sec_det}
\begin{array}{c}
\displaystyle Z_{\pm }(\vartheta )=\pm e^{\vartheta }_{II}+\displaystyle\sum ^{N_{H}^{\pm }}_{k=1}\chi _{II}(\vartheta -h_{k}^{\pm })-\displaystyle\sum ^{N_{S}^{\pm }}_{k=1}\left( \chi _{II}(\vartheta -\hat{y}_{k}^{\pm }+i\eta )+\chi _{II}(\vartheta -\hat{y}_{k}^{\pm }-i\eta )\right) \\
\\
-\displaystyle\sum ^{M_{C}^{\pm }}_{k=1}\chi _{II}(\vartheta -c_{k}^{\pm })-\displaystyle\sum _{k=1}^{M_{W}^{\pm }}\chi _{II}(\vartheta -w_{k}^{\pm })_{II}\pm \theta (p-1)4\chi _{\infty }(S-S^{\pm })+2\pi L_{W}^{\pm }(\textrm{sign}\Im m\vartheta )+\\
\\
+\displaystyle\int dx\, G_{II}(\vartheta -x){\cal Q}(x)+\alpha _{II}\qquad \quad \textrm{for }|\Im m\, \vartheta |>\min (1,\, p)
\end{array}
\end{equation}
where the notation introduced for \( L_{W} \) is so complicated because it
depends on the sign of the imaginary part of \( \vartheta  \):
\begin{equation}
\label{Lw_definizione}
\displaystyle L_{W}^{\pm }(\textrm{sign}\Im m\vartheta )=-\displaystyle\sum _{k=1}^{M_{W}-M_{W}^{\pm }}\lim _{x\, \rightarrow \pm \infty }\left[ \chi _{II}(\vartheta -w_{k}+x)_{II}\mp 4(\chi _{\infty }-\pi )\theta (p-1)\right] =\pi \textrm{ }\cdot \textrm{integer}
\end{equation}
The explicit form is quite complicated. Then this contribution must be computed
case by case. Observe that the term \( e^{\vartheta }_{II} \) vanishes for
\( p>1 \) and \( \alpha _{II} \) vanishes for \( p<1 \).

Now, the ultraviolet limit on the energy and momentum expressions will be done.
To this end, substitute in the energy and momentum expressions (\ref{energia},
\ref{momento}) the ultraviolet behaviour of the roots (\ref{soluzioni-uv})
and retain only terms in \( \displaystyle\frac{1}{L} \). The kinetic terms give contributions
as 
\begin{equation}
\label{UV_cinetico}
{\cal M}\cosh \vartheta _{k}\sim \displaystyle\frac{1}{L}e^{\pm \vartheta _{k}^{\pm }}.
\end{equation}
 The integral terms must be calculated separately. Their form in the energy
and momentum expressions is, respectively:
\[
\begin{array}{c}
\displaystyle -{\cal M}\displaystyle\int \displaystyle\frac{dx}{2\pi }\sinh x\, {\cal Q}(x)=-{\cal M}\displaystyle\int \displaystyle\frac{dx}{2\pi }\displaystyle\frac{1}{2}[e^{x}\, {\cal Q}(x)-e^{-x}\, {\cal Q}(x)]\, ,\\
-{\cal M}\displaystyle\int \displaystyle\frac{dx}{2\pi }\cosh x\, {\cal Q}(x)=-{\cal M}\displaystyle\int \displaystyle\frac{dx}{2\pi }\displaystyle\frac{1}{2}[e^{x}\, {\cal Q}(x)+e^{-x}\, {\cal Q}(x)]\, ,
\end{array}\]
where the contribution depending on \( e^{x} \) is called \emph{``+'' kink
contribution}, and the other term is the \emph{``--'' kink contribution}.
From the definition itself of \( {\cal Q}_{\pm } \) observe that, in the limit
of very small \( l \), it is possible to write for \( {\cal Q} \):
\begin{equation}
\label{qu-asintotico}
{\cal Q}\left( x\pm \log \displaystyle\frac{2}{l},l\right) \sim {\cal Q}_{\pm }\left( x\right) +q_{\pm }(x,l)\: ,
\end{equation}
where the two functions \( q_{\pm }(x,l) \) vanishes in the \( l\, \rightarrow \, 0 \)
limit. The expression (\ref{qu-asintotico}) can be substituted in the integral
form for the two kinks. The integration variable can be shifted by \( x\, \rightarrow \, x+\log (2/l) \)
then
\begin{equation}
\label{kink+}
{\cal M}\displaystyle\int \displaystyle\frac{dx}{2\pi }\displaystyle\frac{1}{2}e^{x}\, {\cal Q}(x)\doteq \displaystyle\frac{1}{L}\displaystyle\int \displaystyle\frac{dx}{2\pi }e^{x}\, {\cal Q}_{+}(x)\, ,
\end{equation}
where the symbol \( \doteq  \) means that only the terms of order \( 1/L \)
are retained. Similarly for the ``--'' kink term
\begin{equation}
\label{kink-}
{\cal M}\displaystyle\int \displaystyle\frac{dx}{2\pi }\displaystyle\frac{1}{2}e^{-x}\, {\cal Q}(x)\doteq \displaystyle\frac{1}{L}\displaystyle\int \displaystyle\frac{dx}{2\pi }e^{-x}\, {\cal Q}_{-}(x)\, .
\end{equation}
 At this point, it is possible to express energy and momentum in a way dependent
only on quantities which are finite in the UV limit. Using then (\ref{UV_cinetico},
\ref{kink+}, \ref{kink-}) in (\ref{andamento-uv}), and rearranging the expression
in terms of \( \Delta ,\, \bar{\Delta } \) (i.e. restoring the light-cone coordinates)
follows:
\[
\begin{array}{c}
\displaystyle \Delta ^{\pm }=\displaystyle\frac{c}{24}+\displaystyle\frac{1}{2\pi }\left( \displaystyle\sum _{j=1}^{N_{H}^{\pm }}e^{\pm h^{\pm }_{j}}-\displaystyle\sum _{j=1}^{N_{S}^{\pm }}\left( e^{\pm \hat{y}^{\pm }_{j}+i\eta }+e^{\pm \hat{y}^{\pm }_{j}-i\eta }\right) +\right. \\
\\
\left. -\displaystyle\sum _{j=1}^{M_{C}^{\pm }}e^{\pm c^{\pm }_{j}}-\displaystyle\sum _{j=1}^{M^{\pm }_{W}}e_{II}^{\pm w_{j}^{\pm }}\mp \displaystyle\int \displaystyle\frac{dx}{2\pi }e^{\pm x}\, {\cal Q}_{\pm }(x)\right) .
\end{array}\]
As a notation, \( \Delta =\Delta ^{+} \) and \( \bar{\Delta }=\Delta ^{-} \).
The NLIE can be used now to express the various terms appearing in the sums
and the integrals. Consider, for example, the quantization condition (\ref{UV_quantum})
for the holes \( h_{j}^{\pm } \). Using the oddity of the \( \chi (x) \) function,
it can be arranged in a more convenient way:
\[
\begin{array}{c}
\pm \displaystyle\sum _{j=1}^{N^{\pm }_{H}}e^{\pm h^{\pm }_{j}}=2\pi \displaystyle\sum ^{N^{\pm }_{H}}_{j=1}I^{\pm }_{h_{j}}+\displaystyle\sum ^{N^{\pm }_{H}}_{j=1}\left( \displaystyle\sum ^{N_{S}^{\pm }}_{k=1}\left( \chi (h^{\pm }_{j}-\hat{y}_{k}^{\pm }+i\eta )+\chi (h^{\pm }_{j}-\hat{y}_{k}^{\pm }-i\eta )\right) +\right. \\
\\
\left. +\displaystyle\sum ^{M_{C}^{\pm }}_{k=1}\chi (h^{\pm }_{j}-c_{k}^{\pm })+\displaystyle\sum _{k=1}^{M_{W}^{\pm }}\chi _{II}(h^{\pm }_{j}-w_{k}^{\pm })\right) -N_{H}^{\pm }\left( \alpha \pm 2\chi _{\infty }(S-S^{\pm })+2\pi l_{W}^{\pm }\right) +\\
\\
-\displaystyle\sum ^{N_{H}^{\pm }}_{j=1}2\, \Im m\, \displaystyle\int \displaystyle\frac{dx}{i}G(h^{\pm }_{j}-x-i\eta )\log \left( 1+(-)^{\delta }e^{iZ_{\pm }(x+i\eta )}\right) 
\end{array}\]
(the nonzero \( \eta  \) has been restored for later convenience). For the
other objects, similar forms hold (remember that the \( \hat{y}_{k} \) are
defined by (\ref{condizione1_special})). All the equations so obtained must
be substituted in the expression for \( \Delta ^{\pm } \). Now, observe that
all the sums of \( \chi (\vartheta ) \), \( \chi _{II}(\vartheta ) \) and
\( \chi _{II}(\vartheta )_{II} \) cancel completely for the oddity of the functions.
Moreover in the integral term the following substitution (for all the type of
sources) can be made
\begin{equation}
\label{source-uv}
2\pi G(h^{\pm }_{j}-x)=\chi '(x-h^{\pm }_{j}).
\end{equation}
 The result is:

\[
\begin{array}{c}
\displaystyle \Delta ^{\pm }=\displaystyle\frac{c}{24}\pm \left( I_{H}^{\pm }-2I^{\pm }_{S}-I^{\pm }_{C}-I^{\pm }_{W}\right) -\displaystyle\frac{2\chi _{\infty }}{\pi }\left( S-S^{\pm }\right) S^{\pm }\mp S^{\pm }\left( \displaystyle\frac{\alpha }{\pi }+l_{W}^{\pm }\right) +\\
\pm {\cal L}^{\pm }_{W}\mp 2\, \Im m\, \displaystyle\int \displaystyle\frac{dx}{(2\pi )^{2}i}\varphi ^{,}_{\pm }(x+i\eta )\log \left( 1+(-)^{\delta }e^{iZ_{\pm }(x+i\eta )}\right) .
\end{array}\]
The following notation has been introduced:
\[
I^{\pm }_{H}=\sum ^{N^{\pm }_{H}}_{j=1}I_{h_{j}}^{\pm }\, ,\, I^{\pm }_{C}=\sum ^{M^{\pm }_{C}}_{j=1}I_{c_{j}}^{\pm }\, ,\, I^{\pm }_{W}=\sum ^{M^{\pm }_{W}}_{j=1}I_{w_{j}}^{\pm }\, \mathrm{and}\, I^{\pm }_{S}=\sum ^{N^{\pm }_{S}}_{j=1}I_{y_{j}}^{\pm }\: .\]
The new constant appearing in that equation takes into account for wide roots,
and is given by (see also (\ref{Z_sec_det})): 
\[
\begin{array}{c}
{\cal L}^{\pm }_{W}=\displaystyle\sum _{j=1}^{M^{\pm }_{W}}L_{W}^{\pm }(\textrm{sign}\Im m\, w_{j}^{\pm })=-\pi \textrm{sign}(p-1)M_{W}^{\pm }2(S-S^{\pm })+\\
-\displaystyle\sum _{j=1}^{M^{\pm }_{W}}\; \displaystyle\sum _{k=1}^{M_{W}-M_{W}^{\pm }}\lim _{x\, \rightarrow \, \pm \infty }\left[ \chi _{II}(w_{j}^{\pm }-w_{k}+x)_{II}\mp 4(\chi _{\infty }-\pi )\theta (p-1)\right] =\pi \textrm{ }\cdot \textrm{integer}
\end{array}\]
It must be computed case by case.

The following function has been introduced:
\begin{equation}
\label{var-fi}
\begin{array}{c}
\displaystyle \varphi _{\pm }(\vartheta )=\pm e^{\pm \vartheta }+\displaystyle\sum ^{N_{H}^{\pm }}_{k=1}\chi (\vartheta -h_{k}^{\pm })-\displaystyle\sum ^{N_{S}^{\pm }}_{k=1}\left( \chi (\vartheta -y_{k}^{\pm }+i\eta )+\chi (\vartheta -y_{k}^{\pm }-i\eta )\right) -\\
\\
-\displaystyle\sum ^{M_{C}^{\pm }}_{k=1}\chi (\vartheta -c_{k}^{\pm })-\displaystyle\sum _{k=1}^{M_{W}^{\pm }}\chi _{II}(\vartheta -w_{k}^{\pm })=Z_{\pm }(\vartheta )-\displaystyle\int dxG(\vartheta -x){\cal Q}_{\pm }(x)-\alpha 
\end{array}
\end{equation}
Now, with the help of the computations shown in appendix \ref{section:Lemma_di_Destri}
the integral becomes:
\[
2\, \Im m\, \int \frac{dx}{(2\pi )^{2}i}\varphi ^{,}_{\pm }(x+i\eta )\log \left( 1+(-)^{\delta }e^{iZ_{\pm }(x+i\eta )}\right) =\pm \left( \frac{1}{24}-\frac{{\cal Q}^{2}_{\pm }(\mp \infty )}{16\pi ^{2}}\frac{p+1}{p}\right) \]
and the UV limit computation ends with a close form for the conformal dimensions:
\begin{equation}
\label{delta}
\begin{array}{c}
\displaystyle \Delta ^{\pm }=\displaystyle\frac{c-1}{24}\mp \displaystyle\frac{\alpha }{\pi }S^{\pm }\pm \left( I_{H}^{\pm }-2I^{\pm }_{S}-I^{\pm }_{C}-I^{\pm }_{W}\right) \mp S^{\pm }2l_{W}^{\pm }+\\
\pm {\cal L}^{\pm }_{W}-\displaystyle\frac{2\chi _{\infty }}{\pi }\left( S-S^{\pm }\right) S^{\pm }+\displaystyle\frac{{\cal Q}^{2}_{\pm }(\mp \infty )}{16\pi ^{2}}\displaystyle\frac{p+1}{p}
\end{array}
\end{equation}
This means that the UV limit admits an exact computation of the spectrum. The
identification of the UV states is one of the fundamental steps to understand
the physical interpretation of the continuum model so far defined. 

An interesting phenomenon is that the conformal weights obtained depend only
on very generic features of the source configuration such as the asymptotics
of the left and right moving sources, the total spin and the number of self-conjugate
roots. This means that if a certain source configuration is given, one can add
new sources separately to the right and the left moving part in such a way that
they are separately neutral (i.e. do not change \( S^{+} \) and \( S^{-} \),
the total spin and the number of self-conjugate roots). In this way the primary
weights do not change, but generally the term \( I^{\pm }_{H}-2I^{\pm }_{S}-I^{\pm }_{C}-I^{\pm }_{W} \)
is increased and so descendents of the initial state are created. An example:
states which have \( S=0=M_{SC} \) and \( S^{\pm }=0 \) are all descendants
of the vacuum, however complicated their actual source configurations are.

It is important also to drive attention to the well known fact (see the section
\ref{section:interpr_fisica}) that (at the moment) there is no proof that the
scaling functions obtained by the method of NLIE span the complete space of
states. This is an extremely difficult problem due to the following two circumstances:
(1) the dependence of the UV conformal weights from the parameters of the source
configuration is rather complicated and (2) to obtain the allowed values of
the complex roots one has to carefully examine the IR limit as suggested in
section \ref{section:IR}. It will be also clear that the same state can be
realized by different root configurations depending on the regime (e.g. the
\( (s\bar{s})_{\pm } \) states: scattering of soliton and antisoliton in even
and odd wave functions; see later).

\section{Sine-Gordon and massive Thirring\label{section:sG-mTH}}

As suggested in section (\ref{section:limite_cont.}), the choice \( \omega =\alpha =0 \)
is supposed to describe the sine-Gordon and massive Thirring models on a cylinder.
In this section the corresponding NLIE will be analyzed, starting with the vacuum
state. Obviously, the coupling \( p \) is expected to be the same that appears
in the s-G and mTh lagrangians via the equation (\ref{parametro_p}). 

\begin{figure}[  htbp]
{\centering \includegraphics{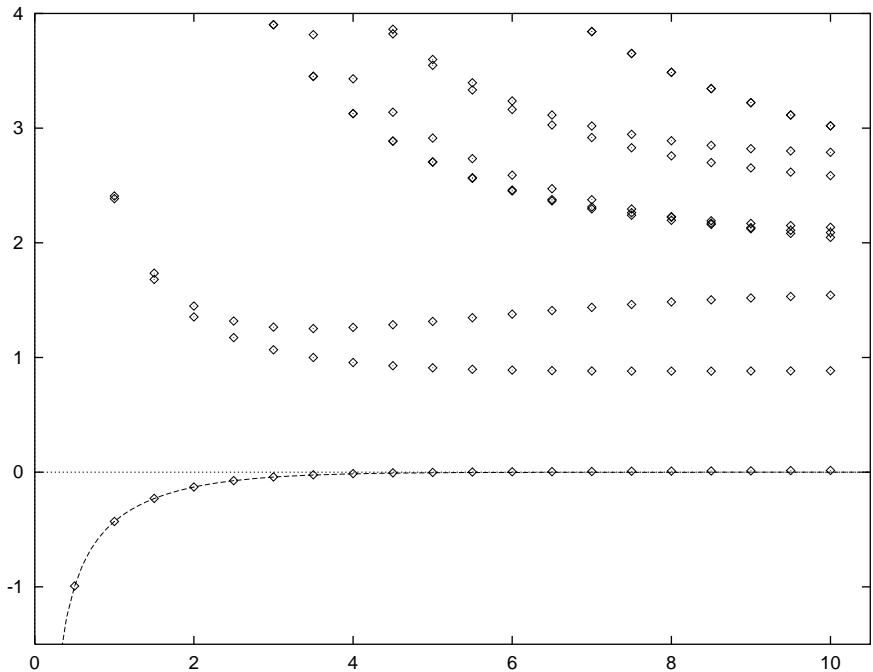} \par}

\caption{\small The first few energy levels (TCSA) in the vacuum (\protect\( m=0\protect \))
sector at \protect\( p=\frac{2}{7}\protect \) (plotted with diamonds) for \protect\( E_{cut}=17.0\protect \)
(the dimension of the space is \protect\( 4141\protect \)) and the NLIE prediction
for the vacuum scaling function (shown with a dotted line).\label{vuoto.eps}}
\end{figure}

Starting from a lattice with \( 2N \) sites one finds that the antiferromagnetic
ground state which has spin \( S=0 \), when written in terms of Bethe vectors
depends on \( M=N \) roots of the Bethe equations, all of which are real, as
indicated in (\ref{section:6_vertex_BA}). This ground state is expected to
correspond to the vacuum of the field theory. However, there are two possibilities:
one for \( N \) even and the other for \( N \) odd, corresponding in the continuum
to the choices \( \delta =0 \) or \( \delta =1 \). As it will be shown in
the sequel, only one of these states can be identified with the vacuum for a
local field theory having a \( c=1 \) UV limiting CFT. The UV dimensions of
the vacuum are \( \Delta ^{\pm }=0 \) (because the theory is assumed unitary).
Choosing \( \delta =0 \), the expression (\ref{delta}) gives a value consistent
with the interpretation proposed, i.e. 
\[
c=1,\]
as obtained in the paper \cite{ddv 95}. At the IR, because no holes neither
complex roots are considered, the energy and momentum (\ref{energia}, \ref{momento})
vanishes, that is what is expected for the vacuum. The numerical iteration of
the NLIE gives the scaling energy shown in figure \ref{vuoto.eps}. In that figure, the value of the predicted bulk energy (\ref{bulk}) has been
subtracted from TCSA data in order to normalize them at the same way as the
NLIE data are. The agreement of the two data is to order \( 10^{-4}-10^{-3} \).

As a consequence of this vacuum analysis, the equations for the asymptotic behaviour
(\ref{Q+-infinito}) and for the conformal dimensions (\ref{delta}) take a
simpler form (\( \alpha =0 \) and \( c=1 \)). If the physical interpretation
is consistent, the expected conformal dimensions must have the form (\ref{delta+-_vertex_op}),
that is a sum of powers of \( R^{2}=\frac{p+1}{2p} \) (as is (\ref{parametro_p})).
Then it is convenient to express \( \Delta ^{\pm } \) in powers of \( \frac{p+1}{p} \),
using also the expression (\ref{Q+-infinito}) for \( {\cal Q}^{2}_{\pm }(\mp \infty ) \).
The final form is 
\begin{equation}
\label{delta_sG}
\begin{array}{c}
\Delta ^{\pm }=\displaystyle\frac{p}{p+1}n^{2}_{\pm }+\displaystyle\frac{m^{2}}{16}\displaystyle\frac{p+1}{p}\pm \displaystyle\frac{n_{\pm }m}{2}+N_{\pm }=\\
=\displaystyle\frac{1}{2}\left( \displaystyle\frac{n_{\pm }}{R}\pm \displaystyle\frac{1}{2}mR^{2}\right) ^{2}+N_{\pm }
\end{array}
\end{equation}
where the following identifications have been made:
\begin{equation}
\label{m,n+-}
\begin{array}{c}
m=2S\\
n_{\pm }=\left( \displaystyle\frac{\delta }{2}-k_{W}^{\pm }+k_{\pm }\right) \mp \left( S-2S^{\pm }\right) 
\end{array}
\end{equation}
and 
\begin{equation}
\label{interiN+-}
N_{\pm }=\pm \left( I_{H}^{\pm }-2I^{\pm }_{S}-I^{\pm }_{C}-I^{\pm }_{W}\right) \mp S^{\pm }2l_{W}^{\pm }\pm {\cal L}^{\pm }_{W}-2\left( S^{\pm }\right) ^{2}\mp 2S^{\pm }\left( \displaystyle\frac{\delta }{2}-k_{W}^{\pm }+k_{\pm }\right) .
\end{equation}
The similarity with (\ref{delta+-_vertex_op}) is manifest. To match an exact
correspondence two conditions are required. The first one requires that 
\begin{equation}
\label{n+-}
|n_{+}|=|n_{-}|.
\end{equation}
The second condition is that one must be able to chose the sign of \( n \)
in such a way that the left and right descendent numbers are integer. Consider
the case where \( n_{+}=n_{-} \): then (using (\ref{delta_sG})) follows that
both \( N_{\pm } \) must be integers. Instead, in the case \( n_{+}=-n_{-} \)
a combination of \( N_{\pm } \) and \( \pm \displaystyle\frac{n_{\pm }m}{2} \) must be
integer. 

Since the \( R^{-2} \) contribution comes exclusively from the last term in
(\ref{delta}), the one-plateau systems (i.e. when \( {\cal Q}_{+}(-\infty )={\cal Q}_{-}(+\infty ) \))
satisfy trivially the condition (\ref{n+-}), but even in that case it is not
clear whether the descendent numbers are integers. At the time of this writing
the analysis of our group cannot exclude the possibility that there exist some
states for which the conformal weights cannot be interpreted within the framework
of \( c=1 \) CFT. However, in the numerous explicit examples we have calculated
so far we have not found any configuration of sources which lead to such behaviour.
If it happens for some configuration of the sources then the corresponding scaling
function must be excluded from the spectrum. 

From the second line of (\ref{m,n+-}) one obtains a useful relation (\( m=2S\in \mathbb Z \))
\[
n_{\pm }=\frac{m+\delta +M_{SC}}{2}+\textrm{integers}.\]
Using now the classification of UV conformal operators for s-G and mTh in the
two algebras \( {\cal A}_{b} \) and \( {\cal A}_{f} \) given in section \ref{section:sG-mT},
one can obtain the following rule:
\begin{equation}
\label{regola_d'oro}
\begin{array}{c}
\displaystyle\frac{m+\delta +M_{SC}}{2}\in \mathbb Z\qquad \textrm{for}\qquad \textrm{sine Gordon}\\
\displaystyle\frac{\delta +M_{SC}}{2}\in \mathbb Z\qquad \textrm{for}\qquad \textrm{massive Thirring}.
\end{array}
\end{equation}
With reference to figure \ref{4sectors.eps}, from this rule follows that all
the four sectors can be accessed by NLIE, because the sectors \textbf{I} and
\textbf{II} describe sine-Gordon, \textbf{I} and \textbf{III} describe massive
Thirring and the sector \textbf{IV} can be obtained by \( m \) even and \( \displaystyle\frac{\delta +M_{SC}}{2} \)
half-integer. If there are no self-conjugate roots, the rule simplifies: \( \delta =0 \)
for mTh (\textbf{I} and \textbf{III}) and \( \delta =1 \) for the sectors \textbf{II}
and \textbf{IV}. The sector \textbf{IV}, that also is accessible by NLIE, contains
non local states. Then NLIE describes also this non local sector.

\subsection{``--`` vacuum}

Consider the case of \( g(\vartheta |\vartheta _{j})=0 \) (i.e. no sources).
There are two possibilities \( \delta =0,\, 1 \). One is the vacuum, considered
previously, the other is \( \delta =1 \). The (\ref{Q+-infinito}) yields
\begin{equation}
\label{vuoto-}
{\cal Q}_{\pm }(\mp \infty )=-2\pi \displaystyle\frac{p}{p+1}(1+2k_{\pm })\, ,
\end{equation}
which admits solution only in the attractive regime when
\begin{equation}
\label{vuoto-attr}
|{\cal Q}_{+}(-\infty )|=|{\cal Q}_{-}(+\infty )|=2\pi \displaystyle\frac{p}{p+1}\, .
\end{equation}
The value is not determined uniquely due to the contribution of wide roots,
but it can be fixed using information from the repulsive regime. 

\begin{figure}[  htbp]
{\centering \includegraphics{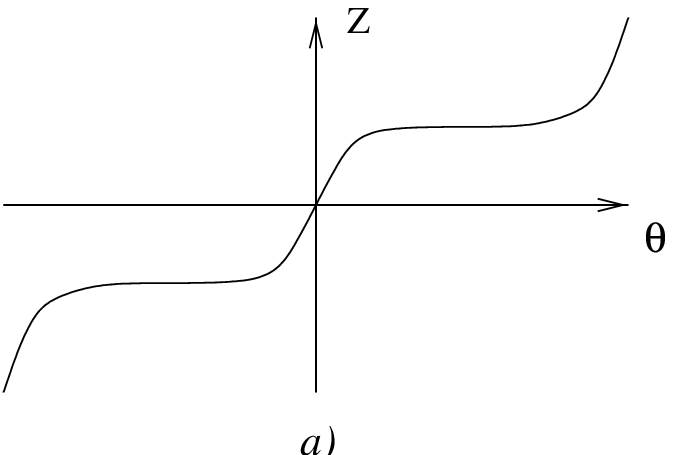}  \includegraphics{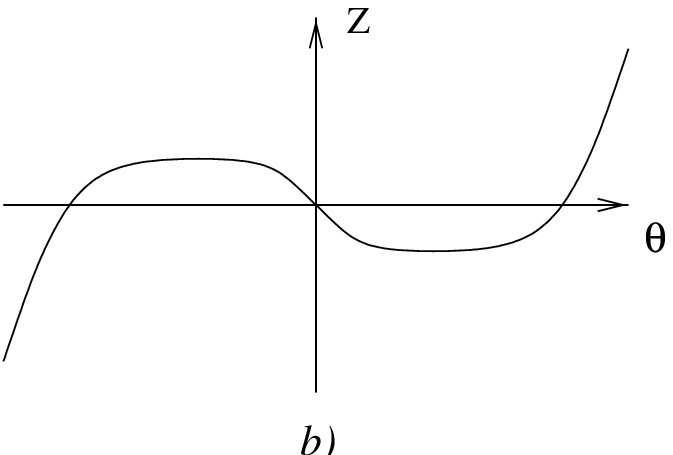} \par}

\caption{\small The UV behaviour of \protect\( Z\protect \)(two plateaus system) for (a) \protect\( p<1\protect \)
and (b) \protect\( p>1\protect \). In (b) the root at the origin turned into
a special one and two normal holes appeared at the points where \protect\( Z\protect \)
crosses the horizontal axis with a positive derivative.\label{special_rep.eps}}
\end{figure}

In that case, by performing the usual iteration procedure described in section
\ref{section:intermedio}, one finds that only for \( l>l_{0} \) the iteration
converges. For smaller values of \( l \) there is no convergence. The ``breakdown''
volume \( l_{0} \) depends by \( p \): for \( p=1.5 \) it is \( l_{0}=0.3 \)
(no breakdown in attractive regime). What happens is that the real root at the
origin is a special one with \( Z'(0)<0 \). By (\ref{continuum_counting}),
because \( S=0 \), also two holes must appear. All of which sources have Bethe
quantum number zero. One of the holes moves to the left and the other one to
the right and so \( S^{\pm }=\displaystyle\frac{1}{2} \), while the special root is central
(the configuration is completely symmetric, then \( Z(-x)=-Z(x) \)). This allows
for a unique solution of the plateau equations:
\[
\displaystyle {\cal Q}_{+}(-\infty )=\pm \frac{2\pi }{p+1}\: ,\qquad Z_{+}(-\infty )=-Z_{-}(+\infty )=\pi \frac{1-p}{p+1}<0\: .\]
Observe that \( Z_{-}(+\infty )>Z_{+}(-\infty ) \). This compared with (\ref{doppio_plateau})
means that there are two plateaus and the left is higher than the right, i.e.
\( Z'(0)<0 \). The picture is completely consistent. This phenomenon can be
thought of as the splitting of the root at the origin into a special root and
two holes and is one of the general mechanisms in which the special sources
are generated (see figure \ref{special_rep.eps}). Another mechanism will be
explored when the UV limit of the soliton-antisoliton states will be examined.Let us suppose that the counting function \( Z \) is analytic as a function
of \( p \) which is plausible because all terms in the NLIE for this state
are analytic in the coupling. This determines which branch to choose for the
plateau values in the attractive regime. One obtains the final result (for both
the attractive and the repulsive regime)
\[
\Delta ^{\pm }=\frac{1}{4}\frac{p}{p+1}=\frac{1}{8R^{2}}=\Delta ^{\pm }_{\pm 1/2,0}\: ,\]
which are the conformal weights of the vertex operators \( V_{(\pm 1/2,0)} \).
The actual UV limit can be a linear combination of these operators; in any case,
it is not contained in the UV spectrum of the sG/mTh theory, but is a state
in the nonlocal sector (sector \textbf{IV} in figure \ref{4sectors.eps}).

\subsection{Pure hole states}

The analysis of pure holes states is the simplest one. In the IR limit one such
state gives (using (\ref{IR_NLIE}):
\begin{equation}
\label{purehole_BA}
Z(\vartheta )=l\sinh \vartheta +\displaystyle\sum _{j=1}^{N_{H}}\chi (\vartheta -h_{j})\quad ,\quad Z(h_{j})=2\pi I_{j}.
\end{equation}
The holes ``rapidities'' are real. Using the observation (\ref{matrice_S_chi})
the equations (\ref{purehole_BA}) show the same structure of the box quantization
equations for physical particles (\ref{quantizz-box}). Then the holes can be
interpreted as particles with rapidity \( h_{j} \), i.e. they are the solitons
of sine-Gordon. Pure holes states are states with only solitons.

The IR limit puts no restriction on the quantum numbers \( I_{k} \) of the
holes; instead the rapidities converge to zero. This seems to be a very general
feature that remains true even when complex roots are allowed. 

\begin{figure}[  htbp]
{\centering \includegraphics{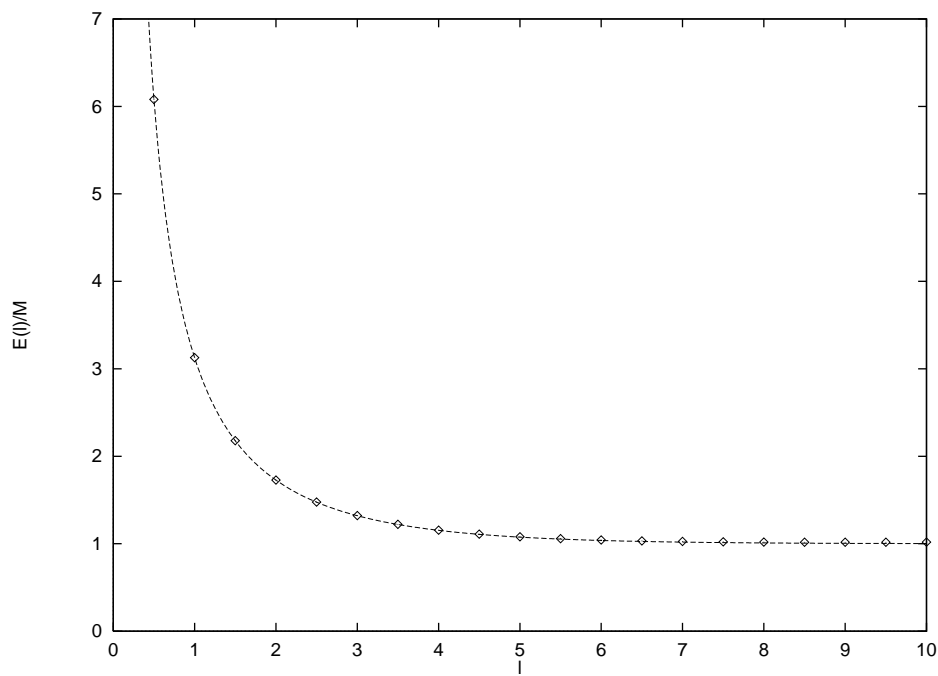} \par}

\caption{\small Comparison of the results for the one-soliton state coming NLIE and TCSA at
\protect\( p=\frac{2}{7}\protect \). Continuous line is the NLIE result, while
the diamonds represent the TCSA data.\label{one_hole.eps}}
\end{figure}

The UV calculations, for \textbf{one hole}, give for \( \delta =1 \) (the hole
can be put fixed in the origin):

\[
\begin{array}{lr}
S=\displaystyle\frac{1}{2}; & S^{\pm }=0.
\end{array}\]
Solving the equation for \( {\cal Q}_{\pm }(\mp \infty ) \) gives:
\[
\begin{array}{c}
m=1\\
n_{+}=n_{-}=0
\end{array}\]
that correspond to the operators \( V_{0,\pm 1} \). They create the sine-Gordon
soliton (see also \cite{kl-me}). For one hole with \( \delta =0 \), using

\[
\begin{array}{lr}
S^{+}=\displaystyle\frac{1}{2}; & S^{-}=S=0,
\end{array}\]
the UV computation gives the dimensions of \( V_{1/2,\, 1} \) or \( V_{-1/2,\, -1} \)
that describe the Thirring fermion. The UV picture is consistent with the IR.
The behaviour at intermediate values of \( l \) is shown in figure \ref{one_hole.eps}where it appears that the interpretation of the hole with the soliton in consistent.
This is the simplest case where the NLIE was used, in \cite{noi PL2}, for an
odd number of particles.\begin{table}[  htbp]
{\centering \begin{tabular}{|c|c|c|c|}
\hline 
l&
NLIE&
TCS&
Relative deviation \\
\hline 
\hline 
.5&
6.080571&
6.08062&
0.000008\\
\hline 
1&
3.126706&
3.12685&
0.00005\\
\hline 
1.5&
2.177411&
2.17791&
0.0002\\
\hline 
2&
1.727224&
1.72776&
0.0003\\
\hline 
2.5&
1.475004&
1.47593&
0.0006\\
\hline 
3&
1.320353&
1.32168&
0.001\\
\hline 
4&
1.153188&
1.15548&
0.002\\
\hline 
5&
1.075376&
1.07908&
0.003\\
\hline 
\end{tabular}\par}

\caption{\small \label{onesoliton_table} Numerical comparison of the energy levels predicted
by the NLIE to the TCS data for the case depicted in figure \ref{one_hole.eps}.
The NLIE data are exact to the precision shown.}
\end{table} Table \ref{onesoliton_table} gives an idea about the numerical magnitude of
the difference. Note that the deviations are extremely small for small values
of \( l \) and they grow with the volume, exactly as expected for truncation
errors. 

\begin{figure}[  htbp]
{\centering \includegraphics{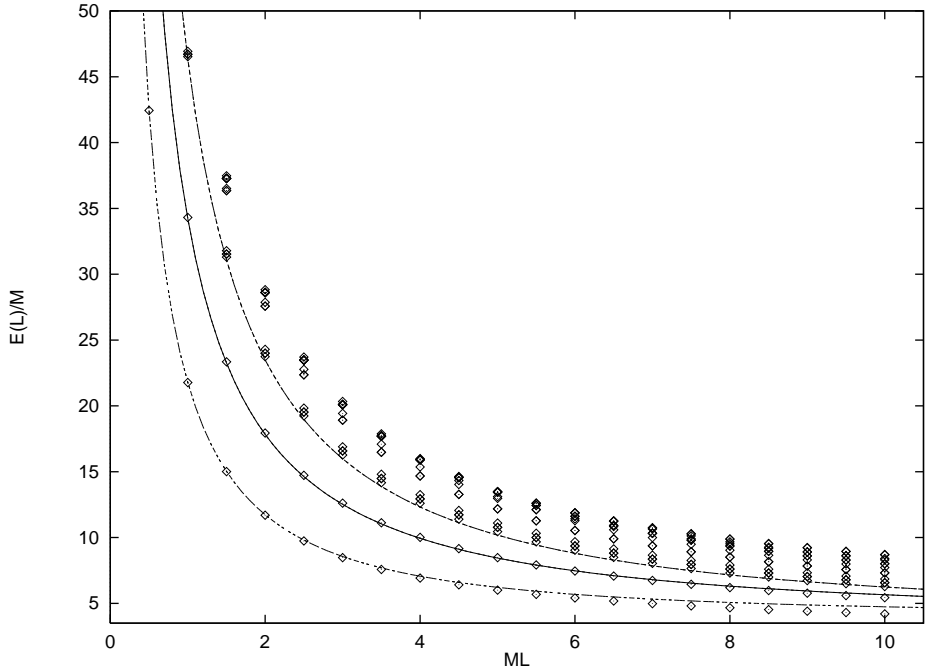} \par}

\caption{\small The first few energy levels (TCSA) for \protect\( m=4\protect \) solitons
at \protect\( p=1.5\protect \) (plotted with diamonds) for \protect\( E_{cut}=22.5\protect \)
(the dimension of the space is 4149) and the NLIE prediction for the four holes
scaling functions with quantum numbers \protect\( (-\frac{3}{2},-\frac{1}{2},\frac{1}{2},\frac{3}{2})\protect \),
\protect\( (-\frac{5}{2},-\frac{1}{2},\frac{1}{2},\frac{5}{2})\protect \)and
\protect\( (-\frac{5}{2},-\frac{3}{2},\frac{3}{2},\frac{5}{2})\protect \) (shown
wit continuous lines).\label{four_holes.eps}}
\end{figure}

For \textbf{two solitons} case the two possibilities, corresponding to \( \delta =0,1 \)
must be checked. The simplest cases are: 

\begin{enumerate}
\item Two holes quantised with \( I_{1,2}=\pm \frac{1}{2} \), \( \delta =0 \). It
yields:
\[
\Delta ^{\pm }=\frac{p+1}{4p}=\frac{R^{2}}{2}=\Delta ^{\pm }_{0,2}\: ,\]
corresponding to the operator \( V_{(0,2)} \), which is the UV limit of the
lowest-lying two-soliton state. as it can also be seen from the TCS data. If
instead of the minimal choice \( I_{1,2} \) we take a nonminimal one \( I_{+}=3/2,\, 5/2,\, ... \),
\( I_{-}=-3/2,\, -5/2,\, ... \), we obtain
\[
\Delta ^{\pm }=\Delta _{0,2}^{\pm }\pm I_{\pm }-\frac{1}{2}\: ,\]
which corresponds to descendents of \( V_{(0,2)} \). This is a general phenomenon:
the ``minimal'' choice of quantum numbers yields the primary state, while
the nonminimal choices give rise to descendents.
\item Two holes quantized with \( I_{1,2}=\pm 1 \), \( \delta =1 \), as proposed
in \cite{ddv 97}. In the repulsive regime a special root is required, as in
the case of the \( \delta =1 \) vacuum state. The result (in both regimes)
is:
\[
\Delta ^{\pm }=1+\frac{1}{4}\frac{1}{p(p+1)}=\Delta ^{+}_{1/2,2}=\Delta ^{-}_{1/2,2}+1\, .\]
This state is a linear combination of \( \bar{a}_{-1}V_{(1/2,2)} \) and \( a_{-1}V_{(-1/2,2)} \)
which means that it is \emph{not} contained in the local operator algebras of
sG/mTh theories. 
\item Two holes quantized with \( I_{1}=0,\quad I_{2}=\pm 1 \), \( \delta =1 \).
Consider in detail the case of \( I_{2}=+1 \) since the other one is similar.
Suppose that the hole with \( I_{1}=0 \) is a left mover and the other one
is a right mover (the other possibilities lead to a contradiction). We find
a solution to the plateau equation only in the attractive regime:
\[
{\cal Q}_{+}(-\infty )=\pm 2\pi \frac{p}{p+1},\quad \: {\cal Q}_{-}(+\infty )=\pm 2\pi \frac{p}{p+1}\: .\]
In the repulsive regime the hole with quantum number \( 0 \) becomes a special
hole \( y \) and emits other two ordinary holes each quantised with zero. We
obtain for the plateau values
\[
{\cal Q}_{+}(-\infty )={\cal Q}_{-}(+\infty )=2\pi \frac{1}{p+1}\: .\]
The conformal weights turn out to be:
\[
\begin{array}{rl}
\displaystyle \Delta ^{+}= & 1+\displaystyle\frac{1}{4}\displaystyle\frac{1}{p(p+1)}=\Delta ^{+}_{1/2,2}\, ,\\
\Delta ^{-}= & \displaystyle\frac{1}{4}\displaystyle\frac{1}{p(p+1)}=\Delta ^{+}_{1/2,2}\, .
\end{array}\]
These are the conformal weights of the vertex operator \( V_{(1/2,2)} \). Performing
a similar calculation for \( I_{2}=-1 \) we obtain the weights of \( V_{(-1/2,2)} \). 
\end{enumerate}
The calculation for a \textbf{generic number of holes} proceeds in complete
analogy with the cases treated until now, we only give a summary of the results.
The conformal families which we obtain depend on how many holes move to the
left and to the right. The primaries are obtained for the minimal choice of
the hole quantum numbers; increasing the quantum numbers we obtain secondary
states, as it was pointed out in the case with two holes. The states we obtain
are in the conformal family of a vertex operator \( V_{n,m} \) with
\[
m=N_{H,eff}\: ,\: n\in {\mathbb Z}+\frac{\delta }{2}\: .\]
As a consequence, all states with \( \delta =0 \) are contained in the UV spectrum
of the sG/mTh theory while the ones with \( \delta =1 \) are not, in agreement
with the relation (\ref{regola_d'oro}). The complete expression for \( n \)
is somewhat complicated, but we give it in the case of symmetric (\( N^{+}_{H}=N^{-}_{H}=N_{H,eff}/2 \))
configurations:
\[
\begin{array}{ll}
n=0\: , & \delta =0\: ,\\
n=\pm \displaystyle\frac{1}{2}\: , & \delta =1\: .
\end{array}\]
The comparison with TCSA completely confirms the analysis made until now. The
example of four solitons states in repulsive regime is in figure \ref{four_holes.eps}.As usual in TCSA, one can see that the truncation errors become larger; at values
of the volume \( l \) close to \( 10 \) the deviation can be observed even
from the figures. For \( l<5 \) the agreement is still within an error of order
\( 10^{-3} \). More comments and examples can be found in \cite{noi PL1, noi PL2}.

\subsection{Holes and close roots\label{section:holes_close}}

Let us now extend our investigation to situations with complex roots and consider
the two particle states in more detail. Forgetting for the moment the breathers,
we have to consider the two-soliton states. The soliton-antisoliton come in
doublets so there are four different polarizations for two particle states of
which the \( ss \) and \( \bar{s}\bar{s} \) have topological charge \( Q=\pm 2 \)
instead \( s\bar{s} \) and \( \bar{s}s \) have zero topological charge. The
first two states are expected to have exactly the same scaling function for
energy and momentum because the sG/mTh theory is charge conjugation invariant.
Instead there are two different situations for the neutral \( s\bar{s} \) state,
which have spatially symmetric and antisymmetric wavefunctions (denoted by \( (s\bar{s})_{+} \)
and \( (s\bar{s})_{-} \), respectively). To separate the symmetric and antisymmetric
part one simply has to diagonalize the \( 4\times 4 \) SG two particle S-matrix
and see that it has 2 coinciding eigenvalues (equal to \( e^{i\chi (\vartheta )} \),
corresponding to \( ss \) and \( \bar{s}\bar{s} \)), and two different eigenvalues
in the \( Q=0 \) channel.

Now we proceed to demonstrate that the IR limit restricts the possible quantum
numbers of the complex roots. To simplify matters we consider only the repulsive
regime \( p>1 \).

In the repulsive regime \( p>1 \), the scaling function \( (s\bar{s})_{-} \)
is realized as the solution to the NLIE with two holes (at positions \( h_{1,2} \))
and a complex pair at the position \( \rho \pm i\sigma  \) . In the IR limit
we have

\noindent 
\begin{equation}
\label{quant._2h-2c}
\begin{array}{l}
\begin{array}{lcl}
Z(\vartheta ) & = & l\sinh \vartheta +\chi (\vartheta -h_{1})+\chi (\vartheta -h_{2})-\chi (\vartheta -\rho -i\sigma )-\chi (\vartheta -\rho +i\sigma ),\\
 & 
\end{array}\\
Z(h_{1,2})=2\pi I_{1,2}\; ,\\
Z(\rho \pm i\sigma )=2\pi I^{\pm }_{C}\; .
\end{array}
\end{equation}
The quantization condition for the complex roots explicitly reads (we write
down the equation only for the upper member of the complex pair, since the other
one is similar)

\begin{equation}
\label{complex_quant}
l\sinh (\rho +i\sigma )+\chi (\rho +i\sigma -h_{1})+\chi (\rho +i\sigma -h_{2})-\chi (2i\sigma )=2\pi I^{+}_{C\: .}
\end{equation}
Now observe that as \( l\: \rightarrow \: \infty  \), the first term on the
left hand side acquires a large imaginary part, but the right hand side is strictly
real. The imaginary contribution should be cancelled by some other term. The
function \( \chi  \) is bounded everywhere except for isolated logarithmic
singularities on the imaginary axis. For the cancellation of the imaginary part
the argument \( 2i\sigma  \) of the last term on the left hand side should
approach one of these singularities (similarly to the analysis in TBA \cite{dorey-tateo}).

In the repulsive regime, taking into account that for a close pair \( \sigma <\pi  \),
the only possible choice for \( \sigma  \) is to approach \( \displaystyle\frac{\pi }{2} \)
as \( l\: \rightarrow \: \infty  \). The soliton-soliton scattering amplitude
has a simple zero at \( \vartheta =i\pi  \) with a derivative which we denote
by \( C \) (the exact value does not matter). To leading order in \( l \),
the cancellation of the imaginary part reads
\begin{equation}
\label{cancell_immagin}
l\cosh \rho +\Re e\log C\left( \sigma -\displaystyle\frac{\pi }{2}\right) =0\: ,
\end{equation}
from which we deduce
\begin{equation}
\label{complex_as}
\left| \sigma -\displaystyle\frac{\pi }{2}\right| \sim \exp \left( -l\cosh \left( \rho \right) \right) \: ,
\end{equation}
so the imaginary part of the complex pair approaches its infrared limit exponentially
fast. This approach is modified by taking into account the finite imaginary
contributions coming from the source terms of the holes and from the convolution
term. These contributions lead to corrections of the order \( e^{-l} \) and
so they modify only the value of the constant \( C \). 

The behaviour of the real part near the singularity is:
\[
\Re e\, \chi (z-i\pi )\, \rightarrow \, 0\qquad \textrm{ if }\qquad \Re e\, z<0.\]
 Assume, for the moment, that \( \sigma  \) moves toward \( \frac{\pi }{2} \)
from below (i.e. \( \sigma -\frac{\pi }{2}<0 \)).

For the real part we get, again to the leading order
\begin{equation}
\label{cancell_reale}
\Re e\chi \left( \rho +i\displaystyle\frac{\pi }{2}-h_{1}\right) +\Re e\chi \left( \rho +i\displaystyle\frac{\pi }{2}-h_{2}\right) =2\pi I^{+}_{C\: .}
\end{equation}
It can be shown that (in the repulsive regime \( p>1 \))
\begin{equation}
\label{scattering_IR}
\xi \left( \vartheta \right) \equiv \Re e\chi \left( \vartheta +i\displaystyle\frac{\pi }{2}\right) =-\displaystyle\frac{i}{2}\log \displaystyle\frac{\sinh \displaystyle\frac{1}{p}\left( i\displaystyle\frac{\pi }{2}-\vartheta \right) }{\sinh \displaystyle\frac{1}{p}\left( i\displaystyle\frac{\pi }{2}+\vartheta \right) }\: ,
\end{equation}
where the the fundamental branch of the logarithm is taken into account. \( \xi  \)
is an odd monotonous function bounded by
\begin{equation}
\label{limite_su_csi}
\left| \xi (\vartheta )\right| \leq \left| \xi (\infty )\right| =\displaystyle\frac{\pi (p-1)}{2p}
\end{equation}
from below and above. This means that for any allowed value of \( I^{\pm }_{C} \)
the real position of the complex pair is determined uniquely and that
\begin{equation}
\label{limite_IR_quantum}
\left| I^{\pm }_{C}\right| <\left| \displaystyle\frac{p-1}{2p}\right| \: ,
\end{equation}
and since in the repulsive regime \( p>1 \), the only possible choice is \( I^{\pm }_{C}=0 \).
Then the solution for \( \rho  \) is
\begin{equation}
\label{posizione_IR_holes}
\rho =\displaystyle\frac{h_{1}+h_{2}}{2}\: ,
\end{equation}
so it approaches the central position between the two holes (the corrections
to this asymptotic are also exponentially small for large \( l \)). In fact,
for the symmetric hole configuration \( I_{1}=-I_{2} \) we expect \( h_{1}=-h_{2} \)
and \( \rho =0 \) to be valid even for finite values \( l \). 

\begin{figure}[  htbp]
{\centering \includegraphics{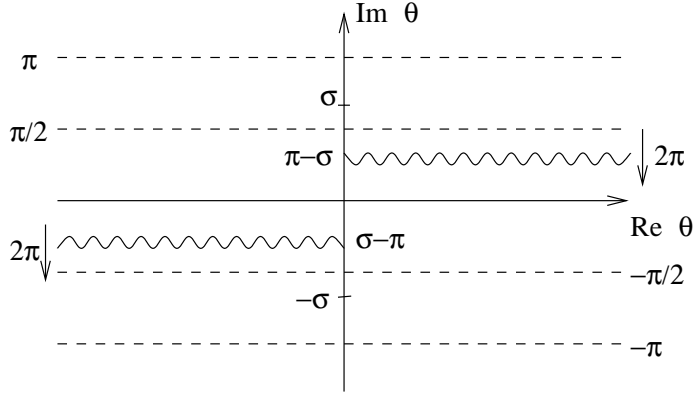} \par}

\caption{\small \label{cuts_figure}The analytic structure of the counting function if \protect\( \sigma >\frac{\pi }{2}\protect \).
The logarithmic cuts are indicated with the wiggly lines. The one lying in the
upper half plane originates from the root in the lower half plane and vice versa.
We also indicated the value of the discontinuity across the cuts.}
\end{figure}

However, the above derivation is valid only if \( \sigma <\displaystyle\frac{\pi }{2} \)
and so we do not cross the boundary of the analyticity strip of the \( \chi (\vartheta ) \)
function which is at \( \Im m\, \vartheta =\pi  \). If \( \sigma >\displaystyle\frac{\pi }{2} \),
we must use a consider that:
\[
\Re e\, \chi (z-i\pi )\, \rightarrow \, \pi \qquad \textrm{ if }\qquad \Re e\, z>0.\]
The conclusion that \( \sigma  \) approaches \( \frac{\pi }{2} \) exponentially
fast as \( l\, \rightarrow \, \infty  \) is unaffected by this change, but
now the real part of \( \chi (2i\sigma ) \) is not zero but instead \( \pm \pi  \)
(modulo \( 2\pi  \)). We choose the branches of the logarithm in such a way
that \( \Re e\, \chi (\pm 2i\sigma )=\mp \pi  \) (the structure of the cuts
is displayed in figure \ref{cuts_figure}). In this way we obtain that
\[
I^{\pm }_{C}=\mp \frac{1}{2}\, \, ,\]
and the asymptotic value of \( \rho  \) is just as before.

The S-matrix of the \( (s\overline{s})_{-} \) configuration can be obtained
by substituting the asymptotic values 
\begin{equation}
\label{scattering_IR_2}
\rho =\displaystyle\frac{h_{1}+h_{2}}{2}\: ,\: \sigma =i\displaystyle\frac{\pi }{2}
\end{equation}
into the expression for \( Z(\vartheta ) \) and considering now the quantization
rules of the holes. We obtain 
\begin{equation}
\label{IRBA_minus}
Z(h_{1})=l\sinh \left( h_{1}\right) -i\log S_{-}(h_{1}-h_{2})=2\pi I_{1}\: ,
\end{equation}
and a similar equation for the second hole, with
\begin{equation}
\label{scattering-}
S_{-}(\vartheta )=-\displaystyle\frac{\sinh \left( \displaystyle\frac{\vartheta +i\pi }{2p}\right) }{\sinh \left( \displaystyle\frac{\vartheta -i\pi }{2p}\right) }S^{++}_{++}(\vartheta )\: ,
\end{equation}
which is the correct scattering amplitude for the antisymmetric configuration
of the soliton-antisoliton system. The amplitude has poles at \( \vartheta =i\pi (1-2kp)\, ,\, k=1,2,\ldots  \)
, corresponding to breathers with mass 
\[
m_{2k}=2{\cal M}\sin kp\pi \: .\]
The equation (\ref{IRBA_minus}) describes an approximation to the full NLIE
valid for large \( l \) which is called the dressed or asymptotic Bethe Ansatz.
There are three levels of approximations to the scaling functions: the full
NLIE (which is in fact exact), the higher level Bethe Ansatz (HLBA) obtained
from the NLIE by dropping the convolution term and the asymptotic BA. The difference
between the BA and the HLBA is that while the former keeps the complex pair
in its asymptotic position, the latter takes into account the corrections coming
from the fact that \( \sigma  \) changes with \( l \) (to first order as given
in (\ref{complex_as}) ). However, since the convolution term is of the same
order in \( l \) as the dependence of \( \sigma  \) derived from the HLBA,
the HLBA is not a self-consistent approximation scheme. Therefore we are left
with the exact NLIE, which is valid for all scales and with the BA as its IR
asymptotic form.

The above conclusions about the \( (s\bar{s})_{-} \) state an be extended into
the attractive regime as long as \( p>\frac{1}{2} \). At the point \( p=\frac{1}{2} \),
however, the pair situated at the asymptotic position \( \pm i\frac{\pi }{2} \)
hits the boundary of the fundamental analyticity strip situated at \( i\pi p \).
This phenomenon has a physical origin: this point is the threshold for the second
breather bound state which is the first pole entering the physical strip in
the \( (s\bar{s})_{-} \) channel. To continue our state further, it requires
to go to configurations having an array of complex roots of the first type (see
\cite{ddv 97}). Without going into further details, let us mention only that
any such array contains a close pair plus some wide pairs depending on the value
of \( p \).

The symmetric state \( (s\bar{s})_{+} \) can be obtained from a configuration
with two holes and one selfconjugate wide root. Considerations similar to the
above lead to the scattering amplitude 
\begin{equation}
\label{scattering+}
S_{+}(\vartheta )=\displaystyle\frac{\cosh \left( \displaystyle\frac{\vartheta +i\pi }{2p}\right) }{\cosh \left( \displaystyle\frac{\vartheta -i\pi }{2p}\right) }S^{++}_{++}(\theta )\: ,
\end{equation}
which agrees with the prediction from the exact S-matrix. The poles are now
at \( \vartheta =i\pi (1-(2k+1)p)\, ,\, k=0,1,\ldots  \) and correspond to
breathers with mass
\[
m_{2k+1}=2{\cal M}\sin \frac{(2k+1)p\pi }{2}\: .\]
This configuration extends down to \( p=1 \), where the selfconjugate root
collides with the boundary of the fundamental analyticity strip at \( i\pi  \).
This is exactly the threshold for the first breather which is the lowest bound
state in the \( (s\bar{s})_{+} \) channel. For \( p<1 \) the \( (s\bar{s})_{+} \)
state contains a degenerate array of the first kind (see \cite{ddv 97}), which
consists of a close pair, a selfconjugate root and some wide pairs, again depending
on the value of \( p \).

We remark that it is easy to extend the above considerations to a state with
four holes and a complex pair. In this case the only essential modification
to the above conclusions is that the limit for the complex pair quantum number
becomes
\begin{equation}
\label{scattering_IR_4h}
\left| I^{\pm }_{c}\right| <\left| \displaystyle\frac{p-1}{p}\right| 
\end{equation}
when \( \sigma <\displaystyle\frac{\pi }{2} \), and
\begin{equation}
\label{scatt_IR_4}
\left| I^{\pm }_{c}\pm \displaystyle\frac{1}{2}\right| <\left| \displaystyle\frac{p-1}{p}\right| \: ,
\end{equation}
when \( \sigma >\displaystyle\frac{\pi }{2} \), since now there are four \( \chi  \) sources
coming from the holes. In the range \( 1<p<2 \) this gives the same constraints
\[
I^{\pm }_{c}=\left\{ \begin{array}{ll}
0\, , & \, \, \sigma <\displaystyle\frac{\pi }{2}\, \, ,\\
\mp \displaystyle\frac{1}{2}\, , & \, \, \sigma >\displaystyle\frac{\pi }{2}
\end{array}\right. \, \, ,\]
as for the \( (s\bar{s})_{-} \) state (for \( p>2 \) it allows for more solutions,
which however we will not need in the sequel). 

From the UV point of view, the results shown in this paragraph about the quantum
numbers of the close roots are quite interesting. Consider the conformal dimensions
(\ref{delta}) or the specific form (\ref{delta_sG}). There is no proof that
the terms \( N_{\pm } \) give always the correct descendent numbers. In particular
there is no proof that the sum of quantum numbers \( \left( I_{H}^{\pm }-2I^{\pm }_{S}-I^{\pm }_{C}-I^{\pm }_{W}\right)  \)
is bounded from below (and nonnegative). Then we don't know if the so called
``vacuum'' is really the hamiltonian ground state. The IR analysis done until
now shows exactly that states with only holes and with holes and close roots
have the correct hamiltonian behaviour. 

We now turn ourselves to the UV limits of the previous states. In particular
we consider the state of two holes and a close pair (in the repulsive regime),
which describes the antisymmetric soliton-antisoliton two-particle state. According
to the results from the IR asymptotic, the quantum numbers of the complex roots
can be \( 0 \) or \( \pm \frac{1}{2} \). Therefore, we are left two possibilities:
either the holes are quantised with integers or with half-integers. 

Let us start with the half-integer choice \( \delta =0 \) and suppose that
one of the holes is right moving (which is the case if its quantum number \( I_{1}=I_{+} \)
is positive) and the other one is left moving (i.e. \( I_{2}=I_{-}<0 \)). Using
the general formula (\ref{Q+-infinito}) and \( S^{\pm }=\frac{1}{2} \), \( S=0 \),
by a simple calculation we obtain
\[
{\cal Q}_{+}(-\infty )=Z_{+}(-\infty )=-2\pi \frac{p-1}{p+1}\: ,\]
which is valid for \( 1<p<4 \) (the complex pair remains central). The other
plateau values follow by oddity of the function \( Z \). For the other half
of the repulsive regime we have to include two new normal holes, one moving
to the left and the other to the right, and two special holes that remain central
(a justification for this will be given shortly) so that \( S^{\pm }=1 \).
We then find
\[
{\cal Q}_{+}(-\infty )=4\pi \frac{1}{p+1}\: ,\: Z_{+}(-\infty )=-2\pi \frac{p-1}{p+1}\: ,\]
which is valid for \( p\geq 4 \). In both cases the result for the UV weights
turns out to be
\[
\Delta ^{\pm }=\frac{p}{p+1}\pm I_{\pm }-\frac{1}{2}=\frac{1}{2R^{2}}\pm I_{\pm }-\frac{1}{2}=\Delta _{(\pm 1,0)}\pm I_{\pm }-\frac{1}{2}\, .\]
The primary state is obtained for the minimal choice \( I_{+}=-I_{-}=\frac{1}{2} \)
and coincides with a linear combination of the vertex operators \( V_{(\pm 1,0)} \)
which is correct for the state to be included in the spectrum of the sG/mTh
theory and agrees with the behaviour of the \( (s\bar{s})_{_{-}} \) state observed
from TCS. If both of the holes move to the right or to the left, we obtain descendents
of the identity operator, i.e. states in the vacuum module of the UV CFT.

\begin{figure}[  htbp]
{\centering \begin{tabular}{ccc}
\includegraphics{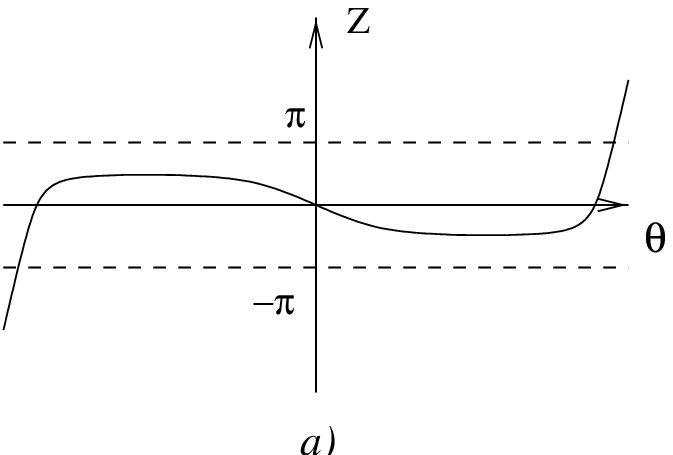} &
&
\includegraphics{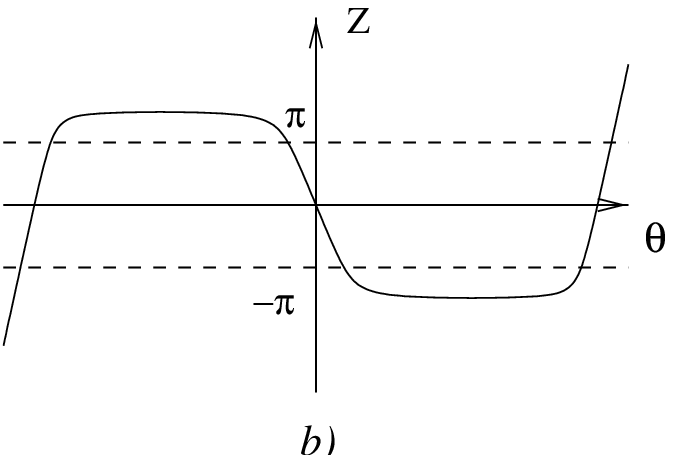} \\
\end{tabular}\par}

\caption{\small \label{twoholescp_specials}The UV behaviour of \protect\( Z\protect \) for
a) \protect\( p<4\protect \) and b) \protect\( p>4\protect \).}
\end{figure}

Figure \ref{twoholescp_specials} shows how the special holes are generated
in this case. Starting from \( p<4 \) and increasing the value of \( p \)
the UV asymptotic form of the counting function varies analytically. Since the
real roots/holes are quantised by half-integers, they are at positions where
the function \( Z \) crosses a value of an odd multiple of \( \pi  \). As
the plots demonstrate, the behaviour of \( Z \) is in fact continuous at the
boundary \( p=4 \): it is our interpretation in terms of the sources that changes,
exactly because we try to keep the logarithm in the integral term of the NLIE
in its fundamental branch. The price we pay is that we have to introduce two
new normal holes (one moves left and the other one moves right) and two special
holes which are central.Let us comment on the extension to the attractive regime. The complex root configuration
changes only by the possible presence of wide pairs, which however have no effect
on the plateau values (\ref{Q+-infinito}) as their contribution can be absorbed
in redefinition of the values \( k_{\pm } \). The other effect the wide roots
have is to shift the terms \( \Sigma _{\pm } \) by multiples of \( 2\pi  \),
but this affects only the integers \( N_{\pm } \) in (\ref{interiN+-}). Therefore
we again obtain states which are descendents of \( V_{(\pm 1,0)} \).

For the state with integer quantization of the holes, we only give the result.
If one of the holes is moving left and the other one to the right, we obtain
\[
\Delta ^{\pm }=\frac{1}{4}\left( \frac{p}{p+1}\right) \pm I_{\pm }=\frac{1}{8R^{2}}\pm I_{\pm }=\Delta _{(\pm 1/2,0)}\pm I_{\pm },\]
i.e. some linear combinations of descendents of the vertex operators \( V_{(\pm 1/2,0)} \).
If both of the holes move to the left or to the right, we obtain other descendants
of the same primary states. Again, this excludes the integer quantised states
from the spectrum of sG/mTh theory.

\subsubsection{Two holes and a selfconjugate complex root}

Taking the integer quantised state, the plateau values are given by the same
formulas as for the case with the close pair above, since the plateau equation
is identical. The only difference is that the numbers \( l^{\pm }_{W} \) (\ref{lW+-})
take a nonzero value
\[
l^{\pm }_{W}=-S^{\pm }.\]
If the one of the holes is a right mover, while the other one is a left mover,
the conformal weights turn out to be
\[
\Delta ^{\pm }=\frac{p}{p+1}\pm I_{\pm }-1=\Delta _{(\pm 1,0)}\pm I_{\pm }-1.\]
If the two holes move in the same direction, we again obtain secondaries of
the vacuum state. 

Concerning the extension to the attractive regime, one can again check that
the plateau solution remains unchanged for the root configuration of two holes
and two close roots, therefore the conformal family to which the state belongs
remains the same, similarly to the case of the \( (s\bar{s})_{-} \) state.

One can ask what happens if we quantize with half-integers. The result is, similarly
to the case with the complex pair, that we obtain descendants of the operators
\( V_{(\pm 1/2,0)} \). Therefore we conclude that in this case the integer
quantised configuration must be accepted, while the half-integer one is ruled
out, in full accordance with the rule (\ref{regola_d'oro}).

\subsection{Breather \protect\( S\protect \)-matrices and IR limit\label{section:IR_breather}}

In the infrared limit \( l\, \rightarrow \, \infty  \) the term \( l\sinh (\vartheta ) \)
develops a large imaginary part in the first determination away from the real
axis, forcing the close complex roots to fall into special configurations called
\emph{arrays} (we use the terminology of \cite{ddv 97})\emph{.} An array is
a set of complex roots in which the roots are placed at specific intervals in
the imaginary direction and have the same real part. In the attractive regime
\( l\sinh (\vartheta )_{II} \) is nonzero and so this is true for nonselfconjugate
wide pairs as well (in the repulsive case wide roots do not have such driving
force), while self-conjugate roots have a fixed imaginary part anyway. The deviation
of the complex roots from their positions in the array decays exponentially
with \( l \) (section \ref{section:holes_close}).

For the rest of this subsection, whenever it is not explicitly stated, we restrict
ourselves to the attractive regime \( p<1 \). The possible arrays fall into
two classes:

\begin{enumerate}
\item \emph{Arrays of the first kind} are the ones containing close roots, which describe
the polarization states of solitons. \\
There are two degenerate cases: \emph{odd degenerate} arrays, which have a self-conjugate
root at
\[
\vartheta _{0}=\vartheta +i\frac{\pi (p+1)}{2}\]
and accompanying complex pairs at
\[
\vartheta _{k}=\vartheta \pm i\frac{\pi (1-(2k+1)p)}{2}\, \, ,\, \, k=0,\ldots ,\left[ \frac{1}{2p}\right] \]
 and \emph{even degenerate} ones, which only contain complex pairs, at the positions
\[
\vartheta _{k}=\vartheta \pm i\frac{\pi (1-2kp)}{2}\, \, ,\, \, k=0,\ldots ,\left[ \frac{1}{2p}\right] \]
These arrays always contain exactly one close pair. The odd degenerate arrays
in the repulsive regime reduce to single self-conjugate roots and the even degenerate
ones to a single close complex pair.  
\item \emph{Arrays of the second kind} describe breather degrees of freedom. The odd
ones contain a self-conjugate root 
\[
\vartheta _{0}=\vartheta +i\pi (p+1)/2\]
and wide pairs as follows:
\[
\Im m\vartheta _{k}=\vartheta \pm i\frac{\pi (1-(2k+1)p)}{2}\, \, ,\, \, k=0,\ldots ,s\, \, ,\]
where
\[
0\leq s\leq \left[ \frac{1}{2p}\right] -1\, \, ,\]
while the even ones only contain wide pairs
\[
\Im m\vartheta _{k}=\vartheta \pm i\frac{\pi (1-2kp)}{2}\, \, ,\, \, k=0,\ldots ,s\, \, ,\]
and \( s \) runs in the same range. They correspond to the \( (2s+1) \)-th
breather \( B_{2s+1} \) and the \( (2s+2) \)-th breather \( B_{2s+2} \),
respectively.
\end{enumerate}
As one can see, arrays of the second kind become degenerate ones of the first
kind, if we analytically continue increasing \( p \). The reason is that breathers
are of course soliton-antisoliton bound states, while degenerate arrays of the
first kind describe scattering states of a soliton and antisoliton, as we will
see shortly.

One can compute the energy and momentum contribution of a array of the second
kind corresponding to the breather \( B_{s} \). The energy-momentum contribution
turns out to be
\begin{equation}
\label{breather-energy}
2{\cal M}\sin \displaystyle\frac{\pi sp}{2}\left( \cosh \vartheta ,\sinh \vartheta \right) \, \, ,
\end{equation}
where \( \vartheta  \) is the common real part of the roots composing the array.
This is just the contribution of a breather \( B_{s} \) moving with rapidity
\( \vartheta  \). Arrays of the first kind do not contribute to the energy-momentum
in the infrared limit, which lends support to their interpretation as polarization
states of solitons.

Now we proceed to show that with the above interpretation the NLIE correctly
reproduces the two-body scattering matrices of sine-Gordon theory including
breathers.

Let us start with breather-soliton matrices. The Bethe quantization conditions
for a state containing a soliton (i.e. a hole) with rapidity \( \vartheta _{1} \)
and a breather \( B_{s} \) with rapidity \( \vartheta _{2} \) take the following
form in the infrared limit. For the hole we get
\[
Z(\vartheta _{1})={\cal M}\sinh \vartheta _{1}-\sum ^{s}_{k=0}\chi _{II}(\vartheta _{1}-\vartheta _{2}-i\rho _{k})=2\pi I_{1},\]
where we denoted the prescribed imaginary parts of the roots in the array \( B_{s} \)
by \( \rho _{k} \). Here we used \( \chi (0)=0 \) to eliminate the source
term for the hole. Now we can compute the second determination of \( \chi  \)
as in (\ref{chi_sec_esplic}). Now it is a matter of elementary algebra to arrive
at
\[
Z(\vartheta _{1})={\cal M}\sinh \vartheta _{1}-i\log S_{SB_{s}}(\vartheta _{1}-\vartheta _{2})=2\pi I_{1}\, \, ,\]
where \( S_{SB_{s}}(\vartheta _{1}-\vartheta _{2}) \) is the soliton-breather
\( S \)-matrix conjectured in \cite{zam79}.

One can start with the breather quantization conditions, too. Writing 
\begin{equation}
\label{quant_br}
Z\left( \vartheta _{2}+i\rho _{k}\right) ={\cal M}\sinh \left( \vartheta _{2}+i\rho _{k}\right) _{II}+\chi _{II}(\vartheta _{2}-\vartheta _{1}+i\rho _{k})+\ldots =2\pi I^{(k)}_{2}\, \, ,k=0,\ldots ,s
\end{equation}
(the dots are terms due to wide root sources themselves, which cancel out up
to multiples of \( 2\pi  \) in the next step). Summing up these equations one
arrives at
\[
2{\cal M}\sin \frac{sp\pi }{2}\sinh (\vartheta _{2})-i\log S_{SB_{s}}(\vartheta _{2}-\vartheta _{1})=2\pi I_{2}\, \, ,\]
where \( I_{2} \) is essentially minus the sum of the quantum numbers of the
wide roots composing the array (shifted by some integer coming from summing
up the terms omitted in equation (\ref{quant_br})).

Using a similar line of argument we also reproduced the breather-breather \( S \)-matrices
by writing down the Bethe quantization conditions for a state with two degenerate
strings \( B_{s} \) and \( B_{r} \) of the second kind. One has to be careful
that when \( Z(\vartheta ) \) contains wide root sources which are expressed
in terms of \( \chi _{II}(\vartheta ) \), the second determinations of these
terms will appear in \( Z(\vartheta ) \)for large \( \Im m\, \vartheta  \),
i.e. terms that can be written roughly like \( \left( \chi _{II}(\vartheta )\right) _{II} \),
as in (\ref{seconda_doppia_det}).

Scattering state of a soliton and an antisoliton can be described by taking
two holes and a degenerate array of the first kind. There are two possibilities
now, corresponding to scattering in the parity-odd and parity-even channels.
Following the procedure outlined in section \ref{section:holes_close}, we were
once again able to reproduce the corresponding scattering amplitudes. The results
presented here together with those of section \ref{section:holes_close} exhaust
all two-particle scattering amplitudes of sine-Gordon theory, both in the repulsive
and attractive regime.

\subsection{Some examples of breather states}

So we proceed to take a look at the simplest neutral excited state, which is
the one containing a first breather \( B_{1} \) at rest. The source term to
be written into the NLIE turns out to be 
\begin{equation}
\label{selfconjugate_source}
g(\vartheta -w_{R})=-\chi _{II}(\vartheta -w)=-i\log \displaystyle\frac{\cos \displaystyle\frac{\pi p}{2}-i\sinh (\vartheta -w_{R})}{\cos \displaystyle\frac{\pi p}{2}+i\sinh (\vartheta -w_{R})}\, \, ,
\end{equation}
where \( w_{R} \) is the real part of the position of the self-conjugate root
(the imaginary part is \( w_{I}=\pi (p+1)/2 \). In this case \( w_{R}=0 \)
since the breather has zero momentum. There is no need to look at the Bethe
quantization condition as the root does not move due to the left-right symmetry
of the problem. This state is quantized with integer Bethe quantum numbers,
i.e. \( \delta =1 \), as in (\ref{regola_d'oro}).

Table \ref{1st_breather} presents the energy values obtained by iterating the
NLIE in comparison to results coming from TCS, at the values \( p=2/7 \) and
\( p=2/9 \).

\begin{table}[  htbp]
{\centering \begin{tabular}{|c|c|c|c|c|}
\hline 
\( l \)&
TCS&
NLIE&
TCS&
NLIE\\
\hline 
1.0&
2.38459&
n/a&
1.84996&
n/a\\
\hline 
1.5&
1.68168&
n/a&
1.30438&
n/a\\
\hline 
2.0&
1.35391&
n/a&
1.05038&
n/a\\
\hline 
2.5&
1.17420&
n/a&
0.91152&
n/a\\
\hline 
3.0&
1.06692&
1.0655810539&
0.82903&
0.8285879853\\
\hline 
3.5&
0.99985&
0.9980153379&
0.77773&
0.7771857432\\
\hline 
4.0&
0.95664&
0.9542867454&
0.74499&
0.7443106400\\
\hline 
4.5&
0.92845&
0.9254766107&
0.72381&
0.7229895177\\
\hline 
5.0&
0.91000&
0.9063029141&
0.71006&
0.7090838262\\
\hline 
\end{tabular}\par}

\caption{\small \label{1st_breather}The first breather state at \protect\( p=\displaystyle\frac{2}{7}\protect \)
and \protect\( p=\displaystyle\frac{2}{9}\protect \). Energies and distances are measured
in units of the soliton mass \protect\( {\cal M}\protect \), and we have subtracted
the predicted bulk energy term from the TCS data.}
\end{table} The table shows that iteration of the NLIE fails for values of \( l \) less
than \( 3 \) (the actual limiting value is around \( 2.5 \)). What is the
reason?

\begin{figure}[  htbp]
{\centering \begin{tabular}{ccc}
\resizebox*{0.4\columnwidth}{!}{\includegraphics{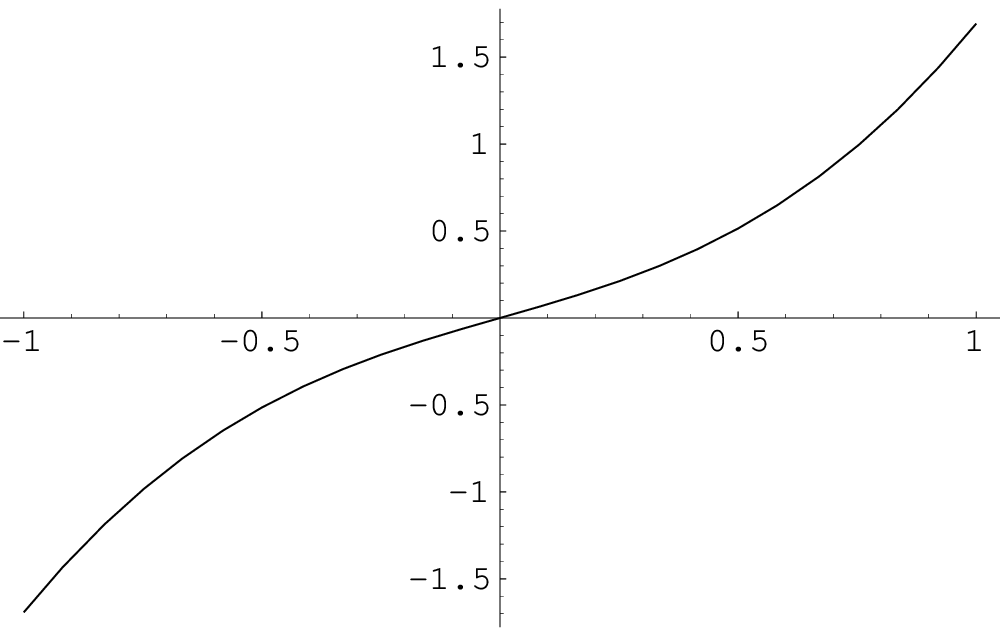}} &
&
\resizebox*{0.4\columnwidth}{!}{\includegraphics{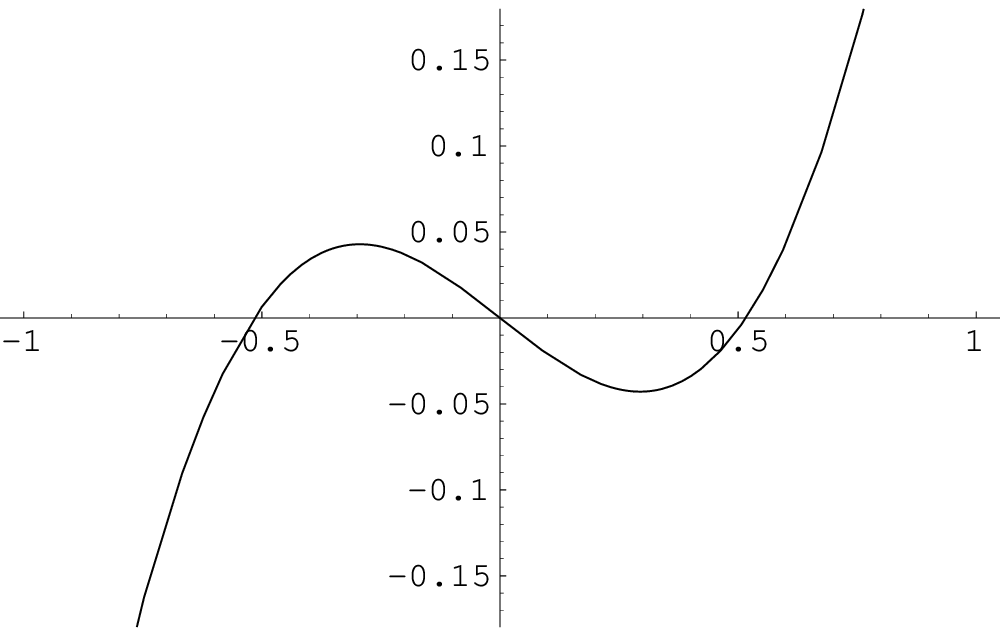}} \\
\end{tabular}\par}

\caption{\small \label{specials} The behavior of the function \protect\( Z(\vartheta )\protect \)
at \protect\( l=3\protect \) and \protect\( l=2\protect \), respectively.}
\end{figure}

We plot the counting function \( Z \) on the real line for \( l=3 \) and \( l=2 \)
in figure \ref{specials}. In a first approximation we can safely neglect the
integral term for these values of the volume to see the qualitative features
that we are interested in. What we see is that the behaviour of the function changes: its derivative changes
sign at the origin. As a result, two new holes appear where the new real zeros
of the function are. But the topological charge remains zero, due to the fact
that now we have a special root at the origin and so \( N_{S}=1 \) and \( N_{H}=2 \).
The two new holes do not give us any new dynamical degrees of freedom: their
quantum numbers are fixed to be \( 0 \) and so their positions are uniquely
determined.

Of course when we calculated the UV dimension, the appearance of the new sources
had to be taken into account. It turns out that for \( \displaystyle\frac{1}{3}<p<1 \) the
two holes are left/right movers, while for \( p<\displaystyle\frac{1}{3} \) they remain
central. Calculating the UV conformal dimensions we obtain
\begin{equation}
\label{1stbreather_UV}
\Delta ^{\pm }=\displaystyle\frac{p}{p+1}\, \, ,
\end{equation}
so the ultraviolet limit of this state is a linear combination of the vertex
operators \( V_{\pm 1,0} \) of the \( c=1 \) UV CFT. This is in perfect agreement
with the TCS calculations performed by us as in table \ref{1st_breather}.

How does this change of sign in the derivative of \( Z \) affect the iteration
scheme for the NLIE (see section \ref{section:intermedio})? The two new zeros
of \( Z(\vartheta ) \), which is a complex analytic function apart from logarithmic
branch cuts, actually correspond to singularities of the logarithmic term in
the NLIE (\ref{nlie-cont}). They come along the imaginary axis in the \( \vartheta  \)
plane as we decrease \( l \), and at a certain point they cross our integration
contour which runs parallel to the real axis at distance \( \eta  \). As they
make the logarithmic term in our NLIE (\ref{nlie-cont}) singular, they blow
up our iteration scheme. After reaching the origin of the \( \vartheta  \)
plane (at exactly the radius where the derivative of \( Z \) becomes \( 0 \)),
they continue to move along the real axis, which corresponds to crossing a square
root branch cut.

This problem of numerically solve NLIE appears in all the cases where special
roots/holes are taken into account (e.g. the ``-`` vacuum, the two holes-close
quantized with ``-`` and so on). 

We do not go into details here as this problem is currently under investigation\footnote{
Work in progress in collaboration with P. E. Dorey and C. Dunning, Durham.
}. We just remark that these issues prove to be highly nontrivial and for the
time being, unfortunately, they prevent us from having a reliable numeric scheme
for the NLIE below the critical volume.

One can estimate the volume where the slope of the counting function changes
sign by neglecting the integral of the logarithmic term. The result is
\begin{equation}
\label{critical_estimate}
l_{critical}=\displaystyle\frac{2}{\cos \displaystyle\frac{p\pi }{2}}
\end{equation}
which gives a value of around \( 2.22 \) for \( p=2/7 \) and \( 2.13 \) for
\( p=2/9 \). The actual limiting value is a bit higher, partly due to the finite
value of \( \eta  \) used in the iteration program and partly because the iteration
already destabilizes when the singularities come close enough to the contours.
It must also be noted that the integral term cannot eventually be neglected
when the singularities are close to the contour, which is an additional reason
why (\ref{critical_estimate}) is just a crude estimate.

We make a short digression to examine the UV limit of the second breather \( B_{2} \).
The second breather at rest is described by a wide pair at positions 
\[
\vartheta =\pm i\frac{\pi }{2}\, \, .\]
Calculating the UV conformal dimension we get 
\[
\Delta ^{+}=\Delta ^{-}=\frac{p}{p+1}\, \, ,\]
which turns out to be the same as that of the first one (\ref{1stbreather_UV}),
i.e. this state must originate from the other linearly independent combination
of the vertex operators \( V_{\pm 1,0} \) in the ultraviolet. This is again
in perfect agreement with TCS and confirms a result by Pallua and Prester \cite{pallua}
who used \( XXZ \) chain in transverse magnetic field to regularize sine-Gordon
theory. They calculated scaling functions numerically on a finite lattice for
several concrete values of \( p \), and arrived at this conclusion by looking
at the numerical data. However, our method to compute UV dimensions gives us
an \emph{exact analytic formula} and therefore much stronger evidence. This
result is interesting because it invalidates a conjecture made previously by
Klassen and Melzer \cite{kl-me} who identified the second breather as a linear
combination of \( V_{\pm 2,0} \). 

To close this section, we present the lowest lying example of a two-breather
state, containing two \( B_{1} \) particles with zero total momentum. It turns
out that this is a state for which the numerical iteration of the NLIE is not
plagued with the problem found above for the first breather. Locality constrains
the state to be quantized with half-integers and for lowest energy the quantum
numbers of the self-conjugate roots must take the values
\[
I_{1}=\frac{1}{2}\, \, ,\, \, I_{2}=-\frac{1}{2}\]
We remark that in contrast to the case of holes, the self-conjugate root with
\( I>0 \) moves to the left, while the one with \( I<0 \) moves to the right.
This is due to the fact that the second determination of \( Z \) is in general
a monotonically decreasing function on the self-conjugate line. In order to
determine the position of the two self-conjugate roots we need the second determination
of \( Z \). The second determination of the self-conjugate root source turns
out to be
\[
\left( \chi _{II}(\vartheta )\right) _{II}=i\log \frac{i\sin \pi p-\sinh \vartheta }{i\sin \pi p+\sinh \vartheta }\, \, .\]
Up to some signs, this is just the phase shift which arises when two breathers
scatter on each other, which is exactly why the IR analysis gives the correct
scattering amplitude. Using this formula, we obtained the numerical data presented
in table \ref{two_breathers}.\begin{table}[  htbp]
{\centering \begin{tabular}{|c|c|c|}
\hline 
\( l \)&
TCS&
NLIE\\
\hline 
1.0&
12.1601&
12.159257\\
\hline 
1.5&
8.20139&
8.2006130\\
\hline 
2.0&
6.24771&
6.2465898\\
\hline 
2.5&
5.09489&
5.0937037\\
\hline 
3.0&
4.34132&
4.3397021\\
\hline 
3.5&
3.81513&
3.8129798\\
\hline 
4.0&
3.43020&
3.4275967\\
\hline 
4.5&
3.13912&
3.1357441\\
\hline 
5.0&
2.91308&
2.9089439\\
\hline 
\end{tabular}\par}

\caption{\small \label{two_breathers} The two-breather state at \protect\( p=\frac{2}{7}\protect \).}
\end{table} The UV limit for this state can be calculated from NLIE to be a symmetric first
level descendent of the vacuum with weights
\[
\Delta ^{+}=\Delta ^{-}=1\, \, ,\]
which agrees with TCS. (Note that this descendent exists due to the fact that
there is a \( \hat{U}(1)_{L}\times \hat{U}(1)_{R} \) Kac-Moody symmetry at
\( c=1 \): this state exactly corresponds to the combination of the left and
right moving currents \( J\bar{J} \).)

\section{\protect\( \alpha \protect \) twist and minimal models (the ground state)\label{section:alfa_gs}}

It is known from \cite{karowsky} that by twisting the Bethe Ansatz Equations
of the six vertex model as indicated in section \ref{section:6_vertex_BA},
the twisted model shows the critical behaviour of conformal minimal models (the
untwisted critical behaviour corresponds to \( c=1 \)). Moreover, in recent
papers, Al. Zamolodchikov has put forward the idea of modifying sine-Gordon
theory by a twist \( \alpha  \) \cite{polymer} to deal with conformal minimal
models. As a consequence of this two ideas, the Bethe equations used to obtain
NLIE contained the twist (\ref{bethe}) \( \omega  \). In the NLIE, as shown
in section \ref{section:NLIE2}, the twist is parametrized by \( \alpha  \)
(\ref{alfa}). 

Looking at the ground state, that in analogy with sine-Gordon is expected to
be a sea of real roots, the source in NLIE is put to zero \( g(\vartheta )=0 \)
(\ref{g_sorgenti}). 

In analogy with the sine-Gordon ground state, we can choose half-integer quantization
rule with \( \delta =0 \). The expression for conformal dimensions (\ref{delta})
gives:
\[
\Delta ^{\pm }=\frac{c-1}{24}+\frac{1}{4}\frac{p}{p+1}\left( \frac{\alpha }{\pi }\right) ^{2}\]
corresponding to an effective central charge
\begin{equation}
\label{alpha_vacuum}
\tilde{c}=1-\displaystyle\frac{6p}{p+1}\left( \displaystyle\frac{\alpha }{\pi }\right) ^{2}
\end{equation}
(the effective central charge is defined in (\ref{andamento-uv})). Only in
the unitary models the \( \tilde{c} \) is the Virasoro central charge. Furthermore,
it is well-known that the perturbation of the Virasoro minimal model \( Vir(r,s) \)
by its relevant primary operator \( \Phi _{(1,3)} \) is integrable and is described
by an RSOS restriction of sine-Gordon theory \cite{reshetikhin_smirnov} with
\begin{equation}
\label{p_min_r_s}
p=\displaystyle\frac{r}{s-r}\, \, .
\end{equation}
We will use for this model the shorthand notation \( Vir(r,s)+\Phi _{(1,3)} \).
Using the rule suggested in (\ref{omega}) for \( \omega  \) and the expression
(\ref{alfa}) for \( \alpha  \) gives: \( \alpha =\pi /r \) (the additional
integer terms are excluded, for the moment). Then
\[
\tilde{c}=1-\frac{6}{rs}\, \, ,\]
which is exactly the \emph{effective central charge} of the minimal model \( Vir(r,s) \).
Therefore one can expect that the twisted equation describes the ground state
of the model \( Vir(r,s)+\Phi _{(1,3)} \). In fact, Fioravanti et al. \cite{fioravanti}
calculated these scaling functions for the unitary case \( s=r+1 \) and showed
that they match perfectly with the TBA predictions already available. Moreover,
choosing the following values for the twist
\begin{equation}
\label{twists}
\alpha =\pm \displaystyle\frac{k\pi }{r}\, \, ,\, \, k=1\ldots r-1\, \, 
\end{equation}
they obtained the conformal weights of the operators \( \Phi _{(k,k)}\, \, ,\, \, k=1\ldots r-1 \)
in the UV limit (the sign choice is just a matter of convention). In our notation,
\( \Phi _{(q,q')} \) denotes the primary field with conformal weights
\begin{equation}
\label{Kac_formula}
h^{+}=h^{-}=\displaystyle\frac{(qs-q'r)^{2}-(s-r)^{2}}{4sr}\, \, .
\end{equation}
The models \( Vir(r,s)+\Phi _{(1,3)} \) have exactly \( r-1 \) ground states.
In fact, one can see from the fusion rules that the matrix of the operator \( \Phi _{(1,3)} \)
is block diagonal with exactly \( r-1 \) blocks in the Hilbert space made up
of states with the same left and right primary weights. In each of these blocks,
there is exactly one ground state and for the unitary series \( s=r+1 \), it
was conjectured in \cite{gr_states} that their UV limits are the states corresponding
to \( \Phi _{(k,k)} \). One can check that in the general nonunitary case the
twists (\ref{twists}) correspond in the UV limit to the lowest dimension operators
among each of the \( r-1 \) different blocks of primaries (see explicit examples
later). 

These ground states are degenerate in infinite volume, but for finite \( l \)
they split; their gaps decay exponentially as \( l\, \rightarrow \, \infty  \).
In the unitary case, they were first analyzed in the context of the NLIE in
\cite{fioravanti} where it was shown that the NLIE predictions perfectly match
with the TBA results already available for the unitary series.

However, ground states for nonunitary models have not been treated so far and
therefore now we proceed to give examples of that. The models we select are
the ones that will be used for comparison in the case of excited states as well.
The first is for the scaling Lee-Yang model \( Vir(2,5)+\Phi _{(1,3)} \), for
which we have also given data from TCS \cite{yurov-zam} and TBA \cite{zam_TBA}
for comparison (table \ref{25_vacuum}).
\[
M_{B}=2{\cal M}\sin \frac{\pi p}{2}=\sqrt{3}{\cal M}\]
is the mass of the fundamental particle of the Lee-Yang model (this is more
natural here than using the mass \( {\cal M} \) of the soliton of the unrestricted
sine-Gordon model as a scale, since the soliton disappears entirely from the
spectrum after RSOS restriction) and call
\[
l_{B}=M_{B}L=\sqrt{3}l\, \]
where \( l \) is the variable appearing in NLIE. There is only one independent
value of the twist, which we choose to be 
\[
\alpha =\frac{\pi }{2}\, \, .\]
Here and in all other subsequent calculations the TCS data were normalized using
the analogue of the coupling-mass gap relation (\ref{TBA_costante}) from \cite{mass_scale}.\begin{table}[  htbp]
{\centering \begin{tabular}{|c|c|c|c|}
\hline 
\( l_{B} \)&
TCS&
NLIE&
TBA\\
\hline 
0.1&
-2.0835015786&
-2.0835015787&
-2.0835015786\\
\hline 
0.5&
-0.3803475256&
-0.3803475281&
-0.3803475281\\
\hline 
1.0&
-0.1532068463&
-0.1532068801&
-0.1532068801\\
\hline 
1.5&
-0.0763483319&
-0.0763484842&
-0.0763484842\\
\hline 
2.0&
-0.0406269362&
-0.0406273676&
-0.0406273676\\
\hline 
2.5&
-0.0222292932&
-0.0222302407&
-0.0222302407\\
\hline 
3.0&
-0.0123492438&
-0.0123510173&
-0.0123510173\\
\hline 
3.5&
-0.0069309029&
-0.0069338817&
-0.0069338817\\
\hline 
4.0&
-0.0039198117&
-0.0039244430&
-0.0039244430\\
\hline 
5.0&
-0.0012721417&
-0.0012816882&
-0.0012816882\\
\hline 
\end{tabular}\par}

\caption{\small \label{25_vacuum} The vacuum of the Virasoro minimal model \protect\( Vir(2,5)\protect \)
perturbed by \protect\( \Phi _{(1,3)}\protect \). The energy and the volume
are normalized to the mass of the lowest excitation \protect\( M_{B}\protect \),
which is the first breather of the unrestricted sine-Gordon model. The TCS data
shown have the predicted bulk energy term subtracted.}
\end{table} There is only one ground state in this model, which corresponds to the primary
field with conformal weights
\[
\Delta ^{+}=\Delta ^{-}=-\frac{1}{5}\, \, ,\]
which is in agreement with TBA and TCS predictions. We have also found a perfect
agreement for the models \( Vir(2,7)+\Phi _{(1,3)} \) and \( Vir(2,9)+\Phi _{(1,3)} \),
but we do not present those data here. We remark that the TCS for the minimal
models converges much better than the one for \( c=1 \) theories: all TCS data
in table \ref{25_vacuum} and subsequent ones were produced by taking a few
hundred states and in some fortunate cases (e.g. the ground state of the scaling
Lee-Yang model for small values of \( l \)) we were able to produce data with
up to \( 9-10 \) digits of accuracy! The better convergence meant that all
the computation could be done with the computer algebra program \emph{Mathematica},
greatly simplifying the programming work.

All the models of the class \( Vir(2,2n+1)+\Phi _{(1,3)} \) have only one ground
state. For models with two ground states, we can take a look at \( Vir(3,5) \)
(\( Vir(3,7) \) was taken into account in \cite{noi NP2}). For \( Vir(3,5) \)
the ultraviolet spectrum is defined by the following Kac table, where the weight
(\ref{Kac_formula}) of the field \( \Phi _{(k,l)} \) is found in the \( k \)-th
row and \( l \)-th column.

\vspace{0.3cm}
{\centering \begin{tabular}{|c|c|c|c|}
\hline 
\( 0 \)&
\( -\frac{1}{20} \)&
\( \frac{1}{5} \)&
\( \frac{3}{4} \)\\
\hline 
\( \frac{3}{4} \)&
\( \frac{1}{5} \)&
\( -\frac{1}{20} \)&
\( 0 \)\\
\hline 
\end{tabular}\par}
\vspace{0.3cm}

The two blocks of the perturbing operator \( \Phi _{(1,3)} \) are defined by
the fields \( \{\Phi _{(1,2)}\, ,\, \Phi _{(1,4)}\} \) and \( \{\Phi _{(1,1)}\, ,\, \Phi _{(1,3)}\} \),
respectively. The ground states correspond in the UV to the operators \( \Phi _{(1,2)} \)
and \( \Phi _{(1,1)} \), as can be checked directly using formulae (\ref{alpha_vacuum}),
(\ref{p_min_r_s}) and (\ref{twists}).

We also have TBA data to compare with, using the TBA equation written by Christe
and Martins \cite{christe-martins}. The lower-lying ground state is obtained
directly from their TBA, while for the other we used the idea of Fendley of
twisting the TBA equation \cite{Fendley}. The numerical results are presented
in tables \ref{vacuum_35_1} and \ref{vacuum_35_2}.

In the case \( Vir(3,7) \) treated in \cite{noi NP2}, it was possible only
to have a comparison with TCS results, but it still looked pretty convincing. \begin{table}[  htbp]
{\centering \begin{tabular}{|c|c|c|c|}
\hline 
\( l \)&
TCS&
NLIE&
TBA\\
\hline 
0.1&
-3.074916&
-3.0749130189&
-3.0749130190\\
\hline 
0.3&
-0.944161&
-0.9441276204&
-0.9441276204\\
\hline 
0.5&
-0.509764&
-0.5096602194&
-0.5096602194\\
\hline 
0.8&
-0.265436&
-0.2651431026&
-0.2651431026\\
\hline 
1.0&
-0.186038&
-0.1855606546&
-0.1855606546\\
\hline 
1.5&
-0.087300&
-0.0861426792&
-0.0861426792\\
\hline 
2.0&
-0.045910&
-0.0437473815&
-0.0437473815\\
\hline 
2.5&
-0.026746&
-0.0232421927&
-0.0232421927\\
\hline 
3.0&
-0.017868&
-0.0126823057&
-0.0126823057\\
\hline 
4.0&
-0.013546&
-0.0039607326&
-0.0039607326\\
\hline 
\end{tabular}\par}

\caption{\small \label{vacuum_35_1} One of the two ground states of the Virasoro minimal model
\protect\( Vir(3,5)\protect \) perturbed by \protect\( \Phi _{(1,3)}\protect \),
corresponding to \protect\( \alpha =\frac{\pi }{3}\protect \). The energy and
the volume are normalized to the mass of the lowest excitation, which is the
soliton of the unrestricted sine-Gordon model. The TCS data shown have the predicted
bulk energy term subtracted.}
\end{table} \begin{table}[  htbp]
{\centering \begin{tabular}{|c|c|c|c|}
\hline 
\( l \)&
TCS&
NLIE&
TBA\\
\hline 
0.1&
3.117844&
3.1178476855&
3.1178476853\\
\hline 
0.3&
0.985360&
0.9853990810&
0.9853990810\\
\hline 
0.5&
0.540427&
0.5405470784&
0.5405470784\\
\hline 
0.8&
0.282725&
0.2830552991&
0.2830552991\\
\hline 
1.0&
0.197143&
0.1976769278&
0.1976769278\\
\hline 
1.5&
0.089277&
0.0905539780&
0.0905539780\\
\hline 
2.0&
0.042960&
0.0453290013&
0.0453290013\\
\hline 
2.5&
0.019978&
0.0238075022&
0.0238075022\\
\hline 
3.0&
0.007209&
0.0128843786&
0.0128843786\\
\hline 
4.0&
-0.006592&
0.0039866371&
0.0039866371\\
\hline 
\end{tabular}\par}

\caption{\small \label{vacuum_35_2} The other ground state of the Virasoro minimal model \protect\( Vir(3,5)\protect \)
perturbed by \protect\( \Phi _{(1,3)}\protect \), corresponding to \protect\( \alpha =\frac{2\pi }{3}\protect \).
The energy and the volume are normalized to the mass of the lowest excitation,
which is the soliton of the unrestricted sine-Gordon model. The TCS data shown
have the predicted bulk energy term subtracted.}
\end{table} To summarize, we now have sufficient evidence to believe that the \( \alpha  \)-twisted
NLIE describes the correct scaling functions for ground states of minimal models
perturbed by \( \Phi _{(1,3)} \) \emph{in unitary and nonunitary case}. However,
the NLIE for sine-Gordon is known to work for excited states as well. But how
do we get the excited state spectrum of the minimal models now?

\section{\protect\( \alpha \protect \) twist and minimal models: excited states}

\subsection{The choice of \protect\( \alpha \protect \)}

From now on we restrict ourselves to the case of neutral (i.e. \( S=0 \)) states.
It is easy to see that even for states with a zero charge the relation between
\( \alpha  \) and \( \omega  \) is highly nontrivial.

Now we proceed to show that choosing the value of \( \omega  \) as
\[
\omega =\frac{k\pi }{p+1}\]
where \( k \) is integer, we can reproduce all the required values of \( \alpha  \)
listed in equation (\ref{twists}). Observe that this expression for \( \omega  \)
is different from (\ref{omega}), but totally equivalent, as will be clear soon.
First of all, we substitute the value of \( p \) from (\ref{p_min_r_s}) to
obtain
\[
\omega =\frac{k(s-r)\pi }{s}\, .\]
Since \( r \) and \( s \) are relative primes, the independent values of \( \omega \, \, \bmod \, \, \pi  \)
can be written as 
\[
\omega =\frac{l\pi }{s}\, \, ,\, \, l=0,\ldots ,s-1\, \, .\]
 For \( S=0 \), we can rewrite the formula (\ref{alfa}) as follows
\[
\alpha =\frac{l\pi }{r}+\frac{2r-s}{2r}\pi \left( \left[ \frac{1}{2}+\frac{l}{s}\right] -\left[ \frac{1}{2}-\frac{l}{s}\right] \right) \, .\]
We are interested only in the value of \( \alpha \, \, \bmod \, \, \pi  \),
since using the parameter \( \delta  \) one can effectively shift \( \alpha  \)
by \( \pi  \). This leaves us with the formula 
\[
\alpha =\frac{l\pi }{r}-\frac{s\pi }{2r}\left( \left[ \frac{1}{2}+\frac{l}{s}\right] -\left[ \frac{1}{2}-\frac{l}{s}\right] \right) \, .\]
The first possibility is that \( l<\frac{s}{2} \), which simply gives us the
values 
\[
\alpha =\frac{l\pi }{r}.\]
 When \( s \) is even, we can have \( l=\frac{s}{2} \), which gives us \( \alpha =0 \).
Finally, when \( l>\frac{s}{2} \), we get the values 
\[
\alpha =\frac{(l-s)\pi }{r}\, \, .\]
It is easy to check that these formulae reproduce every value 
\[
\alpha =\frac{n\pi }{r}\, \, \bmod \, \, \pi \]
at least once, as required by (\ref{twists}), using the fact that \( s>r \)
and that the values above form an uninterrupted sequence of \( s \) numbers
(or when \( s \) is even, of \( s-1 \) numbers, the zero repeated) with equal
distances \( \frac{\pi }{r} \).

As we have already seen in the previous section, all the values
\begin{equation}
\label{valori_alfa}
\alpha =\displaystyle\frac{k\pi }{r}\, \, ,\, \, k=1,\ldots ,r-1
\end{equation}
are necessary to reproduce correctly the \( r-1 \) ground states of the model
\( Vir(r,s)+\Phi _{(1,3)} \). 

The twisted lattice Bethe Ansatz was analyzed by de Vega and Giacomini in \cite{omega-twist}.
On the lattice, passing from the sine-Gordon model to the perturbed Virasoro
model amounts to going from the six-vertex model to a lattice RSOS model. In
\cite{omega-twist} it was shown that to obtain all the states of the RSOS model
it is necessary to take all the twists 
\[
\omega =\frac{k\pi }{p+1}\, \, \bmod \, \, \pi \]
into account. The fact that not all these twists correspond to inequivalent
values of \( \alpha  \) and so to different physical states is a consequence
of the RSOS truncation.

To close this section we remark that the parameter \( \alpha  \) drops out
of the second determination of \( Z \) in the attractive regime. This is important
because as a consequence the IR asymptotics of the breather states does not
depend on \( \alpha  \) and so the \( S \)-matrices involving breathers are
unchanged. In fact, scattering amplitudes between solitons and breathers remain
unchanged too, as can be seen from examining the argument that we used to derive
them in section \ref{section:IR_breather}. This matches with the fact that
the RSOS restriction from sine-Gordon theory to perturbed minimal models does
not modify scattering amplitudes that involve two breathers or a breather and
soliton \cite{reshetikhin_smirnov}.

\subsection{The UV limit}

There is in fact a very simple intuitive argument to show that the states we
get from NLIE with an \( \alpha  \) twist are related to the minimal models
in the UV limit. Let us first recall that the UV limit of the \( \alpha =0 \)
NLIE yields the vertex operators \( V_{(n,m)} \) and their descendants (\ref{delta_sG},
\ref{m,n+-}).

Let us look at the conformal dimensions. First note that because the value of
\( \alpha  \) for a minimal model is never a multiple of \( \pi  \), one does
not expect central sources in the UV limit but only right/movers or left movers.
Indeed, in all the examples of sine-Gordon with central sources, the left-right
symmetry (i.e. \( Z(-\vartheta )=-Z(\vartheta ) \) on the real axis) of the
NLIE was crucial. This symmetry, however, only holds for \( \alpha =0 \) or
\( \pi  \). As a consequence we have
\[
S^{0}=0,\qquad S=S^{+}+S^{-}\, ,\]
and in addition \( {\cal Q}_{+}(-\infty )={\cal Q}_{-}(+\infty ) \), which
means that only \emph{one-plateau systems} are allowed in the twisted case.

Using the formulas (\ref{delta}, \ref{delta_sG}) for the UV limit of the NLIE,
one can see that introducing \( \alpha \neq 0 \) is equivalent to shifting
the quantum number \( n_{\pm } \) to \( n_{\pm }+\frac{\alpha }{2\pi } \).
One has to be careful that since the value of the central charge is shifted
from \( 1 \) to the one of the minimal model, we have to take this shift into
account when computing the conformal weight from the leading UV behaviour of
the energy level (\ref{andamento-uv}). We put in (\ref{delta}) the value
\[
\alpha =\frac{j\pi }{r},\, \, j=1,\ldots ,r-1\, \, ,\]
as in (\ref{valori_alfa}), and the central charge
\[
c=1-\frac{6(r-s)^{2}}{rs}\]
of the minimal model \( Vir(r,s) \), as in section \ref{section:alfa_gs}.
By a computation similar to the one done in section \ref{section:sG-mTH} to
obtain (\ref{delta_sG}, \ref{m,n+-}, \ref{n+-}) one can obtain the following
expression for conformal dimensions:
\begin{equation}
\label{delta_minimal}
\begin{array}{c}
\Delta ^{\pm }=\displaystyle\frac{(ls-l_{\pm }'r)^{2}-(s-r)^{2}}{4rs}+N_{\pm }\\
l=S\\
l_{\pm }'=\pm \left( \delta -\displaystyle\frac{j}{r}+2k_{\pm }-2k_{W}^{\pm }\right) -2\left( S-2S^{\pm }\right) .
\end{array}
\end{equation}
The expression of \( N_{\pm } \) is exactly the same that appears in (\ref{n+-}).
Moreover, from the formulae (\ref{delta}, \ref{delta_minimal}) it is also
clear that in general
\[
2\left( \Delta ^{+}-\Delta ^{-}\right) =2S\frac{\alpha }{\pi }\, \, \bmod \, \, 1\]
because only one plateau systems can take place, and so we see that general
charged states will have fractional Lorentz spin. Actually, it is known that
charged states in the models \( Vir(r,s)+\Phi _{(1,3)} \) generally have fractional
Lorentz spin \cite{felder-leclair}. 

Comparing the first line of (\ref{delta_minimal}) with (\ref{Kac_formula})
one observes that it can represent the conformal dimensions of an operator \( \Phi _{(q,q')} \)
only if some conditions are verified. The most general one is
\[
\left( ls-l'r\right) ^{2}=\left( qs-q'r\right) ^{2}+2rs\cdot \textrm{integer},\qquad 1\leq q\leq r-1,\, 1\leq q'\leq s-1\]
(the \( l,\, l' \) are not integer numbers in general). Observe that it is
an arithmetic (Diofantine) equation in \( q,q' \) and the unknown integer. 

It is convenient to treat some specific cases, because, as for the sine-Gordon
case, there is no proof about the general behaviour of \( l_{\pm }' \) and
\( N_{\pm } \). 

In the unitary \( p=r,\quad s=r+1 \) and neutral case \( S=0 \) the resulting
conformal weights take the form
\begin{equation}
\label{alpha_weights}
\Delta ^{+}=\Delta ^{-}=\displaystyle\frac{(2np+l)^{2}-1}{4p(p+1)}\, \, .
\end{equation}
We see that this is the weight of the field \( \Phi _{(l,l-2n)} \) in the minimal
model \( Vir(p,p+1) \), however, in order not to overflow the Kac table, the
range of \( n \) must be restricted as 
\[
1\leq l-2n\leq p\, \, .\]
For charged states, a similar calculation can be performed. 

By an inspection of the formula (\ref{delta}), for any state with \( S=0 \),
the property to have only one plateau implies that 
\[
\Delta ^{+}-\Delta ^{-}\]
is integer or half-integer. In fact, choosing the quantization rule and the
parameter \( \delta  \) in an appropriate way, one can ensure that this difference
is integer (see \cite{noi NP, noi PL2}). This means that the UV limit of any
neutral state is either a field occurring in the ADE classification of modular
invariants \cite{ADE} or (in case we choose \( \delta  \) so that \( \Delta ^{+}-\Delta ^{-} \)
is half-integer) it is a field from a fermionic version of the minimal model
\cite{kl-me}. 

Until now it is not clear whether the weights actually stay inside the Kac table,
for which in the unitary case one must require
\[
1<l-2k+4S^{+}<p\, .\]
Due to the fact that the configuration of sources in the UV may be very non-trivially
related to the one in the IR, this condition is very hard to check in general,
but no concrete examples that we calculated have ever violated this bound.

\section{Concrete examples of excited states}

\subsection{The \protect\( Vir(2,2n+1)+\Phi (1,3)\protect \) series}

Let us start with examining the scaling Lee-Yang model \( Vir(2,5)+\Phi _{(1,3)} \).
There is only one independent value of the twist which we choose as 
\[
\alpha =\frac{\pi }{2}\, \, ,\]
since we have a single ground state, and as a result there are no kinks in the
spectrum. We fix the value of \( \alpha  \) as above, so we still have a freedom
of choosing \( \delta  \). This can be done by matching to the UV dimensions:
if for a certain state we choose the wrong value of \( \delta  \), we find
a conformal dimension that is not present in the Kac table of the model. 

The excited states are multi-particle states of the first breather of the corresponding
unrestricted sine-Gordon model, which has 
\[
p=\frac{2}{3}.\]
Now one can calculate the state containing one particle at rest. We find the
numerical data presented in table \ref{yl_first}. 

It turns out that as we decrease \( l \), the self-conjugate root starts moving
to the right. It does not remain in the middle like in the \( \alpha =0 \)
case, which is to be expected since for nonzero \( \alpha  \) we have no left/right
symmetry. However, the total momentum of the state still remains zero due to
a contribution from the integral term in momentum equation (\ref{momento}).\begin{table}[  htbp]
{\centering \begin{tabular}{|c|c|c|}
\hline 
\( l_{B} \)&
TCS&
NLIE\\
\hline 
0.1&
23.05277&
n/a\\
\hline 
0.5&
4.679779&
n/a\\
\hline 
1.0&
2.447376&
n/a\\
\hline 
1.5&
1.748874&
n/a\\
\hline 
2.0&
1.430883&
n/a\\
\hline 
2.6&
1.238051&
1.238012(\#)\\
\hline 
3.0&
1.164321&
1.164319\\
\hline 
3.5&
1.105220&
1.105196\\
\hline 
4.0&
1.068256&
1.068237\\
\hline 
5.0&
1.029356&
1.029348\\
\hline 
\end{tabular}\par}

\caption{\small \label{yl_first} The first excited state of the scaling Lee-Yang model. The
energy and the volume are normalized to the mass of the lowest excitation, which
is the first breather of the unrestricted sine-Gordon model. The TCS data shown
have the predicted bulk energy term subtracted. }
\end{table} One can see that once again we have the phenomenon noticed in the case of the
first breather of sine-Gordon theory, namely the appearance of the special root
and its two accompanying holes, so the iteration breaks down again around \( l=2.5 \).
The (\#) in the table \ref{yl_first} written after the NLIE result for \( l=2.6 \)
means that due to the fact that the singularities corresponding to the new holes
and the special root are just about to cross the contour and upset the iteration
scheme, the NLIE result becomes less precise. We will use this notation on later
occasions too. In any case, the agreement still looks quite convincing. 

Let us now look at the UV spectrum of the model. We know that the Lee-Yang model
contains only two primary fields, the identity \( \mathbb I \) and the field
\( \varphi  \) with left/right conformal weights
\[
\Delta ^{+}=\Delta ^{-}=-\frac{1}{5}\, \, .\]
In fact, the ground state of the massive model corresponds to \( \varphi  \)
in the UV limit. One can compute the UV limit of the first particle from the
NLIE too, taking into account the appearance of the special root and the holes.
It turns out that the special root and one of the holes moves to the left together
with the self-conjugate root, while the other hole moves to the right. The result
is
\[
\Delta ^{+}=\Delta ^{-}=0\, \, ,\]
i.e. the identity operator \( \mathbb I \), which fits nicely with the TCS
data (see also \cite{yurov-zam}).

Let us look now at moving breathers. If the self-conjugate root has Bethe quantum
number \( I=1 \), the corresponding state will have momentum quantum number
\( 1 \), i.e. 
\[
P=\frac{2\pi }{R}\, \, ,\]
and in the UV \( \Delta ^{+}-\Delta ^{-}=1 \). One can note from the numerical
data presented in table \ref{yl_first_spin1} that the special root does not
appear here. The reason is that the self-conjugate root moves to the left and
the real part of its position \( \vartheta  \) is given to leading order by
\[
\sinh (\Re e\, \vartheta )\, \sim \, -\frac{2\pi I}{l_{B}}\, \, .\]
As a result, the contribution to the derivative of \( Z \) from the \( l\sinh \vartheta  \)
term remains finite when \( l\, \rightarrow \, 0 \). In the previous example
of the particle at rest the left-moving nature of the self-conjugate root when
\( I=0 \) does not prevent the occurrence of the breakdown in the iteration
scheme: since its Bethe quantum number is zero, it does not move fast enough
to the left in order to balance the negative contribution to derivative of \( Z \)
coming from the self-conjugate root source. At the moment we have no way of
predicting analytically whether or not there will be specials in the UV limit:
we just use the numerical results to establish the configuration for the evaluation
of UV weights, supplemented with a study of the self-consistency of the solution
of the plateau equation (\ref{plateau}).\begin{table}[  htbp]
{\centering \begin{tabular}{|c|c|c|}
\hline 
\( l_{B} \)&
TCS&
NLIE\\
\hline 
0.1&
60.75048&
60.74682\\
\hline 
0.5&
12.20618&
12.20561\\
\hline 
1.0&
6.182516&
6.182363\\
\hline 
1.5&
4.202938&
4.202915\\
\hline 
2.0&
3.231734&
3.231640\\
\hline 
2.5&
2.662186&
2.662110\\
\hline 
3.0&
2.292273&
2.292231\\
\hline 
3.5&
2.035552&
2.035530\\
\hline 
4.0&
1.848892&
1.848849\\
\hline 
5.0&
1.599792&
1.599762\\
\hline 
\end{tabular}\par}

\caption{\small \label{yl_first_spin1} The one-particle states with Lorentz spin 1 of the
scaling Lee-Yang model. The energy and the volume are normalized to the mass
of the lowest excitation, which is the first breather of the unrestricted sine-Gordon
model. The TCS data shown have the predicted bulk energy term subtracted. }
\end{table} The UV dimensions for the moving breather turn out to correspond to the state
\( L_{-1}\varphi  \).

One can similarly compute the UV dimensions for some other excited states. For
example, the two-particle states with half-integer Bethe quantum numbers \( I_{1}>0,\, \, I_{2}<0 \)
for the two self-conjugate roots are found to have 
\[
\Delta ^{+}=-\frac{1}{5}+I_{1}+\frac{1}{2}\, \, ,\, \, \Delta ^{-}=-\frac{1}{5}-I_{2}+\frac{1}{2}\, \, ,\]
in agreement with TCS data which show that they correspond in the UV to descendent
states of \( \varphi  \). The first such state with quantum numbers 
\[
I_{1}=\frac{1}{2}\, \, ,\, \, I_{2}=-\frac{1}{2}\]
corresponds in the UV to \( L_{-1}\bar{L}_{-1}\varphi  \) and is given numerically
in table \ref{yl_second}.

\begin{table}[  htbp]
{\centering \begin{tabular}{|c|c|c|}
\hline 
\( l \)&
TCS&
NLIE\\
\hline 
0.1&
123.583&
123.5693\\
\hline 
0.5&
24.7806&
24.77936\\
\hline 
1.0&
12.4870&
12.48635\\
\hline 
1.5&
8.42926&
8.428693\\
\hline 
2.0&
6.42931&
6.429201\\
\hline 
3.0&
4.48444&
4.484209\\
\hline 
4.0&
3.56367&
3.563519\\
\hline 
5.0&
3.04899&
3.048881\\
\hline 
\end{tabular}\par}

\caption{\small \label{yl_second} The lowest lying zero-momentum two-particle state in the
scaling Lee-Yang model as computed from the NLIE and compared with TCS.}
\end{table} The lowest lying three-particle state of zero momentum, with Bethe quantum numbers
\( (-1,0,1) \) corresponds to the left/right symmetric second descendent of
the identity field, i.e. to the field \( T\bar{T} \), where \( T \) denotes
the energy-momentum tensor. This is very interesting, since from experience
with NLIE UV calculations one would naively expect this to be a first descendent
(descendent numbers are usually linked to the sum of Bethe quantum numbers of
left/right moving particles and this state is the lowest possible descendent
of the identity \( \mathbb I \)). However, the field \( L_{-1}\bar{L}_{-1}\mathbb I \)
is well-known to be a null field in any conformal field theory.

The above correspondences are again confirmed by comparing to TCS (see the wonderful
figures in \cite{yurov-zam}). In general, one can establish the rule that states
with odd number of particles must be quantized by integers (\( \delta =1 \)),
while those containing even number of particles must be quantized by half-integers
(\( \delta =0 \)) in order to reproduce correctly the spectrum of the scaling
Lee-Yang model.

We conducted similar studies for the models \( Vir(2,7)+\Phi _{(1,3)} \) and
\( Vir(2,9)+\Phi _{(1,3)} \) and found similarly good agreement with TCS data.
For the first one-particle state of the model \( Vir(2,7)+\Phi _{(1,3)} \)
we also checked our results against the TBA data in the numerical tables of
\cite{dorey_tateo2} and found agreement with the TBA results.

Given the choice of \( \alpha  \) above, the correct rule of quantization in
all of the models \( Vir(2,2n+1)+\Phi _{(1,3)} \) is 
\[
\delta =M_{sc}\, \, \bmod \, \, 2\, \, ,\]
where \( M_{sc} \) is the number of self-conjugate roots in the source corresponding
to the state. This is exactly the same rule as the one established for pure
sine-Gordon theory in \cite{noi NP}. In the presence of the twist, such a rule
of course has meaning only together with a definite convention for the choice
of \( \alpha  \).

\subsection{One-breather states in the \protect\( Vir(3,7)\protect \) case}

It is interesting to note that in the case of \( Vir(3,n) \) models, all the
neutral states must come in two copies, since they can be built on top of either
of the two ground states. We take the example of the \( Vir(3,7) \) model and
the states corresponding to a breather at rest. We have 
\[
p=\frac{3}{4}\]
 but now there are two inequivalent values for the twist 
\[
\alpha =\frac{\pi }{3}\, \, ,\, \, \frac{2\pi }{3}\, .\]
When \( \alpha =0 \), we can calculate the critical value of \( l \) to be
\( l_{critical}=5.23 \) using (\ref{critical_estimate}). In this case, the
twist helps a bit, because it makes the self-conjugate root a left mover; it
is intuitively clear that the bigger the twist, the more it lowers the eventual
value of \( l_{crtical} \), which is in accord with the numerical results of
table \ref{37_firsts}. From the TCS data one can identify that breather \#1
is really the one-particle state in the sector of ground state \#1 (\( \alpha =\pi /3 \)),
while breather \#2 is in the sector over ground state \#2 (\( \alpha =2\pi /3 \)).

A direct calculation of the conformal weights gives the following results:\( \Delta ^{+}=\Delta ^{-}=\frac{3}{28} \)for
breather \#1 and \( \Delta ^{+}=\Delta ^{-}=0 \)for breather \#2, which are
in complete agreement with the TCS data. We also checked the two different states
containing two breathers with Bethe quantum numbers\( I_{1}=\frac{1}{2}\, \, ,\, \, I_{2}=-\frac{1}{2}\, \, , \)and
found an equally excellent numerical agreement with TCS. Just like in the case
of sine-Gordon and scaling Lee-Yang model, for these states one can continue
the iteration of the NLIE down to any small value of \( l \), although at the
expense of a growing number of necessary iterations to achieve the prescribed
precision.\begin{table}[  htbp]
{\centering \begin{tabular}{|c|c|c|c|c|}
\hline 
&
\multicolumn{2}{|c|}{ breather 1}&
\multicolumn{2}{|c|}{breather 2}\\
\hline 
\( l \)&
TCS&
NLIE&
TCS&
NLIE\\
\hline 
0.1&
32.21645&
n/a&
18.76030&
n/a\\
\hline 
0.5&
6.671964&
n/a&
4.037716&
n/a\\
\hline 
1.0&
3.662027&
n/a&
2.434501&
2.434431(\#)\\
\hline 
1.5&
2.769403&
n/a&
2.027227&
2.027213\\
\hline 
2.0&
2.385451&
n/a&
1.884404&
1.884388\\
\hline 
2.5&
2.190232&
n/a&
1.829459&
1.829456\\
\hline 
3.0&
2.079959&
n/a&
1.809244&
1.809248\\
\hline 
3.5&
2.012596&
n/a&
1.803873&
1.803886\\
\hline 
4.0&
1.968808&
1.968784(\#)&
1.804937&
1.804953\\
\hline 
4.5&
1.938895&
1.938889&
1.808633&
1.808658\\
\hline 
5.0&
1.917648&
1.917635&
1.813191&
1.813224\\
\hline 
\end{tabular}\par}

\caption{\small \label{37_firsts} The two one-breather states of the Virasoro minimal model
\protect\( Vir(3,7)\protect \) perturbed by \protect\( \Phi _{(1,3)}\protect \).
The energy and the volume are normalized to the mass of the kink, which is the
soliton of the unrestricted sine-Gordon model. Breather \#1 has \protect\( \alpha =\frac{\pi }{3}\protect \),
while breather \#2 corresponds to \protect\( \alpha =\frac{2\pi }{3}\protect \).
The TCS data shown have the predicted bulk energy term subtracted. }
\end{table}

\section{Conclusions\label{section:conclusione}}

In this thesis it is studied how the nonlinear integral equation deduced from
the light cone lattice model of \cite{ddv 87} describes the excited states
of the sine-Gordon/massive Thirring theory. The most important results are summarized
as follows:

\begin{enumerate}
\item A derivation of the fundamental NLIE is presented from the light cone lattice
which correctly takes into account the behaviour of the multivalued complex
logarithm function. 
\item By examining the infrared limits of the equation it has been shown that (1)
it leads to the correct two-particle S-matrices for both scattering states and
bounded states; (2) it is in agreement with the predictions of the TCS method
if one chooses the correct quantization conditions for the source terms.
\item By computing the UV conformal weights from the NLIE we have shown that it is
consistent with the UV spectrum of sG/mTh theory only if we choose the parameter
\( \delta  \) (i.e. the quantization rule) as indicated in (\ref{regola_d'oro}). 
\item The predictions of the NLIE have been verified by comparing them to results
coming from the TCS approach (for sG/mTh).
\item The framework required to deal with minimal models perturbed by \( \Phi _{(1,3)} \)
is built up. Many examples and numerical/analytical checks are given for the
ground state in the unitary \cite{fioravanti} and non unitary cases. 
\item All the conformal dimensions can be reproduced (Kac table) with the convenient
choice of the twist (\ref{valori_alfa}). The particular relation suggested
in \cite{zinn-justin} is not enough to describe the whole spectrum. 
\item The IR computations given in section \ref{section:IR_breather} reproduce correctly
the S-matrix of perturbed minimal models in the attractive regime (at list in
the cases involving only solitons), because the attractive second determination
drops the twist.
\item Numerical calculations of concrete examples give a strong evidence for the correctness
of the energy levels derived from the twisted NLIE for excited states. 
\end{enumerate}
The understanding of ``(twisted) sine-Gordon NLIE'' and of the finite size
behaviour of the continuum theory defined from the NLIE (\ref{nlie-cont}) is
not quite complete. Indeed some open questions have not an answer, until now
and also further interesting developments can take place:

\begin{enumerate}
\item Is the set of scaling functions provided by the NLIE complete i.e. can we find
to every sG/mTh or minimal model state a solution of the NLIE describing its
finite volume behaviour? This can be called a ``counting problem''. The main
difficulty is that the structure of the solutions is highly dependent on the
value of the coupling constant -- to see that it is enough to consider e.g.
the appearance of special sources. 
\item The multi-kink states characteristic of the perturbed minimal models (except
for the series \( Vir(2,2n+1) \)) have been omitted in the analysis so far
performed. Although the general treatment of the IR spectrum in section \ref{section:IR_breather}
is valid for those states too, a detailed description is far more complicated
than for states which contain only breathers and is left open to further studies. 
\item There is also an unresolved technical difficulty, namely that the source configuration
of the NLIE may change as we vary the volume parameter \( l \). Typically what
happens is that while the counting function \( Z \) is monotonic on the real
axis for large volume, this may change as we lower the value of \( l \) and
so-called special sources (and accompanying holes) may appear. We do not as
yet have any consistent and tractable numerical iteration scheme to handle this
situation, although the analytic UV calculations and intuitive arguments show
that the appearance of these terms in the NLIE is consistent with all expectations
coming from the known properties of perturbed CFT. In addition, in the range
of \( l \) where we can iterate the NLIE without difficulty, our numerical
results show perfect agreement with TCS. We want to emphasize that these transitions
are not physical: the counting function \( Z \) and the energy of the state
are expected to vary analytically with the volume. It is just their description
by the NLIE with requires a modification of source terms. As it was pointed
out also in \cite{ddv 97, noi NP}, the whole issue is related to the choice
of the branch of the logarithmic term in the NLIE (\ref{nlie-cont}). 
\item The problem pointed out at the previous point is very similar to the behaviour
of singularities encountered in the study of the analytic continuation of the
TBA equation \cite{dorey-tateo} and we can hope that establishing a closer
link between the two approaches can help to clarify the situation. From the
form of the source terms in the NLIE it seems likely that the excited state
equations can be obtained by an analytic continuation procedure analogous to
the one used in TBA \cite{dorey-tateo} to obtain the excited state TBA equations.
Certain features of the arrangement of the complex roots in the attractive regime
and their behaviour at breather thresholds also point into this direction. This
an interesting question to investigate because it can shed light on the organization
of the space of states and can lead closer to solving the counting problem described
above. 
\item Even if on the lattice the Bethe vectors given in (\ref{bethe}) completely
describe the Hilbert space of the XXZ chain, the form of eigenvectors for continuum
energy and momentum is completely unknown. This fact reflects the unresolved
question of the form of energy and momentum eigenvectors for the minkowskian
sine-Gordon theory (the Faddeev-Zamolodchikov algebra is only a phenomenological
picture). Also the determination of correlation functions is in an early stage,
but in recent publications \cite{vev-sG} explicit expressions for the field
\( \left\langle e^{ia\varphi }\right\rangle  \) and its descendent are given.
\item The extension of light-cone approach and NLIE to other QFT models is in progress.
An interesting extension is the description of finite volume spectrum of Ziber-Mikhailov-Shabat
model (imaginary Bullogh-Dodd). A first suggestion in this direction, even if
is obtained in a completely different approach, recently appeared in the \cite{dorey-tateo_ultimo},
for the ground state. 
\end{enumerate}
The second and fourth points are important because their investigation may lead
closer to understanding the relation between the TBA and the NLIE approaches.
It is quite likely that establishing a connection between the two methods would
facilitate the development of both and may point to some common underlying structure.

As a final remark, I want to speak about an application of the NLIE to a chemical
compound, i.e. ``copper benzoate'' (\( \textrm{Cu}(\textrm{C}_{6}\textrm{D}_{5}\textrm{COO})_{2}\cdot 3\textrm{D}_{2}\textrm{O} \)).
Its specific heat has been computed in \cite{copper_benzoate} using the vacuum
untwisted (\( \alpha =0 \)) NLIE, where the finite size \( L \) is the inverse
temperature \( L=(k_{B}T)^{-1} \). This corresponds to do equilibrium thermodynamics
of sine-Gordon model. 

A curiosity: in that contest the NLIE is called Thermal Bethe Ansatz, emphasizing
the well known relation between statistical mechanics and QFT. Exactly in a
similar contest (Heisenberg ferromagnet model) was born the Bethe Ansatz \cite{bethe}. 

\appendix

\chapter{Fourier transformation: some conventions \label{trasf fourier} }

Given a bounded continuous function \( f(x) \) defined on the whole real axis,
its Fourier transform is defined as (all the integrals must be taken on the
whole real axis):
\begin{equation}
\label{fouriertransform}
\tilde{f}(k)=\displaystyle\int dx\, e^{-ikx\displaystyle\frac{1}{p+1}}f(x)
\end{equation}
and the inverse Fourier transform is 
\begin{equation}
\label{inversefourier}
f(x)=\displaystyle\frac{1}{p+1}\displaystyle\int \displaystyle\frac{dk}{2\pi }e^{ikx\displaystyle\frac{1}{p+1}}\tilde{f}(k).
\end{equation}
The convolution of two functions, defined as follows, can be expressed in terms
of the Fourier transforms of the two functions:
\begin{equation}
\label{convolution}
\left( f\ast g\right) _{\theta }=\displaystyle\int dx\, f(\theta -x)\, g(x)=\displaystyle\frac{1}{p+1}\displaystyle\int \displaystyle\frac{dk}{2\pi }e^{ik\theta \displaystyle\frac{1}{p+1}}\tilde{f}(k)\tilde{g}(k)
\end{equation}
The following expression holds between the Fourier transforms of a function
and its derivative:
\[
\tilde{f}(k)=\frac{\tilde{f}'(k)}{ik}(p+1)\]

\chapter{The function \protect\( \phi (\lambda ,\nu )\protect \)\label{funzione fi}}

\begin{figure}[  htbp]
{\centering \includegraphics{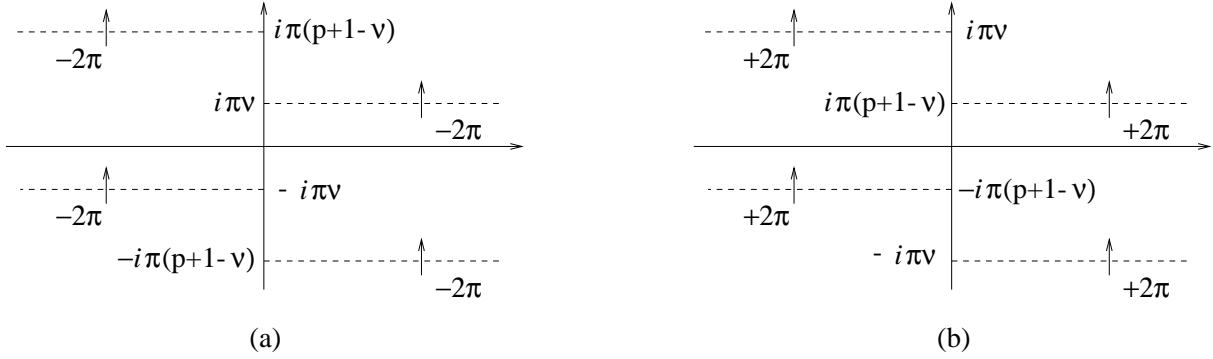} \par}

\caption{\small Positions of singularities and cuts for the function \protect\( \phi .\protect \)
(a) in the case \protect\( \displaystyle\frac{\nu }{p+1}<1/2\protect \); (b) in the case
\protect\( \displaystyle\frac{\nu }{p+1}>1/2.\protect \) The little arrows indicate the
jump in the values of the function. \label{cuts.eps}}
\end{figure}

In this appendix all the important properties of the function \( \phi  \) will
be clarified. Let
\begin{equation}
\label{phifunct}
\phi (\vartheta ,\nu )=i\log \displaystyle\frac{\sinh \displaystyle\frac{1}{p+1}(i\pi \nu +\vartheta )}{\sinh \displaystyle\frac{1}{p+1}(i\pi \nu -\vartheta )},\qquad \phi (-\vartheta ,\nu )=-\phi (\vartheta ,\nu )\quad \forall \quad \vartheta \in \mathbb {C}
\end{equation}
 and continuity is required around the real axis. The interest is for \( \nu =1/2,\; 1 \)
and \( p>0. \) The requirement of oddity implies, for real values of \( \vartheta , \)
that \( \phi (0,\nu )=0. \) Oddity and continuity can be implemented only if
the fundamental determination (FD) of the logarithm is assumed, in a strip containing
the real axis:
\[
-\pi <\Im m(\log _{FD}w)\leq \pi .\]
This choice implies that \( \phi  \) is real on the real axis. The argument
of the log has poles and zeros. They are essential singularities for the \( \phi  \)
function. Their position is given by:
\begin{equation}
\label{singolarita}
\begin{array}{c}
\Re e\, \vartheta =0\\
\Im m\, \vartheta =\pm \pi \, \left( k(p+1)-\nu \right) ,\qquad k\in \mathbb {Z}.
\end{array}
\end{equation}
 The fundamental strip around the real axis (FD) is bounded by the first encountered
singularity:
\begin{equation}
\label{strisciafond}
\vartheta \in \mathbb {R}\times \left] -\min \left\{ \nu \pi ,\, \pi \left( p+1-\nu \right) \right\} ,\min \left\{ \nu \pi ,\, \pi \left( p+1-\nu \right) \right\} \right[ .
\end{equation}
To extend the definition of (\ref{phifunct}) to the whole plane it is necessary
to give a prescription for the position of the cuts (because of the logarithm,
in a small closed trip around one of the singularities (\ref{singolarita}),
the value of \( \phi  \) changes by \( 2\pi  \)). This is done as in figure
\ref{cuts.eps}.

The derivative of \( \phi  \) is the single value meromorphic function:
\begin{equation}
\label{phi_primo}
\phi '(\vartheta ,\nu )=\displaystyle\frac{1}{p+1}\displaystyle\frac{2\sin \displaystyle\frac{2\nu \pi }{p+1}}{\cosh \displaystyle\frac{2\vartheta }{p+1}-\cos \displaystyle\frac{2\nu \pi }{p+1}}.
\end{equation}
This shows that, on the real axis, \( \phi  \) is monotonically increasing
if \( 0<\displaystyle\frac{\nu \pi }{p+1}<\pi /2 \) and decreasing if \( \pi /2<\displaystyle\frac{\nu \pi }{p+1}<\pi . \)
The asymptotic values, for \( \vartheta  \) in the fundamental strip, are:
\begin{equation}
\label{phi-limiti}
\lim _{\vartheta \, \rightarrow \pm \infty }\phi (\vartheta ,\nu )=\pm \pi (1-\displaystyle\frac{2\nu }{p+1}).
\end{equation}
Out of the fundamental strip, the prescription indicated in figure \ref{cuts.eps}
must be used. 

The Fourier transformation of \( \phi ' \) can be obtained from the definition
(\ref{fouriertransform}), using the theorem of residues:
\begin{equation}
\label{Fourierphi'}
\widetilde{\phi '}(k,\nu )=2\pi \displaystyle\frac{\sinh \displaystyle\frac{\pi }{2}\left( \displaystyle\frac{p+1-2\nu }{p+1}\right) k}{\sinh \displaystyle\frac{\pi }{2}k}
\end{equation}
This formula is the Fourier transform of \( \phi '(\vartheta ,\nu ) \) only
for \( \vartheta  \) in the fundamental strip (\ref{strisciafond}), because
out of this strip new singularities of (\ref{phi_primo}) appear in the computation
of Fourier transform.

\chapter{A lemma for UV computations\label{section:Lemma_di_Destri}}

In \cite{ddv 97} the following lemma has been proved. Assume that \( f(x) \)
satisfies the non-linear integral equation
\[
-i\log f(x)=\varphi (x)+2\, \Im m\, \int \frac{dy}{i}G(x-y-i\epsilon )\log (1+f(x+i\epsilon ))\]
where \( \varphi (x) \) is real on the real axis and \( G(x)=G(-x) \) is real
too, with bounded integral and peaked around the origin. From this equation
follows that \( f(x) \) has unit modulus on the real axis. To avoid crossing
of the branch cut of the logarithm, assume that 
\begin{equation}
\label{no_branch-cut}
\textrm{if }\qquad f(x+i\epsilon )\in \mathbb R\qquad \textrm{then}\qquad f(x+i\epsilon )>-1.
\end{equation}
 Then the following expression holds: 
\begin{equation}
\label{Destri's_lemma}
\begin{array}{c}
2\, \Im m\, \displaystyle\int \displaystyle\frac{dx}{i}\varphi '(x+i\epsilon )\log (1+f(x+i\epsilon ))=\\
\\
=-2\, \Re e\, \displaystyle\int _{\Gamma }\displaystyle\frac{du}{u}\log (1+u)-\displaystyle\frac{1}{2}\displaystyle\int dx\, G(x)\left( F^{2}(+\infty )-F^{2}(-\infty )\right) 
\end{array}
\end{equation}
where the curve \( \Gamma  \) is any path in the complex plane that goes from
\( f_{-}=f(-\infty +i\epsilon ) \) to \( f_{+}=f(+\infty +i\epsilon ) \) avoiding
the branch cut, i.e. respecting the condition (\ref{no_branch-cut}) and 
\[
F(x)=2\, \Im m\, \log (1+f(x+i\epsilon )).\]
In the 6-vertex case all the terms can be made manifest by the identification
of \( G \) with that defined in (\ref{funzioneG}), \( f(x)=(-1)^{\delta }e^{iZ_{\pm }(x)} \)
and \( \varphi (x) \) with (\ref{var-fi}). 

Consider the kink ``+''. The integral on the path \( \Gamma  \) goes from
the points
\[
\begin{array}{c}
f_{-}=f(-\infty +i\epsilon )=(-1)^{\delta }e^{iZ_{+}(-\infty )}=e^{i{\cal Q}_{+}(-\infty )}\\
f_{+}=f(+\infty +i\epsilon )=(-1)^{\delta }e^{iZ_{+}(+\infty +i\epsilon )}=0
\end{array}\]
(the first one can be computed with (\ref{Q+-infinito})). Observe that \( |f_{-}|=1 \)
then the path \( \Gamma  \) can be composed by an unit radius arc from \( f_{-} \)
to the point \( 1 \) and a segment from \( 1 \) to \( 0 \). On the arc the
integration variable has unit modulus then it is convenient to do the change
of variables \( u=e^{i\alpha } \). This simple computation gives
\[
-2\, \Re e\, \int _{arc}\frac{du}{u}\log (1+u)=-{\cal Q}^{2}_{+}(-\infty ).\]
The integration on the segment gives a well known dilogarithmic expression:
\[
-2\, \Re e\, \int ^{0}_{1}\frac{du}{u}\log (1+u)=2\, \int _{0}^{1}\frac{du}{u}\log (1+u)=\frac{\pi ^{2}}{6}.\]
The quantities \( F(\pm \infty ) \), from their definition, give:
\[
\begin{array}{c}
F(+\infty )=0\\
F(-\infty )=2\, \Im m\, \log (1+(-1)^{\delta }e^{iZ_{+}(-\infty )})={\cal Q}_{+}(-\infty ).
\end{array}\]
Remembering now that the integral of \( G \) appearing in (\ref{Destri's_lemma})
is expressed in (\ref{chi_infinito}) yields (the computation for the ``--''
kink is completely analogous):
\begin{equation}
\label{UV_integral}
\begin{array}{c}
2\, \Im m\, \displaystyle\int \displaystyle\frac{dx}{i}\varphi _{\pm }'(x+i\eta )\log \left( 1+(-1)^{\delta }e^{iZ_{\pm }(x+i\eta )}\right) =\\
\\
=\pm \displaystyle\frac{\pi ^{2}}{6}\mp \displaystyle\frac{{\cal Q}^{2}_{\pm }(\mp \infty )}{4}\displaystyle\frac{p+1}{p}
\end{array}
\end{equation}

\end{document}